\documentclass[11pt,letterpaper]{JHEP3}
\usepackage{amsmath,epsfig,array}
\usepackage{graphicx}
\usepackage{amssymb}
\usepackage{amsfonts,amstext,latexsym,xspace}
\usepackage{amsfonts}
\usepackage{amsmath}
\usepackage{cite}

\makeatletter
\newif\if@fewtab\@fewtabtrue
\def\moth{\mathsurround=0pt}
\newdimen\zo \zo=0pt

\def\tick{\leaders\hrule height 0.5ex depth 0pt \hskip 0.5pt}
\def\upboxfill{$\moth \setbox\zo\hbox{\tick}%
  \hskip 2pt\hbox to 0pt{$\tick$\hss}\hrulefill \hbox to 2pt{$\tick$\hss}$}

\def\dtick{\leaders\hrule height .34pt depth 0.5ex \hskip 0.5pt}
\def\downboxfill{$\moth \setbox\zo\hbox{\dtick}%
  \hskip 2pt\hbox to 0pt{$\dtick$\hss}\hrulefill%
  \hbox to 2pt{$\dtick$\hss}$}

\usepackage{array}
\usepackage{amsmath}
\usepackage{amssymb}
\usepackage{graphicx}
\usepackage{longtable}

\newcommand{\commute}[2]{\left[ #1 \, , \, #2 \right]}
\newcommand{\anticommute}[2]{\left\{ #1 \, , \, #2 \right\}}
\newcommand{\ket}[1]{\left\lvert #1 \right\rangle}
\newcommand{\eq}{\begin{equation}}
\newcommand{\en}{\end{equation}}
\newcommand{\eqq}{\begin{eqnarray}}
\newcommand{\enn}{\end{eqnarray}}

\newcommand{\stln}{\setlength{\unitlength}{10pt}}  
\newcommand{\lfr}{\framebox(2,1){}}  
\newcommand{\sfr}{\framebox(1,1)[bl]{\begin{picture}(1,1)(0,0)
                                      \put(0,0){\line(1,1){1}}
                                     \end{picture}
                                     }
                  }  
\newcommand{\dotfr}{\stln \lower2.0pt \hbox{\begin{picture}(2,1)(0,0)
                                                 \put(0,0){\lfr}
                                                 \put(0.5,0.5){\dots}
                                                \end{picture}
                                                }
                        } 
\newcommand{\sonebox}{\stln \lower2.0pt \hbox{\begin{picture}(1,1)(0,0)
                                               \put(0,0){\sfr}
                                              \end{picture}
                                              }
                      }
\newcommand{\stwobox}{\stln \lower2.0pt \hbox{\begin{picture}(2,1)(0,0)
                                               \multiput(0,0)(1,0){2}{\sfr}
                                              \end{picture}
                                              }
                      }
\newcommand{\sgenrowbox}{\stln \lower2.0pt \hbox{\begin{picture}(5,1)(0,0)
                                               \multiput(0,0)(1,0){2}{\sfr}
                                               \put(2,0.2){\dotfr}
                                               \put(4,0){\sfr}
                                              \end{picture}
                                              }
                      }

\title{Minimal unitary representation of $SO^*(8) = SO(6,2)$ and its $SU(2)$ deformations as massless $6D$ conformal fields and their supersymmetric extensions}
\author{
Sudarshan Fernando$^{1}$\footnote{fernando@kutztown.edu}  and
Murat G\"{u}naydin$^{2}$\footnote{murat@phys.psu.edu}
\\

$^{1}$\emph{Physical Sciences Department\\Kutztown University\\ Kutztown, PA 19530, USA}  \\

$^{2}$\emph{Center for Fundamental Theory \\ Institute for Gravitation and the Cosmos \\ Physics Department \\
Pennsylvania State University\\
University Park, PA 16802, USA} }

\abstract{ We study the minimal unitary representation (minrep) of $SO(6,2)$
over an Hilbert space of functions of five variables, obtained by quantizing
its quasiconformal realization. The minrep of $SO(6,2)$, which coincides with
the minrep of $SO^*(8)$ similarly constructed, corresponds to a massless
conformal scalar field in six spacetime dimensions. There exists a family of
``deformations'' of the minrep of $SO^*(8)$ labeled by the spin $t$ of an
$SU(2)_T$ subgroup of the little group $SO(4)$ of lightlike vectors. These
deformations labeled by $t$ are positive energy unitary irreducible
representations of $SO^*(8)$ that describe massless conformal fields in six
dimensions. The $SU(2)_T$ spin $t$ is the six dimensional counterpart of $U(1)$ deformations of the minrep of $4D$ conformal group $SU(2,2)$ labeled by helicity. We also construct the supersymmetric extensions of the
minimal unitary representation of $SO^*(8)$ to minimal unitary
representations  of $OSp(8^*|2N)$ that describe massless six dimensional
conformal supermultiplets. The minimal unitary supermultiplet of $OSp(8^*|4)$
is the massless supermultiplet of $(2,0)$ conformal field theory that is
believed to be dual to M-theory on $AdS_7 \times S^4$.}
\keywords{ AdS/CFT, Minimal Unitary Representations, Conformal Group}

\preprint{arXiv:1005.3580 [hep-th]}

\begin{document}


\section{Introduction}
\label{Intro}

Unitary representations of noncompact U-duality groups of
extended supergravity theories were first studied  in early 1980s
\cite{Gunaydin:1981dc,Gunaydin:1981yq,Gunaydin:1981zm}, motivated  by the idea that, in a quantum theory, global
symmetries must be realized unitarily,  as well as by attempts to derive a composite  grand unified theory (GUT) from  $N = 8$
supergravity  \cite{Ellis:1980tf,Ellis:1980cf,Gunaydin:1982gw}. In
the composite model of \cite{Ellis:1980tf},  the local R-symmetry group $SU(8)$
of $N = 8$ supergravity was conjectured to become dynamical at the quantum level.
A similar scenario based on the exceptional supergravity theory
\cite{Gunaydin:1983rk} leads to $E_6$
GUT with a family group $U(1)$. After the discovery of counter terms at higher
loops
in  $N=8$ supergravity and the
Green-Schwarz anomaly cancellation in superstring theory \cite{Green:1984sg},
the work on composite models was all but abandoned.
Recent work proving  cancellations of divergences in $N=8$ supergravity up to four loops  \cite{Bern:2008pv,BjerrumBohr:2008dp,ArkaniHamed:2008gz,Chalmers:2000ks,Green:2006gt,Green:2006yu,Green:2007zzb,Kallosh:2008ru,Bern:2009kd,Kallosh:2009jb} revived the question of finiteness of
$N=8$ supergravity as well as of exceptional supergravity.

The  oscillator method developed in  \cite{Gunaydin:1981yq}, to construct the relevant  unitary representations of noncompact U-duality groups of supergravity theories,  generalized and  unified  previous special  constructions  in the physics literature.   The formulation of  \cite{Gunaydin:1981yq} was later extended to  noncompact supergroups in \cite{Bars:1982ep} using bosonic as well as fermionic oscillators. In these generalized formulations of \cite{Gunaydin:1981yq} and \cite{Bars:1982ep}, one realizes the generators of noncompact groups or supergroups  as bilinears of an arbitrary number  $P$ (``colors'') of  sets of oscillators transforming in an irreducible representation (typically fundamental) of their maximal compact subgroups or subsupergroups.
  For symplectic groups $Sp(2n,\mathbb{R})$ the minimum  value of $P$ turns out to be one, and the resulting unitary representations are simply the singleton representations, which are known as metaplectic representations in the mathematics literature.  In general  the minimum allowed value of $P$ for a noncompact group is two, and the resulting unitary representations were later referred to as doubleton representations. For example, for the groups $SU(n,m)$ and $SO^*(2n)$, with maximal compact subgroups $SU(m)\times SU(n) \times U(1)$ and $U(n)$, respectively, one finds that $P_{min}=2$.  Symplectic groups  $Sp(2n,\mathbb{R})$ admit only two singleton irreducible representations (irreps).   Noncompact groups or supergroups that do not admit singleton representations  have  an infinite number of doubleton irreps. Since the generators are realized as  bilinears of free bosonic and fermionic oscillators,  tensoring of the resulting representations is very straightforward within the oscillator approach. Furthermore the oscillator method is simple and yet very powerful for constructing positive energy unitary representations.
  Even though the positive energy singleton or doubleton irreps  do not belong to the discrete series, by tensoring them one obtains positive energy unitary representations that belong, in general, to the holomorphic discrete series representations  of the respective noncompact group or supergroup.

The oscillator methods  for constructing positive energy unitary representations of non-compact groups and supergroups were applied to spacetime supergroups  beginning in the 1980s.
The Kaluza-Klein spectrum of IIB supergravity spontaneously compactified over the product space  $AdS_5 \times S^5$  was first  obtained via the oscillator method by simple tensoring of the CPT self-conjugate doubleton supermultiplet of  $N=8$ $AdS_5$ superalgebra $PSU(2,2\,|\,4)$ repeatedly with itself and restricting to the CPT self-conjugate short supermultiplets of $PSU(2,2\,|\,4)$ \cite{Gunaydin:1984fk}. The CPT self-conjugate doubleton supermultiplet does not have a Poincar\'e limit in five dimensions and decouples from the Kaluza-Klein spectrum as gauge modes. This led the authors of \cite{Gunaydin:1984fk} to  the proposal that the  field theory of CPT self-conjugate doubleton supermultiplet  of $PSU(2,2\,|\,4)$  lives on the boundary of $AdS_5$, which can be identified with $4D$ Minkowski space on which $SO(4,2)$ acts as a conformal group. Furthermore they pointed out  that the unique candidate for this theory is the  four dimensional $N=4$ super Yang-Mills theory that was known to be conformally invariant.

The spectra of the spontaneous compactifications of eleven dimensional supergravity over $AdS_4 \times S^7$ and $AdS_7 \times S^4$, that had been obtained by other methods previously, were fitted into supermultiplets of the  symmetry superalgebras $OSp(8\,|\,4,\mathbb{R})$ and $OSp(8^*\,|\,4)$ obtained by oscillator methods in \cite{Gunaydin:1985tc} and \cite{Gunaydin:1984wc}, respectively. Furthermore, the entire Kaluza-Klein spectra of eleven dimensional supergravity over these two spaces were obtained by tensoring the singleton and scalar doubleton supermultiplets of $OSp(8\,|\,4,\mathbb{R})$ and $OSp(8^*\,|\,4)$, respectively.
The singleton and doubleton supermultiplets themselves do not have a Poincar\'e limit in four and seven dimensions  and decouple from the respective spectra as gauge modes. Again it was proposed that the field theories of the singleton and scalar doubleton supermutiplets live on the boundaries of $AdS_4$ and $AdS_7$ as super conformally invariant theories \cite{Gunaydin:1984wc,Gunaydin:1985tc}.

 The importance of these results was not fully realized until the work of Maldacena\cite{Maldacena:1997re} and subsequent works  of  Witten \cite{Witten:1998qj}  and of Gubser et al. \cite{Gubser:1998bc}  and have since  become an integral  part of the work on AdS/CFT dualities in M/superstring theory which has seen an exponential growth for over more than a decade now.

 Noncompact groups were also introduced into  physics  as spectrum generating symmetry groups during the 1960s. Inspired by the work of  physicists on spectrum generating symmetry groups, Joseph introduced the  concept of minimal unitary realizations of Lie groups  in \cite{MR0342049}.
These are   unitary
representations of  corresponding noncompact groups over  Hilbert spaces
of functions of  smallest possible (minimal) number of variables.  Joseph gave the
minimal realizations of the complex forms of classical Lie algebras
and of the exceptional Lie algebra $\mathfrak{g}_2$ in a Cartan-Weil basis.   The
  minimal unitary representation of the split exceptional group $E_{8(8)}$ was first identified within  Langland's classification  by Vogan
\cite{MR644845}. In an important paper, Kostant studied  the minimal unitary representation of $SO(4,4)$ and its relation to triality in    \cite{MR1103588}.
A general study of  minimal unitary representations of
simply laced groups   was given  by Kazhdan and Savin\cite{MR1159103}
and by Brylinski and Kostant
\cite{MR1372999,MR1278630}.

The minimal unitary representations of quaternionic real forms of exceptional Lie
groups were  studied by Gross and Wallach \cite{MR1327538} and those
of $SO(p,q)$  in \cite{MR1108044,MR2020550,MR2020551,MR2020552}.
Pioline, Kazhdan and Waldron \cite{Kazhdan:2001nx}
reformulated  the minimal unitary representations of simply laced
groups given in \cite{MR1159103} and  gave
the spherical vectors for the simply laced exceptional groups necessary for the
construction of modular forms. The relation of minimal representations of $SO(p,q)$ to conformal geometry was studied rather recently  in \cite{Gover:2009vc}.

Over the last decade, a great deal of progress was made towards the goal of constructing physically relevant unitary representations  of U-duality groups of extended supergravity theories. An additional motivation towards this goal was provided by the proposals that certain extensions of U-duality groups may act as spectrum generating symmetry groups of these theories.
Work on
orbits of extremal black hole solutions in $N=8$ supergravity and
$N=2$ Maxwell-Einstein supergravity theories with symmetric scalar
manifolds led to the proposal that four dimensional U-duality groups act
as spectrum generating conformal symmetry groups of the corresponding
five dimensional supergravity theories
\cite{Ferrara:1997uz,Gunaydin:2000xr,Gunaydin:2004ku,Gunaydin:2003qm,Gunaydin:2005gd,Gunaydin:2009pk}.   In attempts to find the corresponding spectrum generating symmetry groups of  extremal black hole solutions of four dimensional supergravity theories with symmetric scalar manifolds, geometric quasiconformal realizations  of three dimensional U-duality groups were discovered in \cite{Gunaydin:2000xr}. Based on this novel geometric realization, quasiconformal extensions of four dimensional U-duality groups were proposed   as spectrum generating symmetry groups of the
corresponding supergravity theories with symmetric scalar manifolds
\cite{Gunaydin:2000xr,Gunaydin:2004ku,Gunaydin:2003qm,Gunaydin:2005gd,Gunaydin:2009pk}. A concrete  implementation of the proposal that three
dimensional U-duality groups act as spectrum generating quasiconformal
groups was given in
\cite{Gunaydin:2005mx,Gunaydin:2007bg,Gunaydin:2007qq} using the equivalence of equations of attractor flows of spherically
symmetric stationary BPS black holes of four dimensional supergravity
theories and the geodesic equations of a fiducial particle moving in the target space of the three dimensional supergravity theories obtained by reduction of the $4D$ theories  on a timelike circle \cite{Breitenlohner:1987dg}.

Quasiconformal
realization of three dimensional  U-duality group $E_{8(8)}$ of maximal supergravity
in three dimensions is the first known geometric realization of
$E_{8(8)}$\cite{Gunaydin:2000xr}.  Quasiconformal action of $E_{8(8)}$  leaves invariant
a generalized light-cone with respect to a quartic distance function  in 57
dimensions. Quasiconformal realizations exist for various  real forms of all noncompact groups as well as for their complex forms  \cite{Gunaydin:2000xr,Gunaydin:2005zz}.
Remarkably, the quantization of  geometric quasiconformal action of a noncompact group leads directly to its minimal unitary representation as  was first shown  explicitly for the  maximally split
exceptional group  $E_{8(8)}$ with the maximal compact subgroup $SO(16)$
\cite{Gunaydin:2001bt}.

 The minimal unitary representation  of  three dimensional U-duality group $E_{8(-24)}$  of
the exceptional supergravity \cite{Gunaydin:1983rk}
  was given in \cite{Gunaydin:2004md}.

The minimal unitary representations of  U-duality groups $\mathrm{F}_{4(4)}$, $\mathrm{E}_{6(2)}$,
 $\mathrm{E}_{7(-5)}$ ,  $\mathrm{E}_{8(-24)}$ and $SO(d+2,4)$ of $N=2$ Maxwell-Einstein supergravity theories with symmetric scalar manifolds were studied in
\cite{Gunaydin:2005zz,Gunaydin:2004md}.
In \cite{Gunaydin:2006vz},   a unified
formulation of the minimal unitary representations of certain noncompact real forms of
 groups  of type $A_2$, $G_2$, $D_4$, $F_4$, $E_6$, $E_7$,
$E_8$ and $C_n$ was given. The minimal unitary
representations of $Sp\left(2n,\mathbb{R}\right)$ are simply the
singleton representations. In \cite{Gunaydin:2006vz},  minimal unitary representations of
 noncompact groups $SU\left(m,n\right)$, $SO\left(m,n\right)$,
$SO^*(2n)$ and $SL\left(m,\mathbb{R}\right)$ obtained by quasiconformal methods  were also given
explicitly.
Furthermore, this unified  approach was generalized to define and construct the corresponding
minimal representations of non-compact supergroups $G$ whose even
subgroups are of the form $H\times SL(2,\mathbb{R})$ with $H$
compact.
The unified construction with $H$
simple or Abelian leads to the minimal unitary representations of supergroups  $G(3),
F(4)$ and $OSp\left(n|2,\mathbb{R}\right)$. The minimal unitary representations of
$OSp\left(n|2,\mathbb{R}\right)$ with  even subgroups $SO(n)\times
Sp(2,\mathbb{R})$ are the singleton supermultiplets. The minimal realization of the one parameter family of Lie superalgebras $D\left(2,1;\sigma\right)$ with even subgroup $ SU(2)\times SU(2) \times SU(1,1)$  was also presented in \cite{Gunaydin:2006vz}.

In mathematics literature, the term minimal unitary representation refers, in general, to a unique  representation of the respective noncompact group. The symplectic group $Sp(2N,\mathbb{R})$ admits two singleton irreps whose quadratic Casimirs take on the same value. Both of these singleton representations are minimal unitary representations, even though in some of the mathematics literature only the scalar singleton is referred to as the minrep.   Similarly one finds that the supergroups $OSp(M|2N,\mathbb{R}) $ with the even subgroup $SO(M) \times Sp(2N,\mathbb{R})$ admit two inequivalent singleton supermultiplets \cite{Gunaydin:1985tc,Gunaydin:1988kz,Gunaydin:1987hb}.
For noncompact groups or supergroups that admit only doubleton irreps, this raises the question as to whether any of the doubleton unitary representations can be identified with the  minimal representation, and if so, how the infinite set of doubletons are related to the minrep.
More recently, we investigated this issue for $5D$ anti-de Sitter  or $4D$ conformal group $SU(2,2)$ and  corresponding supergroups $SU(2,2|N)$ \cite{Fernando:2009fq}. We
gave a detailed study of the minimal unitary representation of the group $SU(2,2)$  by quantization of its quasiconformal realization and showed that it coincides with the scalar doubleton representation corresponding to a massless scalar field in four dimensions. Furthermore we showed that the minrep of $SU(2,2)$ admits a one-parameter family ($\zeta$) of deformations, and for a positive
(negative)  integer value of the deformation parameter $\zeta$,  one obtains
a positive energy
unitary irreducible representation of  $SU(2,2)$ corresponding to a
massless conformal field in four dimensions  transforming in  $\left( 0
\,,\, \frac{\zeta}{2} \right)$ $\left( \left( -\frac{\zeta}{2} \,,\, 0
\right) \right)$ representation of the Lorentz subgroup, $SL(2,\mathbb{C})$ of $SU(2,2)$. These are simply the doubleton representations of $SU(2,2)$  that describe massless conformal fields in four dimensions \cite{Gunaydin:1998sw,Gunaydin:1998jc}. They were referred to as ladder (or most degenerate discrete series) unitary representations by Mack and Todorov, who showed that they remain irreducible under restriction to the Poincar\'{e} subgroup \cite{Mack:1969dg}. Hence the deformation parameter can be identified with twice the helicity $h$ of the corresponding massless representation of the Poincar\'e group.
We  extended  these results to the minimal unitary representations of supergroups $SU(2,2\,|\,N)$ with the even subgroup $SU(2,2)\times U(N)$  and their deformations.  The minimal unitary supermultiplet of $SU(2,2|N)$ coincides with the CPT self-conjugate (scalar) doubleton supermultiplet, and for $PSU(2,2|4)$ it is simply the four dimensional $N=4$  Yang-Mills supermultiplet.
 Again in the supersymmetric case, one finds a one-parameter family of deformations of the minimal unitary supermultiplet of $SU(2,2|N)$.  Each integer value of the  deformation parameter
$\zeta$ leads to  a unique unitary supermultiplet of $SU(2,2\,|\,N)$.    The minimal unitary supermultiplet of $SU(2,2\,|\,N)$ and its deformations turn out to be precisely the doubleton supermultiplets that were constructed and studied using the oscillator method earlier \cite{Gunaydin:1984fk,Gunaydin:1998sw,Gunaydin:1998jc}.  These  results extend to the minreps of $SU(m,n)$ and of $SU(m,n\,|N)$ and their deformations in a straightforward manner.

In this paper we give a detailed study of the minimal unitary representation of  $7D$ anti-de Sitter or $6D$ conformal group $SO^*(8)=SO(6,2)$, obtained by quantizing its realization as a quasiconformal group that leaves invariant a quartic light-cone in nine dimensions, its deformations and their
supersymmetric extensions to supermultiplets of $OSp(8^*|2N)$.
The oscillator construction of the positive energy unitary supermultiplets of $OSp(8^*|2N)$ were first given in \cite{Gunaydin:1984wc}. These unitary supermultiplets were further studied in \cite{Gunaydin:1999ci,Fernando:2001ak} where it was shown that the doubleton supermultiplets correspond to massless conformal supermultiplets in six dimensions.  A classification of the positive energy unitary supermultiplets of $6D$ superconformal algebras using other methods was given in \cite{Minwalla:1997ka,Dobrev:2002dt}. The oscillator construction of positive energy representations of general supergroups $OSp(2M^*|2N) $ with maximal compact subgroup $SO^*(2M)\times USp(2N)$ was given in \cite{Gunaydin:1990ag}.

The plan of our paper is as follows. In section \ref{quasiconf}  we  review
the geometric quasiconformal realizations of  groups $SO(d+2,2)$  as
 invariance groups of a light-cone with respect to a quartic distance
function in $(2d+1)$ dimensional space. The quantization of this geometric realization leads to the  minimal unitary representation
of $SO(d+2,2)$
 over an Hilbert space of functions in $d+1$ variables.  We then specialize and study the case of $SO(6,2)$ in great detail in section \ref{minrepSO(6,2)}.
 In section \ref{minrepSO*(8)}, we study the minimal unitary
realization of $SO^*(8)$  which is isomorphic to $SO(6,2)$. The transformations relating the $SO^*(8)$ basis to that of $SO(6,2)$ is given in Appendix \ref{app:bogoliubov}. Section \ref{SU(1,1)ofSO*(8)} discusses the properties of a distinguished $SU(1,1)$ subgroup of $SO^*(8)$ generated by singular (isotonic) oscillators.
  We then give the K-type decomposition\footnote{K-type decomposition is the decomposition with respect to the maximal compact subgroup.} of the minrep of $SO^*(8)$
in section \ref{SU2SU2U1} and show that it coincides with the K-type
decomposition of scalar doubleton representation
corresponding to a massless conformal scalar field in six dimensions that were studied in \cite{Gunaydin:1984wc,Gunaydin:1999ci,Fernando:2001ak}.
Section \ref{USp(2N)} reviews the fermionic construction of relevant representations of $USp(2N)$.

In section \ref{minrepOSp(8*|2N)-5Gr} and section \ref{minrepOSp(8*|2N)-3Gr}, we give the minimal unitary realization of the superalgebra $OSp(8^*|2N)$ with even subgroup $SO^*(8) \times USp(2N)$ obtained from quantizing its quasiconformal realization. Section \ref{minrepsupermultiplet} presents the minimal unitary supermultiplets of $OSp(8^*|2N)$. We devote a special subsection to the minimal unitary supermultiplet of $OSp(8^*|4)$, which is the symmetry supergroup of M-theory compactified  over $AdS_7 \times S^4$ and show that it coincides with the $(2,0)$ doubleton supermultiplet studied in \cite{Gunaydin:1984wc,Gunaydin:1999ci,Fernando:2001ak}. M-theory compactified
over $AdS_7\times S^4$ is believed to be dual to a $(2,0)$ six dimensional superconformal theory based on this supermultiplet.

We then study the general deformations of
  the minrep of $SO^*(8)$, independently of supersymmetry, in section \ref{SO*(8)deformations}, and show that there exist an infinite family of deformations labeled by the spin $t$ of an $SU(2)_{\buildrel _\circ \over {T}}$ subgroup of the semi-simple part of the little group $SO(4)$ of massless states in six dimensions.
 For every spin value $t$, one obtains
a positive energy
unitary irreducible representation of  $SO^*(8)$ corresponding to a
massless conformal field in six dimensions with Dynkin labels  $(2t,0,0)$ with respect to the covering group $SU^*(4)$ of the six dimensional Lorentz group $SO(5,1)$. The $SU(2)$ spin label $t$ for deformations is the $6D$ analog of the helicity label for deformations of the minrep of $4D$ conformal group\cite{Fernando:2009fq}.

\section{Quasiconformal Realizations of $SO\left(d+2,2\right)$ and Their Minimal Unitary Representations}
\label{quasiconf}

\renewcommand{\theequation}{\arabic{section}.\arabic{equation}}
\setcounter{equation}{0}


\subsection{Geometric realizations of $SO\left(d+2,2\right)$ as
quasiconformal groups}
\label{geomSO(d+2,2)}

Lie algebra of the $(d+2)$ dimensional conformal group $SO\left(d+2,2\right)$
can be given a 5-graded decomposition with respect to its subalgebra
$\mathfrak{so}(d) \oplus \mathfrak{so}(1,1)$ \cite{Gunaydin:2005zz}
\begin{equation}
\mathfrak{so}\left(d+2,2\right)
= \mathbf{1}^{(-2)} \oplus
  \left(\mathbf{d}, \mathbf{2} \right)^{(-1)} \oplus
  \left[
   \Delta \oplus \mathfrak{sp}\left(2,\mathbb{R}\right) \oplus
   \mathfrak{so}\left(d\right)
  \right] \oplus
  \left(\mathbf{d}, \mathbf{2} \right)^{(+1)} \oplus
  \mathbf{1}^{(+2)}
\end{equation}
where $\Delta$ is the $SO(1,1)$ generator that determines the five grading.
The superscript  $m$ labels the grade of a generator:
\begin{equation}
\commute{\Delta}{\mathfrak{g}^{(m)}} = m \, \mathfrak{g}^{(m)}
\end{equation}
In the above decomposition, $\left(\mathbf{d}, \mathbf{2} \right)^{(m)}$
labels the generators transforming in the $(d,2)$ representation of $SO(d)
\times Sp(2,\mathbb{R})$ with grade $m$. Generators of quasiconformal action
are realized as differential operators acting on a $(2d+1)$ dimensional space
$\mathcal{T}$ corresponding to the Heisenberg subalgebra generated by the
elements of $\mathfrak{g}^{(-2)} \oplus \mathfrak{g}^{(-1)}$ subspace. We
shall denote the coordinates of the space $\mathcal{T}$ as $\mathcal{X} =
\left( X^{i,\alpha} , x \right)$, where $X^{i,\alpha}$ transform in the
$(d,2)$ representation of $SO(d) \times Sp(2,\mathbb{R})$, with $i =
1,2,\dots,d$ and $\alpha = 1,2$, and $x$ is a singlet coordinate.

Let $\epsilon_{\alpha\beta}$ be the symplectic metric of $Sp(2,\mathbb{R})$
and $\eta_{ij}$ the $SO(d)$ invariant metric ($\eta_{ij} = - \delta_{ij}$).
Then the quartic polynomial in $X^{i,\alpha}$
\begin{equation}
\mathcal{I}_4 (X)
= \eta_{ij} \eta_{kl} \epsilon_{\alpha\gamma} \epsilon_{\beta\delta}
  X^{i,\alpha} X^{j,\beta} X^{k,\gamma} X^{l,\delta}
\end{equation}
is invariant under $SO(d) \times Sp(2,\mathbb{R})$ subgroup.

We shall label the generators belonging to various grade subspaces as follows
\begin{equation}
\mathfrak{so}(d+2,2)
= K_- \oplus U_{i,\alpha} \oplus
  \left[ \Delta \oplus J_{\alpha\beta} \oplus M_{ij} \right] \oplus
  \widetilde{U}_{i,\alpha} \oplus K_+
\end{equation}
where $J_{\alpha\beta}$ and $M_{ij}$ are the generators of $Sp(2,\mathbb{R})$
and $SO(d)$ subgroups, respectively. The infinitesimal generators of the
quasiconformal action of $SO(d+2,2)$ take on the form
\begin{equation}
\begin{split}
K_+
&= \frac{1}{2} \left( 2 x^2 - \mathcal{I}_4 \right)
   \frac{\partial}{\partial x}
   - \frac{1}{4} \frac{\partial \mathcal{I}_4}{\partial X^{i,\alpha}}
     \eta^{ij} \epsilon^{\alpha\beta} \frac{\partial}{\partial X^{j,\beta}}
   +  x \, X^{i,\alpha} \frac{\partial}{\partial X^{i,\alpha}}
\\
U_{i,\alpha}
&= \frac{\partial}{\partial X^{i,\alpha}}
   - \eta_{ij} \epsilon_{\alpha\beta} X^{i,\beta}
     \frac{\partial}{\partial x}
\\
M_{ij}
&= \eta_{ik} X^{k,\alpha} \frac{\partial}{\partial X^{j,\alpha}}
   - \eta_{jk} X^{k,\alpha} \frac{\partial}{\partial X^{i,\alpha}}
\\
J_{\alpha\beta}
&= \epsilon_{\alpha\gamma} X^{i,\gamma}
   \frac{\partial}{\partial X^{i,\beta}}
   + \epsilon_{\beta\gamma} X^{i,\gamma}
     \frac{\partial}{\partial X^{i,\alpha}}
\\
K_-
&= \frac{\partial}{\partial x}
\\
\Delta
& = 2 \, x \frac{\partial}{\partial x}
    + X^{i,\alpha} \frac{\partial}{\partial X^{i,\alpha}}
\\
\widetilde{U}_{i,\alpha}
&= \commute{U_{i,\alpha}}{K_+}
\end{split}
\end{equation}
where $\epsilon^{\alpha\beta}$ is the inverse symplectic metric such that
$\epsilon^{\alpha\beta} \epsilon_{\beta\gamma} = \delta^\alpha_{~\gamma}$.
Using
\begin{equation}
\frac{\partial \mathcal{I}_4}{\partial X^{i,\alpha}}
= - 4 \, \eta_{ij} \, \eta_{kl} \,
  X^{j,\beta} X^{k,\gamma} X^{l,\delta} \,
  \epsilon_{\beta\gamma} \epsilon_{\alpha\delta}
\end{equation}
one obtains the explicit form of grade $+1$ generators
$\widetilde{U}^{i,\alpha}$:
\begin{equation}
\begin{split}
\widetilde{U}_{i,\alpha}
&= \eta_{ij} \epsilon_{\alpha\delta}
   \left( \eta_{kl} \epsilon_{\beta\gamma}
          X^{j,\beta} X^{k,\gamma} X^{l,\delta}
          - x X^{j,\delta}
   \right)
   \frac{\partial}{\partial x}
   +  x \frac{\partial}{\partial X^{i,\alpha}}
\\
& \quad
   - \eta_{ij} \epsilon_{\alpha\beta} X^{j,\beta} X^{k,\gamma}
     \frac{\partial}{\partial X^{k,\gamma}}
   - \eta_{jk} \epsilon_{\alpha\delta} X^{k,\delta} X^{j,\gamma}
     \frac{\partial}{\partial X^{i,\gamma}}
\\
& \quad
   + \eta_{ij} \epsilon_{\alpha\gamma} X^{k,\gamma} X^{j,\beta}
     \frac{\partial}{\partial X^{k,\beta}}
   + \eta_{ij} \epsilon_{\beta\gamma} X^{j,\beta} X^{k,\gamma}
     \frac{\partial}{\partial X^{k,\alpha}}
\end{split}
\end{equation}

The above generators satisfy the following commutation relations:
\begin{subequations}
\begin{equation}
\begin{split}
\commute{M_{ij}}{M_{kl}}
&= \eta_{jk} \, M_{il} - \eta_{ik} \, M_{jl}
   - \eta_{jl} \, M_{ik} + \eta_{il} \, M_{jk}
\\
\commute{J_{\alpha\beta}}{J_{\gamma\delta}}
&= \epsilon_{\gamma\beta} \, J_{\alpha\delta}
   + \epsilon_{\gamma\alpha} \, J_{\beta\delta}
   + \epsilon_{\delta\beta} \, J_{\alpha\gamma}
   + \epsilon_{\delta\alpha} \, J_{\beta\gamma}
\end{split}
\end{equation}
\begin{equation}
\begin{split}
\commute{\Delta}{K_\pm}
&= \pm 2 \, K_\pm
\qquad \qquad \qquad
\commute{K_-}{K_+}
= \Delta
\\
\commute{\Delta}{U_{i,\alpha}}
&= - U_{i,\alpha}
\qquad \qquad \qquad
\commute{\Delta}{\widetilde{U}_{i,\alpha}}
= \widetilde{U}_{i,\alpha}
\\
\commute{U_{i,\alpha}}{K_+}
&= \widetilde{U}_{i,\alpha}
\qquad \qquad \qquad \quad
\commute{\widetilde{U}_{i,\alpha}}{K_-}
= - U_{i,\alpha}
\\
\commute{U_{i,\alpha}}{U_{j,\beta}}
&= 2 \, \eta_{ij} \epsilon_{\alpha\beta} \, K_-
\qquad \qquad
\commute{\widetilde{U}_{i,\alpha}}{\widetilde{U}_{j,\beta}}
= 2 \, \eta_{ij} \epsilon_{\alpha\beta} \, K_+
\end{split}
\end{equation}
\begin{equation}
\begin{split}
\commute{M_{ij}}{U_{k,\alpha}}
&= \eta_{jk} \, U_{i,\alpha} - \eta_{ik} \, U_{j,\alpha}
\qquad \qquad
\commute{M_{ij}}{\widetilde{U}_{k,\alpha}}
= \eta_{jk} \, \widetilde{U}_{i,\alpha}
  - \eta_{ik} \, \widetilde{U}_{j,\alpha}
\\
\commute{J_{\alpha\beta}}{U_{i,\gamma}}
&= \epsilon_{\gamma\beta} \, U_{i,\alpha}
   + \epsilon_{\gamma\alpha} \, U_{i,\beta}
\qquad \qquad
\commute{J_{\alpha\beta}}{\widetilde{U}_{i,\gamma}}
 = \epsilon_{\gamma\beta} \, \widetilde{U}_{i,\alpha}
   + \epsilon_{\gamma\alpha} \, \widetilde{U}_{i,\beta}
\\
\end{split}
\end{equation}
\begin{equation}
\commute{U_{i,\alpha}}{\widetilde{U}_{j,\beta}}
 = \eta_{ij} \epsilon_{\alpha\beta} \, \Delta
   - 2 \, \epsilon_{\alpha\beta} \, M_{ij} - \eta_{ij} \, J_{\alpha\beta}
\end{equation}
\end{subequations}

One defines the quartic norm $\mathcal{N}_4 (\mathcal{X})$ of a vector
$\mathcal{X}= \left( X^{i,\alpha} , x \right)$ in $\mathcal{T}$ as
\begin{equation}
\mathcal{N}_4 \left(\mathcal{X} \right)
:= \mathcal{I}_4\left(X\right) + 2 \, x^2
\end{equation}
and the ``quartic distance'' between any two points with coordinate vectors
$\mathcal{X}$ and $\mathcal{Y}$ as
\begin{equation}
d \left( \mathcal{X} , \mathcal{Y} \right)
:= \mathcal{N}_4
   \left( \delta \left( \mathcal{X} , \mathcal{Y} \right) \right)
\end{equation}
where $\delta \left( \mathcal{X} , \mathcal{Y} \right)$ is the ``symplectic''
difference of two vectors $\mathcal{X}$ and $\mathcal{Y}$ in the $(2d+1)$
dimensional space $\mathcal{T}$ given by
\cite{Gunaydin:2000xr,Gunaydin:2005zz}
\begin{equation}
\delta \left( \mathcal{X} , \mathcal{Y} \right)
:= \left(
    X^{i,\alpha} - Y^{i,\alpha} \,,\,
    x - y - \eta_{ij} \epsilon_{\alpha\beta} \, X^{i,\alpha} Y^{j,\beta}
   \right) \,.
\end{equation}

Under the quasiconformal action of the generators of $SO(d+2,2)$ the quartic
distance function transforms as:
\begin{equation}
\begin{split}
\Delta  d \left( \mathcal{X} , \mathcal{Y} \right)
&= 4 \, d \left( \mathcal{X} , \mathcal{Y} \right)
\\
\widetilde{U}_{i,\alpha} d \left( \mathcal{X} , \mathcal{Y} \right)
&= - 2 \, \eta_{ij} \epsilon_{\alpha\beta}
   \left( X^{j,\beta} + Y^{j,\beta} \right)
   d \left( \mathcal{X} , \mathcal{Y} \right)
\\
K_+ d \left( \mathcal{X} , \mathcal{Y} \right)
&= 2 \, \left( x + y \right) \,
   d \left( \mathcal{X} , \mathcal{Y} \right)
\\
M_{ij} \, d \left( \mathcal{X} , \mathcal{Y} \right)
&= 0
\\
J_{\alpha\beta} \, d \left( \mathcal{X} , \mathcal{Y} \right)
&= 0
\\
U_{i,\alpha} \, d \left( \mathcal{X} , \mathcal{Y} \right)
&= 0
\\
K_- \, d \left( \mathcal{X} , \mathcal{Y} \right)
&= 0
\end{split}
\end{equation}
They imply that light-like separations
\[ d \left( \mathcal{X} , \mathcal{Y} \right) = 0 \]
are left invariant under the quasiconformal action. In other words, the
quasiconformal action of $SO(d+2,2)$ leaves the ``light-cone'' in
$\mathcal{T}$ with respect to the {\it quartic} distance function invariant.


\subsection{Minimal unitary representations of $SO\left(d+2,2\right)$ from
quantization of their quasiconformal realizations}
\label{minrepSO(d+2,2)}

Minimal unitary representations of noncompact groups can be obtained by the
quantization of their geometric realizations as quasiconformal groups \cite{Gunaydin:2001bt,Gunaydin:2004md,Gunaydin:2005zz,Gunaydin:2006vz,Gunaydin:2007qq}.
In this section we shall review the minimal unitary representations of
orthogonal groups $SO(d+2,2)$ thus obtained following
\cite{Gunaydin:2005zz,Gunaydin:2006vz}. Let $X^i$ and $P_i$ be the quantum
mechanical coordinate and momentum operators on $\mathbb{R}^{(d)}$ satisfying
the canonical commutation relations
\begin{equation}
\commute{X^i}{P_j} = i \, {\delta^i_j} \,.
\end{equation}
The generators belonging to the subspace  $\mathfrak{g}^{(-2)} \oplus
\mathfrak{g}^{(-1)}$ of the Lie algebra of $SO(d+2,2)$ form an Heisenberg
algebra
\begin{equation}
\commute{U_{i,\alpha}}{U_{j,\beta}}
= 2 \, \eta_{ij} \epsilon_{\alpha\beta} \, K_-
\label{heisenberg}
\end{equation}
with $K_-$ playing the role of the central charge. We shall relabel the
generators and define
\begin{equation}
U_{i,1} \equiv U_{i}
\qquad \qquad \qquad
U_{i,2} \equiv V_{i}
\end{equation}
and realize the Heisenberg algebra (equation (\ref{heisenberg})) in terms of
coordinate and momentum operators $X^i$, $P_i$ and an extra ``central charge
coordinate'' $x$ as
\begin{equation}
\begin{split}
U_i = x P_i
& \qquad \qquad
V^i = x X^i
\\
&K_- = \frac{1}{2} x^2
\end{split}
\end{equation}

\begin{equation}
\commute{V^i}{U_j} = 2 i \, \delta^i_j \, K_-
\label{heisenberg2}
\end{equation}

By introducing the quantum mechanical momentum operator $p$, conjugate to the
central charge coordinate $x$, such that
\begin{equation}
\commute{x}{p} = i
\end{equation}
one can realize the generators of $SO(d) \times Sp(2,\mathbb{R}) \times
SO(1,1)$ subgroup belonging to the grade zero subalgebra of
$\mathfrak{so}(d+2,2)$ as bilinears of canonically conjugate pairs of
coordinate and momentum operators \cite{Gunaydin:2005zz,Gunaydin:2006vz}:
\begin{equation}
\begin{split}
M_{ij}
&= - i \, \delta_{ik} X^k P_j + i \, \delta_{jk} X^k P_i
\\
J_0
&= \frac{1}{2} \left( X^i P_i + P_i X^i \right)
\\
J_-
&= - \delta_{ij} X^i X^j
\\
J_+
&= - \delta^{ij} P_i P_j
\\
\Delta
&= \frac{1}{2} \left( x p + p x \right)
\end{split}
\label{gradezeroSO(d+2,2)}
\end{equation}
The generators $M_{ij}$ of $SO(d)$ satisfy the commutation relations
\begin{equation}
\commute{M_{ij}}{M_{kl}}
= - \delta_{jk} \, M_{il} + \delta_{ik} \, M_{jl}
  + \delta_{jl} \, M_{ik} - \delta_{il} \, M_{jk}
\end{equation}
and the generators $J_0$ and $J_{\pm}$ of $Sp(2,\mathbb{R})$ satisfy
\begin{equation}
\commute{J_0}{J_\pm} = \pm 2 i \, J_\pm
\qquad \qquad \qquad
\commute{J_-}{J_+} = 4 i \, J_0 \,.
\end{equation}
Note that the compact generator of this $Sp(2,\mathbb{R})$ is $\left( J_+ + 
J_- \right)$.

The coordinate $X^i$ and momentum $P_i$ operators transform in the vector
representation of $SO(d)$ subgroup generated by $M_{ij}$ and form doublets of
the symplectic group $Sp(2,\mathbb{R}$):
\begin{equation}
\begin{aligned}
\commute{J_0}{V^i}
&= - i \, V^i
\\
\commute{J_0}{U_i}
&= + i \, U_i
\end{aligned}
\qquad
\begin{aligned}
\commute{J_-}{V^i}
&= 0
\\
\commute{J_-}{U_i}
&= - 2 i \, \delta_{ij} \, V^j
\end{aligned}
\qquad
\begin{aligned}
\commute{J_+}{V^i}
&= + 2 i \, \delta^{ij} \, U_j
\\
\commute{J_+}{U_i}
&= 0
\end{aligned}
\end{equation}

There is a normal ordering ambiguity in defining the quantum operator
corresponding to the quartic invariant. We shall choose the quantum quartic
invariant given in \cite{Gunaydin:2005zz}:
\begin{equation}
\begin{split}
\mathcal{I}_4
&= \left( \delta_{ij} X^i X^j \right)
   \left( \delta^{ij} P_i P_j \right)
   + \left( \delta^{ij} P_i P_j \right)
     \left( \delta_{ij} X^i X^j \right)
\\
& \qquad
   - \left( X^i P_i \right) \left( P_j X^j \right)
   - \left( P_i X^i \right) \left( X^j P_j \right)
\end{split}
\end{equation}
In terms of the quartic invariant, the grade +2 generator $K_+$ of
$SO(d+2,2)$ takes the form
\begin{equation}
K_+ = \frac{1}{2} p^2
      + \frac{1}{4 \, x^2} \left( \mathcal{I}_4 + \frac{d^2+3}{2} \right) \,.
\end{equation}
Then grade $+1$ generators are obtained by the commutation of grade $-1$
generators with $K_+$:
\begin{equation}
\widetilde{U}_i = - i \commute{U_i}{K_+}
\qquad \qquad
\widetilde{V}^i = - i \commute{V^i}{K_+}
\end{equation}
which explicitly read as follows:
\begin{equation}
\begin{split}
\widetilde{U}_i
&= p \, P_i
   - \frac{1}{2 \, x} \, \delta_{ij} \delta^{kl}
     \left( X^j P_k P_l  +  P_k P_l X^j \right)
\\
& \quad
   + \frac{1}{4 \, x}
     \left[ P_i \left( X^j P_j + P_j X^j \right)
            + \left(X^j P_j + P_j X^j \right) P_i
     \right]
\\
\widetilde{V}^i
&= p \, X^i
   + \frac{1}{2 \, x} \, \delta^{ij} \delta_{kl}
     \left(  P_j X^k X^l + X^k X^l P_j \right)
\\
& \quad
   - \frac{1}{4 \, x}
     \left[ X^i \left( X^j P_j + P_j X^j \right)
            + \left( X^j P_j + P_j X^j \right) X^i
     \right]
\end{split}
\end{equation}
Conversely we also have
 \begin{equation}
V^i = i \commute{\widetilde{V}^i}{K_-}
\qquad
U_i = i \commute{\widetilde{U}_i}{K_-} \,.
\end{equation}
The generators in $\mathfrak{g}^{(+1)} \oplus \mathfrak{g}^{(+2)}$ subspace
form an Heisenberg algebra isomorphic to equation (\ref{heisenberg2}):
\begin{equation}
\commute{\widetilde{V}^i}{\widetilde{U}_j} = 2 i \, \delta^i_j \, K_+
\end{equation}
Commutators $\commute{\mathfrak{g}^{(-1)}}{\mathfrak{g}^{(+1)}}$ close into
grade zero subspace $\mathfrak{g}^{(0)}$:
\begin{equation}
\begin{split}
\commute{U_i}{\widetilde{U}_j}
&= - i \, \delta_{ij} \, J_+
\qquad
\commute{V^i}{\widetilde{V}^j}
= - i \, \delta^{ij} \, J_-
\\
\commute{V^i}{\widetilde{U}_j}
&= - 2 \, \delta^{ik} \, M_{kj}
   + i \, \delta^i_j \left( J_0 + \Delta \right)
\\
\commute{U_i}{\widetilde{V}^j}
&= + 2 \, \delta^{jk} \, M_{ik}
   + i \, \delta^j_i \left( J_0 - \Delta \right)
\end{split}
\end{equation}
$\Delta$ is the generator that determines the 5-grading:
\begin{equation}
\begin{aligned}
\commute{K_-}{K_+}
&= i \, \Delta
\\
\commute{\Delta}{U_i}
&= - i \, U_i
\\
\commute{\Delta}{\widetilde{U}_i}
&= + i \, \widetilde{U}_i
\end{aligned}
\qquad \qquad
\begin{aligned}
\commute{\Delta}{K_\pm}
&= \pm 2 i \, K_\pm
\\
\commute{\Delta}{V^i}
&= - i \, V^i
\\
\commute{\Delta}{\widetilde{V}^i}
&= + i \, \widetilde{V}^i
\end{aligned}
\end{equation}

We note that in this realization, the generators $M_{ij}$ are anti-hermitian 
and all the other generators of $SO(d+2,2)$ are hermitian.

The quadratic Casimir operators of subalgebras $\mathfrak{so}(d)$ and
$\mathfrak{sp}(2,\mathbb{R})_J$ of grade zero subspace, and
$\mathfrak{sp}(2,\mathbb{R})_K$ generated by $K_{\pm}$ and $\Delta$ are given
by
\begin{equation}
\begin{split}
M_{ij} M^{ij}
&= - \mathcal{I}_4 - 2 d
\\
J_- J_+ +  J_+ J_- - 2  \left(J_0\right)^2
&= \mathcal{I}_4 + \frac{d^2}{2}
\\
K_- K_+ + K_+ K_- - \frac{1}{2} \Delta^2
&= \frac{1}{4} \mathcal{I}_4 + \frac{d^2}{8} \,.
\end{split}
\end{equation}
They all reduce to the quartic invariant operator $\mathcal{I}_4$ modulo some
additive constants. Furthermore, grade $\pm1$ generators belonging to the
coset
\[ \frac{SO(d+2,2)}{SO(d)\times SO(2,2)} \]
satisfy the identity
\begin{equation}
U_i \widetilde{V}^i + \widetilde{V}^i U_i
- V^i \widetilde{U}_i - \widetilde{U}_i V^i
= 2 \mathcal{I}_4 + d \left( d + 4 \right)
\end{equation}
in the above realization. The above relations prove the existence of a family
of degree 2 polynomials in the enveloping algebra of $\mathfrak{so}(d+2,2)$
that degenerate to a $c$-number for the minimal unitary realization, in
accordance with Joseph's theorem \cite{MR0342049}:
\begin{equation}
\label{eq:JosephIdeal}
\begin{split}
&M_{ij} M^{ij}
+ \kappa_1 \left( J_- J_+ +  J_+ J_- - 2  \left(J_0\right)^2 \right)
+ 4 \, \kappa_2 \left( K_- K_+ + K_+ K_- - \frac{1}{2} \Delta^2 \right)
\\
&- \frac{1}{2} \left( \kappa_1 + \kappa_2 - 1 \right)
 \left( U_i \widetilde{V}^i + \widetilde{V}^i U_i
        - V^i \widetilde{U}_i - \widetilde{U}_i V^i \right)
\\
& \qquad \qquad
= \frac{1}{2} d \left[ d  - 4 \left( \kappa_1 + \kappa_2 \right) \right]
\end{split}
\end{equation}

The quadratic Casimir of $\mathfrak{so}(d+2,2)$ corresponds to the choice
$2 \kappa_1 = 2 \kappa_2 = - 1$ in \eqref{eq:JosephIdeal}. Hence the
eigenvalue of the quadratic Casimir for the minimal unitary representation is
equal to $\frac{1}{2} d \left( d + 4 \right)$. This minimal unitary
representation is realized over the Hilbert space of square integrable
functions in $(d+1)$ variables.


\section{Minimal Unitary Realization of $SO(6,2)$ over the Hilbert Space of
$L^2$ Functions in  Five Variables}
\label{minrepSO(6,2)}

We shall specialize the minimal unitary realization of $SO(d+2,2)$ given
above to the case of $SO(6,2)$. The corresponding 5-grading of the Lie
algebra of $SO(6,2)$ is with respect to its subalgebra $\mathfrak{g}^{(0)} =
\mathfrak{so}(4) \oplus \mathfrak{sp}(2,\mathbb{R}) \oplus
\mathfrak{so}(1,1)$:
\begin{equation}
\begin{split}
\mathfrak{so}(6,2)
&= \mathfrak{g}^{(-2)} \oplus
   \mathfrak{g}^{(-1)} \oplus
   \left[
    \mathfrak{so}(4) \oplus
    \mathfrak{sp}(2,\mathbb{R}) \oplus
    \mathfrak{so}(1,1)
   \right] \oplus
   \mathfrak{g}^{(+1)} \oplus
   \mathfrak{g}^{(+2)}
\\
&= K_- \oplus
   \left[ U_i \oplus V^i \right] \oplus
   \left[
    M_{ij} \oplus
    J_{\pm,0} \oplus
    \Delta
   \right] \oplus
   \left[
    \widetilde{U}_i \oplus
    \widetilde{V}^i
   \right] \oplus
   K_+
\end{split}
\end{equation}
where $i,j,\dots = 1,2,3,4$.


\subsection{The noncompact 3-grading of $SO(6,2)$ with respect to the
subgroup $SO(5,1) \times SO(1,1)$}
\label{SO(5,1)xSO(1,1)}

Considered as the six dimensional conformal group, $SO(6,2)$ has a natural
3-grading with respect to the generator $\mathcal{D}$ of dilatations whose
eigenvalues determine the conformal dimensions of operators and states.
Let us denote the corresponding 3-graded decomposition of
$\mathfrak{so}(6,2)$ as
\begin{equation}
\mathfrak{so}(6,2)
= \mathfrak{N}^- \oplus
  \mathfrak{N}^0 \oplus
  \mathfrak{N}^+
\end{equation}
where $\mathfrak{N}^0 = \mathfrak{so}(5,1) \oplus
\mathfrak{so}(1,1)_{\mathcal{D}}$ with the subalgebra $\mathfrak{so}(5,1)$ in
$\mathfrak{N}^{0}$ representing the Lorentz algebra in six dimensions. The
\emph{noncompact} dilatation generator $\mathfrak{so}(1,1)_{\mathcal{D}}$ is
given by
\begin{equation}
\mathcal{D}
= \frac{1}{2} \left( \Delta + J_0 \right)
= \frac{1}{4} \left( x p + p x + X^i P_i + P_i X^i \right)
\end{equation}
and the generators belonging to $\mathfrak{N}^{\pm}$ and $\mathfrak{N}^0$ are
as follows:
\begin{equation}
\begin{split}
\mathfrak{N}^-
&= K_- \oplus
   J_-  \oplus
   V^i
\\
\mathfrak{N}^0
&= \mathcal{D} \oplus
   \frac{1}{2} \left( \Delta - J_0 \right) \oplus
   M_{ij} \oplus
   U_i  \oplus
   \widetilde{V}^i
\\
\mathfrak{N}^+
&= K_+ \oplus
   J_+ \oplus
   \widetilde{U}_i
\end{split}
\end{equation}
The Lorentz generators $\mathcal{M}_{\mu\nu}$ ($\mu,\nu,\dots =
0,1,2,\dots,5$) are given by
\begin{equation}
\begin{aligned}
\mathcal{M}_{0i}
&= \frac{1}{2} \left( U_i + \delta_{ij} \widetilde{V}^j \right)
\\
\mathcal{M}_{ij}
&= - i \, M_{ij}
\end{aligned}
\qquad \qquad
\begin{aligned}
\mathcal{M}_{05}
&= \frac{1}{2} \left( \Delta - J_0 \right)
\\
\mathcal{M}_{i5}
&= \frac{1}{2} \left( U_i - \delta_{ij} \widetilde{V}^j \right)
\end{aligned}
\end{equation}
and satisfy the $\mathfrak{so}(5,1)$ commutation relations
\begin{equation}
\commute{\mathcal{M}_{\mu\nu}}{\mathcal{M}_{\rho\tau}}
= i \left(
     \eta_{\nu\rho} \mathcal{M}_{\mu\tau}
     - \eta_{\mu\rho} \mathcal{M}_{\nu\tau}
     - \eta_{\nu\tau} \mathcal{M}_{\mu\rho}
     + \eta_{\mu\tau} \mathcal{M}_{\nu\rho}
    \right)
\end{equation}
where $\eta_{\mu\nu} = \mathrm{diag} (-,+,+,+,+,+)$. The six translation
generators $\mathcal{P}_\mu$ ($\mu = 0,1,2,\dots,5$) of the conformal group
$SO(6,2)$ are given by
\begin{equation}
\mathcal{P}_0 = K_+ - \frac{1}{2} J_+
\qquad \quad
\mathcal{P}_i = \widetilde{U}_i \quad (i = 1,2,3,4)
\qquad \quad
\mathcal{P}_5 = K_+ + \frac{1}{2} J_+
\end{equation}
and the  special conformal generators $\mathcal{K}_\mu$ ($\mu =
0,1,2,\dots,5$) are given by
\begin{equation}
\mathcal{K}_0 = - \frac{1}{2} J_- + K_-
\qquad \quad
\mathcal{K}_i = - V^i \quad (i = 1,2,3,4)
\qquad \quad
\mathcal{K}_5 = - \frac{1}{2} J_- - K_- \,.
\end{equation}
These generators satisfy the commutation relations of $SO(6,2)$ as the six
dimensional conformal algebra:
\begin{equation}
\begin{split}
\commute{\mathcal{M}_{\mu\nu}}{\mathcal{M}_{\rho\tau}}
&= i \left(
     \eta_{\nu\rho} \mathcal{M}_{\mu\tau}
     - \eta_{\mu\rho} \mathcal{M}_{\nu\tau}
     - \eta_{\nu\tau} \mathcal{M}_{\mu\rho}
     + \eta_{\mu\tau} \mathcal{M}_{\nu\rho}
    \right)
\\
\commute{\mathcal{P}_\mu}{\mathcal{M}_{\nu\rho}}
&= i \left( \eta_{\mu\nu} \, \mathcal{P}_\rho
            - \eta_{\mu\rho} \, \mathcal{P}_\nu
     \right)
\\
\commute{\mathcal{K}_\mu}{\mathcal{M}_{\nu\rho}}
&= i \left( \eta_{\mu\nu} \, \mathcal{K}_\rho
            - \eta_{\mu\rho} \, \mathcal{K}_\nu
     \right)
\\
\commute{\mathcal{D}}{\mathcal{M}_{\mu\nu}}
&= \commute{\mathcal{P}_\mu}{\mathcal{P}_\nu}
 = \commute{\mathcal{K}_\mu}{\mathcal{K}_\nu}
 = 0
\\
\commute{\mathcal{D}}{\mathcal{P}_\mu}
&= + i \, \mathcal{P}_\mu
\qquad \qquad
\commute{\mathcal{D}}{\mathcal{K}_\mu}
 = - i \, \mathcal{K}_\mu
\\
\commute{\mathcal{P}_\mu}{\mathcal{K}_\nu}
&= 2 i \left( \eta_{\mu\nu} \, \mathcal{D} + \mathcal{M}_{\mu\nu} \right)
\end{split}
\end{equation}

We should note that the $6D$ Poincar\'e mass operator vanishes identically
\begin{equation}
\mathcal{M}^2 = \eta_{\mu\nu} \mathcal{P}^\mu \mathcal{P}^\nu
              = 0
\end{equation}
for the minimal unitary realization given above. Hence the minimal unitary
representation of $SO(6,2)$ corresponds to a massless representation as a six
dimensional conformal group. We shall refer to the above 3-graded
decomposition as the \emph{noncompact} 3-grading.


\subsection{The compact 3-grading of $SO(6,2)$ with respect to the subgroup
$SO(6) \times SO(2)$}
\label{SO(6)xSO(2)}

The Lie algebra $\mathfrak{so}(6,2)$ has a 3-grading with
respect to its maximal compact subalgebra $\mathfrak{C}^0 = \mathfrak{so}(6)
\oplus \mathfrak{so}(2)$, determined by the $\mathfrak{so}(2)$ generator
\begin{equation}
H = \frac{1}{2}
    \left[
     \left( K_+ + K_- \right)
     - \frac{1}{2} ( J_+ + J_- )
    \right]
\end{equation}
such that
\begin{equation}
\mathfrak{so}(6,2)
= \mathfrak{C}^- \oplus
  \left[ \mathfrak{so}(6) \oplus \mathfrak{so}(2) \right] \oplus
  \mathfrak{C}^+
\end{equation}
and  satisfy
\begin{equation}
\commute{H}{\mathfrak{C}^+} = + \, \mathfrak{C}^+
\qquad \qquad \qquad
\commute{H}{\mathfrak{C}^-} = - \, \mathfrak{C}^- \,.
\end{equation}

In this decomposition:
\begin{equation}
\begin{split}
\mathfrak{C}^0
 &= \mathfrak{so}(6) \oplus \mathfrak{so}(2)
  = \left[
     M_{ij} \oplus
     \left( \left( K_+ + K_- \right)
      + \frac{1}{2} \left( J_+ + J_- \right) \right) \oplus
     \left( U_i - \delta_{ij} \widetilde{V}^j \right)
    \right.
\\
 & \qquad \qquad \qquad \qquad \quad
   \left.
    \left( \widetilde{U}_i + \delta_{ij} V^j \right)
   \right]
   \oplus
   \frac{1}{2}
   \left[ \left( K_+ + K_- \right)
    - \frac{1}{2} \left( J_+ + J_- \right)
   \right]
\\
\mathfrak{C}^+
 &= \left[ \Delta - i \left( K_+ - K_- \right) \right] \oplus
    \left[ J_0 + \frac{i}{2} \left( J_+ - J_- \right) \right] \oplus
    \left[
     \frac{1}{2} \left( U_i + \delta_{ij} \widetilde{V}^j \right)
      - \frac{i}{2} \left( \widetilde{U}_i - \delta_{ij} V^j \right)
    \right] \\
\mathfrak{C}^-
 &= \left[ \Delta + i \left( K_+ - K_- \right) \right] \oplus
    \left[ J_0 - \frac{i}{2} \left( J_+ - J_- \right) \right] \oplus
    \left[
     \frac{1}{2} \left( U_i + \delta_{ij} \widetilde{V}^j \right)
      + \frac{i}{2} \left( \widetilde{U}_i - \delta_{ij} V^j \right)
    \right]
\end{split}
\end{equation}
Note that in the above 3-grading, the operators belonging to $\mathfrak{C}^+$
are Hermitian conjugates of those belonging to $\mathfrak{C}^-$. In the
corresponding minimal unitary realization one takes only the hermitian linear
combinations of these operators as generators of $\mathfrak{so}(6,2)$. The
generator $H$ is the conformal Hamiltonian or the $AdS$ energy depending on
whether one is considering $SO(6,2)$ as six dimensional conformal group or
the seven dimensional $AdS$ group. We shall refer to this grading as the
\emph{compact} 3-grading.

The $\mathfrak{so}(6)$ generators $\widetilde{M}_{MN}$ ($M,N,\dots =
1,2,\dots,6$) in grade zero subspace $\mathfrak{C}^0$ are given by
\begin{equation}
\begin{aligned}
\widetilde{M}_{ij}
&= i \, M_{ij}
\\
\widetilde{M}_{i6}
&= \frac{1}{2} \left( \widetilde{U}_i + \delta_{ij} V^j \right)
\end{aligned}
\qquad \qquad
\begin{aligned}
\widetilde{M}_{i5}
&= \frac{1}{2} \left( U_i - \delta_{ij} \widetilde{V}^j \right)
\\
\widetilde{M}_{56}
&= \frac{1}{2}
   \left[
    \left( K_+ + K_- \right) + \frac{1}{2} \left( J_+ + J_- \right)
   \right]
\end{aligned}
\end{equation}
and satisfy the $\mathfrak{so}(6)$ algebra
\begin{equation}
\commute{\widetilde{M}_{MN}}{\widetilde{M}_{PQ}}
= i \left(
     \delta_{NP} \widetilde{M}_{MQ} - \delta_{MP} \widetilde{M}_{NQ}
     - \delta_{NQ} \widetilde{M}_{MP} + \delta_{MQ} \widetilde{M}_{NP}
    \right) \,.
\end{equation}

To give the decomposition of the minrep of $SO(6,2)$ into its  K-finite vectors of its
maximal compact subgroup $SO(6) \times SO(2)$ we shall define the oscillators
\begin{equation}
c_i = \frac{1}{\sqrt{2}} \left( X^i + i \, P_i \right)
\qquad \qquad \qquad
c_i^\dag = \frac{1}{\sqrt{2}} \left( X^i - i \, P_i \right)
\end{equation}
or conversely
\begin{equation}
X^i = \frac{1}{\sqrt{2}} \left( c_i^\dag + c_i \right)
\qquad \qquad \qquad
P_i = \frac{i}{\sqrt{2}} \left( c_i^\dag - c_i \right) \,.
\end{equation}
These oscillators satisfy the commutation relations
\begin{equation}
\commute{c_i}{c_j^\dag} = \delta_{ij} \,.
\end{equation}

The quartic invariant operator $\mathcal{I}_4$ takes on a simple form in
terms of these oscillators:
\begin{equation}
\begin{split}
\mathcal{I}_4
&= - \left( c_i^\dag c_j - c_j^\dag c_i \right)^2 - 8
\\
& = - M_{ij} M_{ij} - 8
\end{split}
\end{equation}

The $\mathfrak{so}(2)$ generator in $\mathfrak{C}^0$, that determines the
3-grading and plays the role of the $AdS$ energy
\cite{Gunaydin:1984wc,Gunaydin:1999ci,Fernando:2001ak}, is given in terms of $x$, $p$ and
oscillators $c_i$, $c_i^\dag$ as:
\begin{equation}
\begin{split}
H
&= \frac{1}{2}
   \left[
    \left( K_+ + K_- \right) - \frac{1}{2} \left( J_+ + J_- \right)
   \right]
\\
&= \frac{1}{4} \left( x^2 + p^2 \right)
   + \frac{1}{2} c_i^\dag c_i
   - \frac{1}{8 \, x^2} \left( c_i^\dag c_j - c_j^\dag c_i \right)^2
   + \frac{3}{16 \, x^2}
   + 1
\end{split}
\label{SO2generator}
\end{equation}
We can also write the $\mathfrak{so}(6)$ generators $\widetilde{M}_{MN}$ in
terms of these oscillators as follows:
\begin{equation}
\begin{split}
\widetilde{M}_{ij}
&= i M_{ij}
\\
&= i \left( c_i^\dag c_j - c_j^\dag c_i \right)
\\
\widetilde{M}_{i5}
&= \frac{1}{2} \left( U_i - \delta_{ij} \widetilde{V}^j \right)
\\
&= \frac{i}{2 \sqrt{2}} \left( x + i \, p \right) c_i^\dag
   - \frac{i}{2 \sqrt{2}} \left( x - i \, p \right) c_i
   - \frac{i}{2 \sqrt{2} \, x}
     \left( c_i^\dag c_j - c_j^\dag c_i \right) \left( c_j^\dag + c_j \right)
   + \frac{3 i}{4 \sqrt{2} \, x} \left( c_i^\dag + c_i \right)
\\
\widetilde{M}_{i6}
&= \frac{1}{2} \left( \widetilde{U}_i + \delta_{ij} V^j \right)
\\
&= \frac{1}{2 \sqrt{2}} \left( x + i \, p \right) c_i^\dag
   + \frac{1}{2 \sqrt{2}} \left( x - i \, p \right) c_i
   - \frac{1}{2 \sqrt{2} \, x}
     \left( c_i^\dag c_j - c_j^\dag c_i \right) \left( c_j^\dag - c_j \right)
   + \frac{3}{4 \sqrt{2} \, x} \left( c_i^\dag - c_i \right)
\\
\widetilde{M}_{56}
&= \frac{1}{2}
   \left[
    \left( K_+ + K_- \right) + \frac{1}{2} \left( J_+ + J_- \right)
   \right]
\\
&= \frac{1}{4} \left( x^2 + p^2 \right)
   - \frac{1}{2} c_i^\dag c_i
   - \frac{1}{8 \, x^2}
     \left( c_i^\dag c_j - c_j^\dag c_i \right)^2
   + \frac{3}{16 \, x^2}
   - 1
\end{split}
\end{equation}

Six operators that belong to the grade $+1$ subspace $\mathfrak{C}^+$ have
the following form in terms of these oscillators:
\begin{equation}
\begin{split}
\frac{1}{2} \left[
             \left( U_i + \delta_{ij} \widetilde{V}^j \right)
             - i \left( \widetilde{U}_i - \delta_{ij} V^j \right)
            \right]
&= \frac{i}{\sqrt{2}} \, \left( x - i \, p \right) c_i^\dag
\\
&\quad
   + \frac{i}{2 \sqrt{2} \, x}
     \left[ c_j^\dag \left( c_i^\dag c_j - c_j^\dag c_i \right)
            + \left( c_i^\dag c_j - c_j^\dag c_i \right) c_j^\dag
     \right]
\\
J_0 + \frac{i}{2} \left( J_+ - J_- \right)
 &= i \, c_i^\dag c_i^\dag
\\
\Delta - i \left( K_+ - K_- \right)
&= \frac{i}{2} \, \left( x - i \, p \right)^2
   + \frac{i}{4 \, x^2}
     \left[ \left( c_i^\dag c_j - c_j^\dag c_i \right)^2
            + \frac{9}{2}
     \right]
\end{split}
\end{equation}
and those that belong to the grade $-1$ subspace $\mathfrak{C}^-$ are given
by
\begin{equation}
\begin{split}
\frac{1}{2} \left[
             \left( U_i + \delta_{ij} \widetilde{V}^j \right)
             + i \left( \widetilde{U}_i - \delta_{ij} V^j \right)
            \right]
&= - \frac{i}{\sqrt{2}} \, \left( x + i \, p \right) c_i
\\
&\quad
   + \frac{i}{2 \sqrt{2} \, x}
     \left[ c_j \left( c_i^\dag c_j - c_j^\dag c_i \right)
            + \left( c_i^\dag c_j - c_j^\dag c_i \right) c_j
     \right]
\\
J_0 - \frac{i}{2} \left( J_+ - J_- \right)
 &= - i \, c_i c_i
\\
\Delta + i \left( K_+ - K_- \right)
&= - \frac{i}{\sqrt{2}} \, \left( x + i \, p \right)^2
   - \frac{i}{4 \, x^2}
     \left[ \left( c_i^\dag c_j - c_j^\dag c_i \right)^2
            + \frac{9}{2}
     \right] \,.
\end{split}
\end{equation}

One could also give a decomposition of $SO(6,2)$ with respect to the subgroup 
$SO(4) \times SO(2,2)$, which we present in appendix \ref{SO(4)xSO(2,2)}.


\section{Minimal Unitary Representation of $SO^*(8)$}
\label{minrepSO*(8)}

The groups $SO(d+2,2)$ have supersymmetric extensions which are in general
supergroups of the form $OSp(d+2,2\,|\,2n,\mathbb{R})$ with even subgroups
$SO(d+2,2) \times Sp(2n,\mathbb{R})$. The supergroups whose even subgroups
are products of two simple noncompact groups do not, in general, admit any
unitary representations. Furthermore, if the group $SO(d+2,2)$ is considered
as a conformal group in $(d+2)$ dimensions or as anti-de Sitter group in
$(d+3)$ dimensions, its factor group in its supersymmetric extension is the
$R$-symmetry group which must be compact \cite{Nahm:1977tg}. Remarkably,
either the existence of exceptional superalgebras or certain special
isomorphisms allow such possibilities for special values of $d$. The group
$SO(5,2)$ has an extension to the exceptional supergroup $F(4)$ with even
subgroup $SO(5,2) \times SU(2)$ which admits positive energy unitary
representations. The covering group of $SO(4,2)$ is the group $SU(2,2)$ which
extends to an infinite family of supergroups $SU(2,2|N)$ with even subgroups
$SU(2,2) \times U(N)$ that admit positive energy unitary representations.
Similarly isomorphism of $SO(3,2)$ to $Sp(4,\mathbb{R})$ allows extension to
supergroups $OSp(N|4,\mathbb{R})$ with even subgroups $Sp(4,\mathbb{R})
\times SO(N)$ that admit positive energy unitary representations. Since
$SO(2,2)$ is not simple, one finds a rich family of supersymmetric extensions
that admit positive energy unitary unitary representations that were studied
in \cite{Gunaydin:1986fe}. Similarly, the Lie algebra of $SO(6,2)$ is
isomorphic to that of $SO^*(8)$ which have extensions to supergroups
$OSp(8^*|2N)$ with even subgroups $SO^*(8) \times USp(2N)$ that admit
positive energy unitary representations.\footnote{As such it belongs to an
infinite family of supergroups $OSp(2M^*|2N)$ with even subgroups $SO^*(2M)
\times USp(2N)$ whose positive energy unitary representations were studied in
\cite{Gunaydin:1990ag}.} Hence we will now study the minimal unitary
realizations of $SO^*(8)$ and their supersymmetric extensions in the
subsequent sections.


\subsection{The 5-grading of $SO^*(8)$ with respect to the subgroup $SO^*(4)
\times SU(2) \times SO(1,1)$}
\label{5GrSO*(8)}

The noncompact Lie algebra $\mathfrak{so}^*(8)$ has a 5-grading with respect
to its subalgebra $\mathfrak{so}^*(4) \oplus \mathfrak{su}(2) \oplus
\mathfrak{so}(1,1)$, where the $\mathfrak{so}(1,1)$ generator $\Delta$
defines the 5-grading \cite{Gunaydin:2006vz}:
\begin{equation}
\mathfrak{so}^*(8)
 = \mathfrak{g}^{(-2)} \oplus
   \mathfrak{g}^{(-1)} \oplus
   \left[
    \mathfrak{so}^*(4) \oplus \mathfrak{su}(2) \oplus \Delta
   \right] \oplus
   \mathfrak{g}^{(+1)} \oplus
   \mathfrak{g}^{(+2)}
\end{equation}
such that
\begin{equation}
\commute{\Delta}{\mathfrak{g}^{(m)}} = m \, \mathfrak{g}^{(m)}
\end{equation}
In this decomposition, $\mathfrak{g}^{(\pm 2)}$ subspaces are
one-dimensional, and $\mathfrak{g}^{(\pm 1)}$ subspaces transform in the
$\left( \mathbf{4} , \mathbf{2} \right)$ dimensional representation of
$SO^*(4) \times SU(2)$. Since $SO^*(4) = SU(1,1)\times SU(2)$, the grade zero
subalgebra $SU(1,1) \times SU(2) \times SU(2) \times SO(1,1)$ is also
isomorphic to that of $SO(6,2)$.

For the study of the minrep of $SO^*(8)$ we shall relabel  the oscillators introduced in the previous sections as
$a_m$, $b_m$ and their hermitian conjugates $a^m = \left( a_m \right)^\dag$, $b^m = \left( b_m \right)^\dag$ ($m,n,\dots = 1,2$)
\begin{equation}
\begin{aligned}
a_m
&= \frac{1}{\sqrt{2}} \left( X^m + i \, P_m \right)
\\
b_m
&= \frac{1}{\sqrt{2}} \left( X^{2+m} + i \, P_{2+m} \right)
\end{aligned}
\qquad \qquad \qquad
\begin{aligned}
a^m
&= \frac{1}{\sqrt{2}} \left( X^m - i \, P_m \right)
\\
b^m
&= \frac{1}{\sqrt{2}} \left( X^{2+m} - i \, P_{2+m} \right)
\end{aligned}
\end{equation}
so that
\begin{equation}
\begin{aligned}
X^m
&= \frac{1}{\sqrt{2}} \left( a^m + a_m \right)
\\
X^{2+m}
&= \frac{1}{\sqrt{2}} \left( b^m + b_m \right)
\end{aligned}
\qquad \qquad \qquad
\begin{aligned}
P_m
&= \frac{i}{\sqrt{2}} \left( a^m - a_m \right)
\\
P_{2+m}
&= \frac{i}{\sqrt{2}} \left( b^m - b_m \right) \,.
\end{aligned}
\end{equation}
They satisfy the commutation relations
\begin{equation}
\commute{a_m}{a^n} = \delta^n_m
\qquad \qquad \qquad
\commute{b_m}{b^n} = \delta^n_m \,.
\end{equation}
Then the generators of $\mathfrak{su}(2)$ of $\mathfrak{g}^{(0)}$ that
commute with $\mathfrak{so}^*(4)$ can be realized as follows:
\begin{equation}
S_+ = a^m b_m
\qquad \qquad
S_- = \left( S_+ \right)^\dag
    = a_m b^m
\qquad \qquad
S_0 = \frac{1}{2} \left( N_a - N_b \right)
\label{SU(2)S_generators}
\end{equation}
where $N_a = a^m a_m$ and $N_b = b^m b_m$ are the respective number
operators. We denote this subalgebra as $\mathfrak{su}(2)_S$. Its generators satisfy:
\begin{equation}
\commute{S_+}{S_-} = 2 \, S_0
\qquad \qquad \qquad
\commute{S_0}{S_\pm} = \pm S_\pm
\end{equation}
The quadratic Casimir of  $\mathfrak{su}(2)_S$ is given by
\begin{equation}
\begin{split}
\mathcal{C}_2 \left[ \mathfrak{su}(2)_S \right]
 = \mathcal{S}^2
&= {S_0}^2 + \frac{1}{2} \left( S_+ S_- + S_- S_+ \right)
\\
&= \frac{1}{2} \left(N_a + N_b \right)
   \left[ \frac{1}{2} \left( N_a + N_b \right) + 1 \right]
   - 2 a^{[m} b^{n]} \, a_{[m} b_{n]}
\end{split}
\end{equation}
where square bracketing $a_{[m} b_{n]} = \frac{1}{2} \left( a_m b_n - a_n b_m
\right)$ represents antisymmetrization of weight one.

We shall label the simple factors of $\mathfrak{so}^*(4)$ subalgebra that
commutes with $\mathfrak{su}(2)_S$ as
\begin{equation}
\mathfrak{so}^*(4) = \mathfrak{su}(2)_A \oplus \mathfrak{su}(1,1)_N
\end{equation}
and  denote the generators of  $\mathfrak{su}(2)_A$ and
$\mathfrak{su}(1,1)_N$ as $A_{\pm,0}$ and $N_{\pm,0}$, respectively.
In terms of the above $a$-type and $b$-type oscillators, these generators
have the following realization:
\begin{equation}
\begin{aligned}
A_+ &= a^1 a_2 + b^1 b_2
\\
A_- &= \left( A_+ \right)^\dag = a_1 a^2 + b_1 b^2
\\
A_0 &= \frac{1}{2} \left( a^1 a_1 - a^2 a_2 + b^1 b_1 - b^2 b_2 \right)
\end{aligned}
\qquad  \qquad
\begin{aligned}
N_+ &= a^1 b^2 - a^2 b^1
\\
N_- &= \left( N_+ \right)^\dag = a_1 b_2 - a_2 b_1
\\
N_0 &= \frac{1}{2} \left( N_a + N_b \right) + 1
\end{aligned}
\label{SU(2)AN_generators}
\end{equation}
They satisfy the commutation relations:
\begin{equation}
\begin{aligned}
\commute{A_+}{A_-} &= 2 \, A_0
\\
\commute{A_0}{A_\pm} &= \pm A_\pm
\end{aligned}
\qquad \qquad \qquad
\begin{aligned}
\commute{N_-}{N_+} &= 2 \, N_0
\\
\commute{N_0}{N_\pm} &= \pm N_\pm
\end{aligned}
\end{equation}
The quadratic Casimirs of these subalgebras
\begin{equation}
\begin{split}
\mathcal{C}_2 \left[ \mathfrak{su}(2)_A \right]
= \mathcal{A}^2 &= {A_0}^2 + \frac{1}{2} \left( A_+ A_- + A_- A_+ \right)
\\
\mathcal{C}_2 \left[ \mathfrak{su}(1,1)_N \right]
= \mathcal{N}^2 &= {N_0}^2 - \frac{1}{2} \left( N_+ N_- + N_- N_+ \right)
\end{split}
\end{equation}
coincide and are equal to that of $\mathfrak{su}(2)_S$:
\begin{equation}
\mathcal{S}^2 = \mathcal{A}^2 = \mathcal{N}^2
\end{equation}

The transformations relating the $SO^*(8)$ oscillators $a_m$ and
$b_m$ to the $SO(6,2)$ oscillators $c_i$ are given in Appendix
\ref{app:bogoliubov}.

The generator that defines the 5-grading can be written as
\begin{equation}
\Delta = \frac{1}{2} \left( x p + p x \right)
\end{equation}
and the $\mathfrak{g}^{(\pm 2)}$ generators are realized as
\begin{equation}
K_-
= \frac{1}{2} x^2
\qquad \qquad \qquad
K_+
= \frac{1}{2} p^2
  + \frac{1}{4 \, x^2} \left( 8 \, \mathcal{S}^2 + \frac{3}{2} \right) \,.
\end{equation}
These three generators form another $\mathfrak{su}(1,1)_K$ subalgebra
\begin{equation}
\commute{K_-}{K_+} = i \, \Delta
\qquad \qquad \qquad
\commute{\Delta}{K_\pm} = \pm 2 i \, K_\pm
\end{equation}
with the quadratic Casimir operator
\begin{equation}
\mathcal{C}_2 \left[ \mathfrak{su}(1,1)_K \right]
= \mathcal{K}^2 = \frac{1}{2} (K_+ K_- + K_- K_+) - \frac{1}{4} \Delta^2
              =  \, \mathcal{S}^2 \,.
\end{equation}

The eight generators that are in $\mathfrak{g}^{(-1)}$ subspace take the form
\begin{equation}
\begin{aligned}
U_m &= x \, a_m
\\
V_m &= x \, b_m
\end{aligned}
\qquad \qquad \qquad \qquad
\begin{aligned}
U^m &= x \, a^m
\\
V^m &= x \, b^m
\end{aligned}
\end{equation}
and together with $K_-$ form an Heisenberg algebra:
\begin{equation}
\begin{split}
\commute{U_m}{U^n}
&= \commute{V_m}{V^n}
 = 2 \, \delta^n_m \, K_-
\\
\commute{U_m}{U_n}
&= \commute{V_m}{V_n}
 = 0
\end{split}
\end{equation}
The  generators  in $\mathfrak{g}^{(+1)}$ are obtained from the commutators
$\commute{\mathfrak{g}^{(-1)}}{\mathfrak{g}^{(+2)}}$:
\begin{equation}
\begin{split}
\widetilde{U}_m = i \commute{U_m}{K_+}
& \qquad \qquad \qquad \qquad
\widetilde{U}^m = \left( \widetilde{U}_m \right)^\dag
                = i \commute{U^m}{K_+}
\\
\widetilde{V}_m = i \commute{V_m}{K_+}
& \qquad \qquad \qquad \qquad
\widetilde{V}^m = \left( \widetilde{V}_m \right)^\dag
                = i \commute{V^m}{K_+}
\end{split}
\end{equation}
Explicitly they are given by
\begin{equation}
\begin{split}
\widetilde{U}_m
&= - p \, a_m
   + \frac{2i}{x}
      \left[ \left( S_0 + \frac{3}{4} \right) a_m + S_- b_m \right]
\\
\widetilde{U}^m
&= - p \, a^m
   - \frac{2i}{x}
      \left[ \left( S_0 - \frac{3}{4} \right) a^m + S_+ b^m \right]
\\
\widetilde{V}_m
&= - p \, b_m
   - \frac{2i}{x}
      \left[ \left( S_0 - \frac{3}{4} \right) b_m - S_+ a_m \right]
\\
\widetilde{V}^m
&= - p \, b^m
   + \frac{2i}{x}
      \left[ \left( S_0 + \frac{3}{4} \right) b^m - S_- a^m \right]
\end{split}
\label{g+1bosonic}
\end{equation}
and also form another Heisenberg algebra with $K_+$ as its ``central
charge'':
\begin{equation}
\begin{split}
\commute{\widetilde{U}_m}{\widetilde{U}^n}
&= \commute{\widetilde{V}_m}{\widetilde{V}^n}
 = 2 \, \delta^n_m \, K_+
\\
\commute{\widetilde{U}_m}{\widetilde{U}_n}
&= \commute{\widetilde{V}_m}{\widetilde{V}_n}
 = 0
\end{split}
\end{equation}
The commutators $\commute{\mathfrak{g}^{(-2)}}{\mathfrak{g}^{(+1)}}$ take the
following form:
\begin{equation}
\begin{aligned}
\commute{\widetilde{U}_m}{K_-} &= i \, U_m
\\
\commute{\widetilde{V}_m}{K_-} &= i \, V_m
\end{aligned}
\qquad \qquad \qquad \qquad
\begin{aligned}
\commute{\widetilde{U}^m}{K_-} &= i \, U^m
\\
\commute{\widetilde{V}^m}{K_-} &= i \, V^m
\end{aligned}
\end{equation}
Finally, the non-vanishing commutators of the form
$\commute{\mathfrak{g}^{(-1)}}{\mathfrak{g}^{(+1)}}$ are as follows:
\begin{equation}
\begin{split}
\commute{U_m}{\widetilde{U}^n}
&= - \delta^n_m \, \Delta
   - 2 i \, \delta^n_m \, N_0
   - 2 i \, \delta^n_m \, S_0
   - 2 i \, A^n_{~m}
\\
\commute{V_m}{\widetilde{V}^n}
&= - \delta^n_m \, \Delta
   - 2 i \, \delta^n_m \, N_0
   + 2 i \, \delta^n_m \, S_0
   - 2 i \, A^n_{~m}
\\
\commute{U_m}{\widetilde{V}^n}
&= - 2 i \, \delta^n_m \, S_-
\qquad \qquad
\commute{V_m}{\widetilde{U}^n}
 = - 2 i \, \delta^n_m \, S_+
\\
\commute{U_m}{\widetilde{V}_n}
&= - 2 i \, \epsilon_{mn} \, N_-
\qquad \qquad
\commute{V_m}{\widetilde{U}_n}
 = + 2 i \, \epsilon_{mn} \, N_-
\end{split}
\end{equation}
where we have labeled the generators of $\mathfrak{su}(2)_A$ as $A^m_{~n}$:
\begin{equation}
A^1_{~1} = - A^2_{~2} = A_0
\qquad \qquad \qquad
A^1_{~2} = A_+
\qquad \qquad \qquad
A^2_{~1} = \left( {A^1_{~2}} \right)^\dag = A_-
\end{equation}
and  denoted the completely antisymmetric tensor by $\epsilon_{mn}$
($\epsilon_{12} = +1$).

With the generators defined above, the 5-grading of the Lie algebra
$\mathfrak{so}^*(8)$, defined by $\Delta$, takes the form:
\begin{equation}
\begin{split}
\mathfrak{so}^*(8)
&= ~ \mathbf{1} ~~ \oplus
   ~~ \left( \mathbf{4} , \mathbf{2} \right) ~ \oplus
   \left[ \mathfrak{su}(2)_A \oplus
          \mathfrak{su}(1,1)_N \oplus
          \mathfrak{su}(2)_S \oplus
          \mathfrak{so}(1,1)_{\Delta}
   \right] \oplus
   ~ \left( \mathbf{4} , \mathbf{2} \right) ~ \oplus
   ~ \mathbf{1}
\\
&= K_-
   \oplus
   \left[ U_m \,,\, U^m \,,\, V_m \,,\, V^m \right]
   \oplus
   \left[ ~ A_{\pm,0} ~ \oplus ~ N_{\pm,0} ~ \oplus ~ S_{\pm,0} ~ \oplus
          ~ \Delta ~ \right]
\\
& \qquad \qquad \qquad \qquad \qquad \qquad \qquad \qquad \qquad \qquad \quad
   \oplus
   \left[ \widetilde{U}_m \,,\, \widetilde{U}^m \,,\,
          \widetilde{V}_m \,,\, \widetilde{V}^m
   \right]
   \oplus
   K_+ \,
\end{split}
\label{so*(8)5-grading}
\end{equation}

As expected, the quadratic Casimir of $\mathfrak{so}^*(8)$, given by
\begin{equation}
\mathcal{C}_2 \left[ \mathfrak{so}^*(8) \right]
= \mathcal{C}_2 \left[ \mathfrak{su}(2)_S \right]
  + \mathcal{C}_2 \left[ \mathfrak{su}(2)_A \right]
  + \mathcal{C}_2 \left[ \mathfrak{su}(1,1)_N \right]
  + \mathcal{C}_2 \left[ \mathfrak{su}(1,1)_K \right]
  - \frac{i}{4} \mathcal{F} \left( U , V \right)
\end{equation}
where
\begin{equation}
\begin{split}
\mathcal{F} \left( U , V \right)
&= \left( U_m \widetilde{U}^m + V_m \widetilde{V}^m
          + \widetilde{U}^m U_m + \widetilde{V}^m V_m \right)
\\
& \qquad
   - \left( U^m \widetilde{U}_m + V^m \widetilde{V}_m
         + \widetilde{U}_m U^m + \widetilde{V}_m V^m \right)
\end{split}
\end{equation}
reduces to a $c$-number, $-4$.


\subsection{The noncompact 3-grading of $SO^*(8)$ with respect to the
subgroup $SU^*(4) \times SO(1,1)$}
\label{NC3GrSO*(8)}

Considered as the six dimensional conformal group, $SO^*(8)$ has a noncompact
3-grading determined by the dilatation generator $\mathcal{D}$:
\begin{equation}
\mathfrak{so}^*(8)
= \mathfrak{N}^- \oplus
  \mathfrak{N}^0 \oplus
  \mathfrak{N}^+
\end{equation}
where $\mathfrak{N}^0 = \mathfrak{su}^*(4) \oplus
\mathfrak{so}(1,1)_{\mathcal{D}}$ and
\begin{equation}
\mathcal{D}
= \frac{1}{2} \left[ \Delta - i \left( N_+ - N_- \right) \right] \,.
\end{equation}
The generators that belong to $\mathfrak{N}^{\pm}$ and $\mathfrak{N}^0$
subspaces are as follows:
\begin{equation}
\begin{split}
\mathfrak{N}^-
&= K_-
   \oplus \left[ N_0 - \frac{1}{2} \left( N_+ + N_- \right) \right]
\\
& \quad
   \oplus \left( U^1 - V_2 \right)
   \oplus \left( U^2 + V_1 \right)
   \oplus \left( V^1 + U_2 \right)
   \oplus \left( V^2 - U_1 \right)
\\
\mathfrak{N}^0
&= \mathcal{D}
   \oplus \frac{1}{2} \left[ \Delta + i \left( N_+ - N_- \right) \right]
   \oplus S_{\pm,0}
   \oplus A_{\pm,0}
\\
& \quad
   \oplus \left( U^1 + V_2 \right)
   \oplus \left( U^2 - V_1 \right)
   \oplus \left( V^1 - U_2 \right)
   \oplus \left( V^2 + U_1 \right)
\\
& \quad
   \oplus \left( \widetilde{U}^1 - \widetilde{V}_2 \right)
   \oplus \left( \widetilde{U}^2 + \widetilde{V}_1 \right)
   \oplus \left( \widetilde{V}^1 + \widetilde{U}_2 \right)
   \oplus \left( \widetilde{V}^2 - \widetilde{U}_1 \right)
\\
\mathfrak{N}^+
&= K_+
   \oplus \left[ N_0 + \frac{1}{2} \left( N_+ + N_- \right) \right]
\\
& \quad
   \oplus \left( \widetilde{U}^1 + \widetilde{V}_2 \right)
   \oplus \left( \widetilde{U}^2 - \widetilde{V}_1 \right)
   \oplus \left( \widetilde{V}^1 - \widetilde{U}_2 \right)
   \oplus \left( \widetilde{V}^2 + \widetilde{U}_1 \right)
\end{split}
\end{equation}

Since $\mathfrak{su}^*(4) \simeq \mathfrak{so}(5,1)$, we find that the
Lorentz generators $\mathcal{M}_{\mu\nu}$ ($\mu,\nu,\dots = 0,1,2,\dots,5$)
in six dimensions are given by:
\begin{subequations}
\begin{equation}
\begin{aligned}
\mathcal{M}_{01}
&= \frac{1}{4} \left[ \left( U^1 + V_2 \right)
                      + \left( V^2 + U_1 \right)
                      + i \left( \widetilde{U}^1 - \widetilde{V}_2 \right)
                      + i \left( \widetilde{V}^2 - \widetilde{U}_1 \right)
               \right]
\\
\mathcal{M}_{02}
&= \frac{i}{4} \left[ \left( U^1 + V_2 \right)
                      - \left( V^2 + U_1 \right)
                      + i \left( \widetilde{U}^1 - \widetilde{V}_2 \right)
                      - i \left( \widetilde{V}^2 - \widetilde{U}_1 \right)
               \right]
\\
\mathcal{M}_{03}
&= \frac{i}{4} \left[ \left( U^2 - V_1 \right)
                      + \left( V^1 - U_2 \right)
                      + i \left( \widetilde{U}^2 + \widetilde{V}_1 \right)
                      + i \left( \widetilde{V}^1 + \widetilde{U}_2 \right)
               \right]
\\
\mathcal{M}_{04}
&= - \frac{1}{4} \left[ \left( U^2 - V_1 \right)
                        - \left( V^1 - U_2 \right)
                        + i \left( \widetilde{U}^2 + \widetilde{V}_1 \right)
                        - i \left( \widetilde{V}^1 + \widetilde{U}_2 \right)
                 \right]
\end{aligned}
\end{equation}
\begin{equation}
\begin{aligned}
\mathcal{M}_{15}
&= \frac{1}{4} \left[ \left( U^1 + V_2 \right)
                      + \left( V^2 + U_1 \right)
                      - i \left( \widetilde{U}^1 - \widetilde{V}_2 \right)
                      - i \left( \widetilde{V}^2 - \widetilde{U}_1 \right)
               \right]
\\
\mathcal{M}_{25}
&= \frac{i}{4} \left[ \left( U^1 + V_2 \right)
                      - \left( V^2 + U_1 \right)
                      - i \left( \widetilde{U}^1 - \widetilde{V}_2 \right)
                      + i \left( \widetilde{V}^2 - \widetilde{U}_1 \right)
               \right]
\\
\mathcal{M}_{35}
&= \frac{i}{4} \left[ \left( U^2 - V_1 \right)
                      + \left( V^1 - U_2 \right)
                      - i \left( \widetilde{U}^2 + \widetilde{V}_1 \right)
                      - i \left( \widetilde{V}^1 + \widetilde{U}_2 \right)
               \right]
\\
\mathcal{M}_{45}
&= - \frac{1}{4} \left[ \left( U^2 - V_1 \right)
                        - \left( V^1 - U_2 \right)
                        - i \left( \widetilde{U}^2 + \widetilde{V}_1 \right)
                        + i \left( \widetilde{V}^1 + \widetilde{U}_2 \right)
                 \right]
\end{aligned}
\end{equation}
\begin{equation}
\begin{aligned}
\mathcal{M}_{12}
&= S_0 + A_0
\\
\mathcal{M}_{14}
&= \frac{i}{2} \left( S_+ - S_- - A_+ + A_- \right)
\\
\mathcal{M}_{24}
&= - \frac{1}{2} \left( S_+ + S_- - A_+ - A_- \right)
\end{aligned}
\qquad \qquad
\begin{aligned}
\mathcal{M}_{13}
&= \frac{1}{2} \left( S_+ + S_- + A_+ + A_- \right)
\\
\mathcal{M}_{23}
&= \frac{i}{2} \left( S_+ - S_- + A_+ - A_- \right)
\\
\mathcal{M}_{34}
&= S_0 - A_0
\end{aligned}
\end{equation}
\begin{equation}
\mathcal{M}_{05}
= \frac{1}{2} \left[ \Delta + i \left( N_+ - N_- \right) \right]
\end{equation}
\end{subequations}

These Lorentz generators satisfy the $\mathfrak{so}(5,1)$ commutation
relations
\begin{equation}
\commute{\mathcal{M}_{\mu\nu}}{\mathcal{M}_{\rho\tau}}
= i \left(
     \eta_{\nu\rho} \mathcal{M}_{\mu\tau}
     - \eta_{\mu\rho} \mathcal{M}_{\nu\tau}
     - \eta_{\nu\tau} \mathcal{M}_{\mu\rho}
     + \eta_{\mu\tau} \mathcal{M}_{\nu\rho}
    \right)
\end{equation}
where $\eta_{\mu\nu} = \mathrm{diag} (-,+,+,+,+,+)$.
The six generators of grade $+1$ space are the momentum generators
$\mathcal{P}_\mu$, and the six generators of grade $-1$ space are the special
conformal transformations $\mathcal{K}_\mu$ ($\mu = 0,1,2,\dots,5$):
\begin{equation}
\begin{aligned}
\mathcal{P}_0
&= K_+ + \left[ N_0 + \frac{1}{2} \left( N_+ + N_- \right) \right]
\\
\mathcal{P}_1
&= - \frac{1}{2} \left[
                  \left( \widetilde{U}^1 + \widetilde{V}_2 \right)
                  + \left( \widetilde{V}^2 + \widetilde{U}_1 \right)
                 \right]
\\
\mathcal{P}_2
&= - \frac{i}{2} \left[
                  \left( \widetilde{U}^1 + \widetilde{V}_2 \right)
                  - \left( \widetilde{V}^2 + \widetilde{U}_1 \right)
                 \right]
\\
\mathcal{P}_3
&= - \frac{i}{2} \left[
                  \left( \widetilde{U}^2 - \widetilde{V}_1 \right)
                  + \left( \widetilde{V}^1 - \widetilde{U}_2 \right)
                 \right]
\\
\mathcal{P}_4
&= \frac{1}{2} \left[
                \left( \widetilde{U}^2 - \widetilde{V}_1 \right)
                - \left( \widetilde{V}^1 - \widetilde{U}_2 \right)
               \right]
\\
\mathcal{P}_5
&= K_+ - \left[ N_0 + \frac{1}{2} \left( N_+ + N_- \right) \right]
\end{aligned}
\qquad \qquad
\begin{aligned}
\mathcal{K}_0
&= \left[ N_0 - \frac{1}{2} \left( N_+ + N_- \right) \right] + K_-
\\
\mathcal{K}_1
&= \frac{i}{2} \left[
                \left( U^1 - V_2 \right) + \left( V^2 - U_1 \right)
               \right]
\\
\mathcal{K}_2
&= - \frac{1}{2} \left[
                  \left( U^1 - V_2 \right) - \left( V^2 - U_1 \right)
                 \right]
\\
\mathcal{K}_3
&= - \frac{1}{2} \left[
                  \left( U^2 + V_1 \right) + \left( V^1 + U_2 \right)
                 \right]
\\
\mathcal{K}_4
&= - \frac{i}{2} \left[
                  \left( U^2 + V_1 \right) - \left( V^1 + U_2 \right)
                 \right]
\\
\mathcal{K}_5
&= \left[ N_0 - \frac{1}{2} \left( N_+ + N_- \right) \right] - K_-
\end{aligned}
\label{SO*(8)PKgenerators}
\end{equation}
Together with the  generators $\mathcal{M}_{\mu\nu}$ and $\mathcal{D}$, they
satisfy the commutation relations:
\begin{equation}
\begin{split}
\commute{\mathcal{D}}{\mathcal{P}_\mu}
&= + i \, \mathcal{P}_\mu
\qquad \qquad
\commute{\mathcal{D}}{\mathcal{K}_\mu}
 = - i \, \mathcal{K}_\mu
\\
\commute{\mathcal{D}}{\mathcal{M}_{\mu\nu}}
&= \commute{\mathcal{P}_\mu}{\mathcal{P}_\nu}
 = \commute{\mathcal{K}_\mu}{\mathcal{K}_\nu}
 = 0
\\
\commute{\mathcal{P}_\mu}{\mathcal{M}_{\nu\rho}}
&= i \left( \eta_{\mu\nu} \, \mathcal{P}_\rho
            - \eta_{\mu\rho} \, \mathcal{P}_\nu
     \right)
\\
\commute{\mathcal{K}_\mu}{\mathcal{M}_{\nu\rho}}
&= i \left( \eta_{\mu\nu} \, \mathcal{K}_\rho
            - \eta_{\mu\rho} \, \mathcal{K}_\nu
     \right)
\\
\commute{\mathcal{P}_\mu}{\mathcal{K}_\nu}
&= 2 i \left( \eta_{\mu\nu} \, \mathcal{D} + \mathcal{M}_{\mu\nu} \right)
\end{split}
\end{equation}

It is also important to note that the six dimensional Poincar\'e mass operator vanishes identically
\begin{equation}
\mathcal{M}^2 = \eta_{\mu\nu} \mathcal{P}^\mu \mathcal{P}^\nu =0
\end{equation}
for the minimal unitary realization of $SO^*(8)$ given above.

In Appendix \ref{C3GrSO*(8)}, we give the compact 3-grading of $SO^*(8)$ with respect to the maximal compact subgroup $SU(4) \times U(1)_E$:
\begin{equation}
\mathfrak{so}^*(8)
= \mathfrak{C}^- \oplus \mathfrak{C}^0 \oplus \mathfrak{C}^+
\end{equation}
where the  $\mathfrak{u}(1)_E$ generator $H$ that determines the compact
3-grading is
\begin{equation}
H = N_0 + \frac{1}{2} \left( K_+ + K_- \right) \,.
\label{BosonicHamiltonian}
\end{equation}
It is the $AdS$ energy operator or the conformal Hamiltonian when $SO^*(8)$
is taken as the seven-dimensional $AdS$ group or the six-dimensional
conformal group, respectively. It should also be pointed out that, in the 
earlier noncompact 3-grading with respect to $\mathfrak{N}^0 = 
\mathfrak{su}^*(4) \oplus \mathfrak{so}(1,1)_{\mathcal{D}}$, this $AdS$ 
energy corresponds to $\frac{1}{2} \left( \mathcal{K}_0 + \mathcal{P}_0 
\right)$. (See equation (\ref{SO*(8)PKgenerators}).)

In Appendix \ref{C3GrSO*(8)}, we also give the decomposition of the Lie subalgebra $\mathfrak{su}(4)$ 
of $\mathfrak{C}^0$ with respect to its subalgebra 
$ \mathfrak{su}(2)_S \oplus \mathfrak{su}(2)_A \oplus \mathfrak{u}(1)_J$ where the  $U(1)$ charge
\begin{equation}
J = N_0 -\frac{1}{2} \left( K_+ + K_- \right) \,.
\end{equation}
is equal to  $\frac{1}{2} \left( \mathcal{K}_5 - 
\mathcal{P}_5 \right)$ in the above noncompact 3-grading with respect to 
$\mathfrak{N}^0 = \mathfrak{su}^*(4) \oplus 
\mathfrak{so}(1,1)_{\mathcal{D}}$. (See equation (\ref{SO*(8)PKgenerators}).)


\section{Distinguished $SU(1,1)_K$ subgroup of $SO^*(8)$ generated by the
isotonic (singular) oscillators}
\label{SU(1,1)ofSO*(8)}

Note that in terms of the oscillators $a_m$, $b_m$ (and their respective
hermitian conjugates $a^m$, $b^m$) and the coordinate $x$ and its conjugate
momentum $p$, the $\mathfrak{u}(1)$ generator $H$ (as given in equation
(\ref{BosonicHamiltonian})) has the following form:
\begin{equation}
\begin{split}
H &= N_0 + \frac{1}{2} \left( K_+ + K_- \right)
\\
  &= \frac{1}{2} \left( N_a + N_b \right) + 1
     + \frac{1}{4} \left( x^2 + p^2 \right)
     + \frac{1}{8 \, x^2} \left( 8 \, \mathcal{S}^2 + \frac{3}{2} \right)
\\
  &= H_a + H_b + H_\odot
\end{split}
\end{equation}
This $H$ plays the role of the seven dimensional $AdS$ energy operator or the
six dimensional conformal Hamiltonian. $H_a$ and $H_b$ are the contributions
to the Hamiltonian from $a$-type and $b$-type oscillators, that correspond to
standard non-singular harmonic oscillators. On the other hand, $H_\odot$ is
the Hamiltonian of a singular harmonic oscillator with a potential function
\begin{equation}
V \left( x \right) = \frac{G}{x^2}
\qquad \mbox{where} \quad
G = \frac{1}{4} \left( 8 \, \mathcal{S}^2 + \frac{3}{2} \right) \,.
\end{equation}
$H_\odot$ also arises as the Hamiltonian of conformal quantum mechanics
\cite{de Alfaro:1976je} with $G$ playing the role of coupling constant
\cite{Gunaydin:2001bt}. In some literature it is also referred to as the
isotonic oscillator \cite{Casahorran:1995vt,carinena-2007}.

Let us now consider this singular harmonic oscillator Hamiltonian
\begin{equation}
\begin{split}
H_\odot
&= \frac{1}{2} \left( K_+ + K_- \right)
 = \frac{1}{4} \left( x^2 + p^2 \right)
   + \frac{1}{8 \, x^2} \left( 8 \, \mathcal{S}^2 + \frac{3}{2} \right)
\\
&= \frac{1}{4} \left( x^2 - \frac{\partial^2}{\partial x^2} \right)
   + \frac{1}{8 \, x^2} \left( 8 \, \mathcal{S}^2 + \frac{3}{2} \right) \,.
\end{split}
\label{SingularHamiltonian}
\end{equation}
Together with the following generators $B_-$ and $B_+$ of $\mathfrak{C}^-$
and $\mathfrak{C}^+$ subspaces of $SO^*(8)$ (see Appendix \ref{C3GrSO*(8)}):
\begin{equation}
\begin{split}
B_-
  = \frac{i}{2} \left[ \Delta + i \left( K_+ - K_- \right) \right]
 &= \frac{1}{4} \left( x + i p \right)^2
    - \frac{1}{2 \, x^2} \left( 2 \, \mathcal{S}^2 + \frac{3}{8} \right)
\\
 &= \frac{1}{4} \left( x + \frac{\partial}{\partial x} \right)^2
    - \frac{1}{2 \, x^2} \left( 2 \, \mathcal{S}^2 + \frac{3}{8} \right)
\\
B_+
  = - \frac{i}{2} \left[ \Delta - i \left( K_+ - K_- \right) \right]
 &= \frac{1}{4} \left( x - i p \right)^2
    - \frac{1}{2 \, x^2} \left( 2 \, \mathcal{S}^2 + \frac{3}{8} \right)
\\
 &= \frac{1}{4} \left( x - \frac{\partial}{\partial x} \right)^2
    - \frac{1}{2 \, x^2} \left( 2 \, \mathcal{S}^2 + \frac{3}{8} \right)
\end{split}
\end{equation}
$H_\odot$ generates the distinguished  $\mathfrak{su}(1,1)_K$
subalgebra\footnote{This is the $SU(1,1)$ subgroup generated by the longest
root vector.}
\begin{equation}
\commute{B_-}{B_+} = 2 \, H_\odot
\qquad \qquad
\commute{H_\odot}{B_+} = + \, B_+
\qquad \qquad
\commute{H_\odot}{B_-} = - \, B_- \,.
\end{equation}
For a given eigenvalue $\mathfrak{s} \left( \mathfrak{s} + 1 \right)$ of the 
quadratic Casimir operator $\mathcal{S}^2 $ of $SU(2)_S$, the wave functions 
corresponding to the lowest energy eigenvalue of this singular harmonic 
oscillator Hamiltonian $H_\odot$ will be superpositions of functions of the 
form $\psi_0^{(\alpha_\mathfrak{s})} \left( x \right) \Lambda 
\left( \mathfrak{s} , m_\mathfrak{s} \right)$, where $\Lambda \left( 
\mathfrak{s} . m_\mathfrak{s} \right)$ is an eigenstate of $\mathcal{S}^2$ 
and $S_0$, independent of $x$:
\begin{equation}
\mathcal{S}^2 \Lambda \left( \mathfrak{s} , m_\mathfrak{s} \right)
= \mathfrak{s} \left( \mathfrak{s} + 1 \right)
  \Lambda \left( \mathfrak{s} , m_\mathfrak{s} \right)
\qquad \qquad
S_0 \, \Lambda \left( \mathfrak{s} , m_\mathfrak{s} \right)
= m_\mathfrak{s} \, \Lambda \left( \mathfrak{s} , m_\mathfrak{s} \right)
\end{equation}
and $\psi_0^{(\alpha_\mathfrak{s})} \left( x \right)$ is a function that 
satisfies
\begin{equation}
B_- \, \psi_0^{(\alpha_\mathfrak{s})} \left( x \right) 
\Lambda \left( \mathfrak{s} , m_\mathfrak{s} \right)
= 0
\end{equation}
whose solution is given by \cite{MR858831}
\begin{equation}
\psi_0^{(\alpha_\mathfrak{s})} \left( x \right)
= C_0 \, x^{\alpha_\mathfrak{s}} e^{-x^2/2}
\label{singularwavefunctions}
\end{equation}
where $C_0$ is a normalization constant and
\begin{equation}
\alpha_\mathfrak{s}
= \frac{1}{2} + \sqrt{1 + 4 \, \mathfrak{s} \left( \mathfrak{s} + 1 \right)}
= 2 \mathfrak{s} + \frac{3}{2} \,.
\end{equation}

The normalizability of the state imposes the constraint
\begin{equation}
\alpha_\mathfrak{s} \geq \frac{1}{2} \,.
\end{equation}

Clearly, $\psi_0^{(\alpha_\mathfrak{s} = 2 \mathfrak{s} + 3/2)} \left( x 
\right) \Lambda \left( \mathfrak{s} , m_\mathfrak{s} \right)$ is an 
eigenstate of $H_\odot$ with eigenvalue $E_{\odot,0}^{(\alpha_\mathfrak{s})} 
= \left( \mathfrak{s} + 1 \right)$:
\begin{equation}
H_\odot \, \psi_0^{(2 \mathfrak{s} + 3/2)} \left( x \right)
\Lambda \left( \mathfrak{s} , m_\mathfrak{s} \right)
= \left( \mathfrak{s} + 1 \right) \, 
  \psi_0^{(2 \mathfrak{s} + 3/2)} \left( x \right)
  \Lambda \left( \mathfrak{s} , m_\mathfrak{s} \right) \,.
\end{equation}
The lowest energy normalizable eigenstate of $H_\odot$ corresponds to the
case $\mathfrak{s} = 0$ (therefore $\alpha_\mathfrak{s} = \frac{3}{2}$). Note 
that $\Lambda \left( 0 , 0 \right)$ is simply the Fock vacuum $\ket{0}$ of 
$a$- and $b$-type oscillators. This ``ground state'' has energy
\begin{equation}
E_{\odot,0}^{(3/2)} = 1 \,.
\end{equation}
The higher energy eigenstates of $H_\odot$ can be obtained from
$\psi_0^{(3/2)} \left( x \right) \Lambda \left( 0 , 0 \right)$ by acting on 
it repeatedly with the raising operator $B_+$:
\begin{equation}
\psi_n^{(3/2)} \left( x \right) \Lambda \left( 0 , 0 \right)
 = C_n \, \left( B_+ \right)^n 
   \psi_0^{(3/2)} \left( x \right) \Lambda \left( 0 , 0 \right)
\label{isotonicgroundstate}
\end{equation}
where $C_n$ are normalization constants. They correspond to energy
eigenvalues:
\begin{equation}
H_\odot \, \psi_n^{(3/2)} \left( x \right) \Lambda \left( 0 , 0 \right)
 = E_{\odot,n}^{(3/2)} \, 
   \psi_n^{(3/2)} \left( x \right) \Lambda \left( 0 , 0 \right)
 \end{equation}
where
\begin{equation}
E_{\odot,n}^{(3/2)}
 = E_{\odot,0}^{(3/2)} + n
 = 1 + n \,.
\end{equation}
We shall denote the corresponding states as $\ket{\psi_n^{(3/2)} \left( x
\right) \Lambda \left( 0 , 0 \right)} = \ket{\psi_n^{(3/2)} \left( x \right)} 
\otimes \ket{\Lambda \left( 0 , 0 \right)}$ and refer to them as the particle 
basis of the state space of the (isotonic) singular oscillator.


\section{$SU(2)_S \times SU(2)_A \times U(1)_J$  Basis  of the
Minimal Unitary Representation  of $SO^*(8)$ }
\label{SU2SU2U1}

Consider the tensor product of Fock spaces of the oscillators $a^m$ and
$b^m$. The vacuum state $\ket{0}$ is annihilated by all $a_m$ and $b_m$:
\begin{equation}
a_m \ket{0} = b_m \ket{0} = 0
\end{equation}
where $m = 1,2$. Note that $\ket{\Lambda \left( 0 , 0 \right)} = \ket{0}$. A 
``particle basis'' of states in this  Fock space is provided by tensor 
products of the following states
\begin{equation}
\ket{n_{a,m}}
= \frac{1}{\sqrt{n_{a,m} !}} \left( a^m \right)^{n_{a,m}} \ket{0}
\qquad \qquad \qquad
\ket{n_{b,m}}
= \frac{1}{\sqrt{n_{b,m} !}} \left( b^m \right)^{n_{b,m}} \ket{0}
\end{equation}
where $m = 1,2$ and $n_{a,m}$ and  $n_{a,m}$ are non-negative integers.

To construct a "particle basis" of the  Hilbert space of the minimal unitary
representation of $SO^*(8)$, consider  the following tensor products of 
the above states with the state space of the singular (isotonic) oscillator:
\begin{equation*}
\ket{n_{a,1}} \otimes \ket{n_{a,2}} \otimes \ket{n_{b,1}} \otimes
\ket{n_{b,2}} \otimes
\ket{\psi_n^{(\alpha_\mathfrak{s})} \left( x \right)
\Lambda \left( 0 , 0 \right)}
\end{equation*}
and denote them as
\begin{equation*}
\left( a^1 \right)^{n_{a,1}} \left( a^2 \right)^{n_{a,2}}
\left( b^1 \right)^{n_{b,1}} \left( b^2 \right)^{n_{b,2}}
\ket{\psi_n^{(\alpha_\mathfrak{s})} \left( x \right)}
\end{equation*}
or simply as
\begin{equation*}
\ket{\psi_n^{(\alpha_\mathfrak{s})} \left( x \right) \,;\, 
     n_{a,1} , n_{a,2} , n_{b,1} , n_{b,2}} \,.
\end{equation*}

For a fixed $N = n_{a,1} + n_{a,2} + n_{b,1} + n_{b,2}$, these states 
transform in the $\left( \frac{N}{2} , \frac{N}{2} \right)$ representation 
under the $SU(2)_S \times SU(2)_A$ subgroup:
\begin{equation}
\begin{aligned}
\mathfrak{s}
&= \frac{N}{2}
\\
\mathfrak{s}_3
&= \frac{1}{2} \left( n_{a,1} + n_{a,2} - n_{b,1} - n_{b,2} \right)
\end{aligned}
\qquad \qquad
\begin{aligned}
\mathfrak{a}
&= \frac{N}{2}
\\
\mathfrak{a}_3
&= \frac{1}{2} \left( n_{a,1} - n_{a,2} + n_{b,1} - n_{b,2} \right)
\end{aligned}
\end{equation}
However they are, in general, not eigenstates of $U(1)_J$ or the energy 
operator $H$ ($AdS_7$ energy or $6D$ conformal Hamiltonian) that determines 
the compact 3-grading of $SO^*(8)$, given in appendix \ref{C3GrSO*(8)}, and 
commutes with $SU(4)$ generators.

There  exists a unique lowest energy state in this Hilbert space, namely
\begin{equation}
\ket{\psi_0^{(3/2)} \left( x \right) \,;\, 0 , 0 , 0 , 0}
\end{equation}
that is annihilated by the following six operators in $\mathfrak{C}^-$ 
subspace of $\mathfrak{so}^*(8)$:
\begin{equation}
\begin{aligned}
Y_m &= \frac{1}{2} \left( U_m - i \, \widetilde{U}_m \right)
\\
Z_m &= \frac{1}{2} \left( V_m - i \, \widetilde{V}_m \right)
\end{aligned}
\qquad \qquad \qquad
\begin{aligned}
N_- &= a_1 b_2 - a_2 b_1
\\
B_- &= \frac{i}{2} \left[ \Delta + i \left( K_+ - K_- \right) \right]
\end{aligned}
\end{equation}
in the compact three grading, and transforms as a singlet of $SU(2)_S \times 
SU(2)_A$ and is an eigenstate of $H$ and $J$ with  eigenvalue $E = 2$ and 
$\mathfrak{J}=0$, respectively. This state is also a singlet of $SU(4)$ 
subgroup. Hence the minimal unitary representation of $SO^*(8)$ is a unitary 
lowest weight representation. All the other states of the particle basis of 
the minrep with higher energies can be obtained from $\ket{\psi_0^{(3/2)} 
\left( x \right) \,;\, 0 , 0 , 0 , 0}$ by repeatedly acting on it with the 
following operators in $\mathfrak{C}^+$ subspace of $\mathfrak{so}^*(8)$:
\begin{equation}
\begin{aligned}
Y^m &= \frac{1}{2} \left( U^m + i \, \widetilde{U}^m \right)
\\
Z^m &= \frac{1}{2} \left( V^m + i \, \widetilde{V}^m \right)
\end{aligned}
\qquad \qquad \qquad
\begin{aligned}
N_+ &= a^1 b^2 - a^2 b^1
\\
B_+ &= - \frac{i}{2} \left[ \Delta - i \left( K_+ - K_- \right) \right]
\end{aligned}
\end{equation}

The above six operators in $\mathfrak{C}^+$ transform under 
$\mathfrak{su}(2)_S \oplus \mathfrak{su}(2)_A \oplus \mathfrak{u}(1)_J$ as 
follows:
\begin{equation*}
6 = ( 1/2 , 1/2 )^0 \oplus
    ( 0 , 0 )^{+1} \oplus
    ( 0 , 0 )^{-1} \,.
\end{equation*}
The operators $\left( Y^1 , Z^1 \right)$ and $\left( Y^2 , Z^2 \right)$ form 
two doublets under $\mathfrak{su}(2)_S$. The operators $\left( Y^1 , Y^2 
\right)$ and $\left( Z^1 , Z^2 \right)$ form two doublets under 
$\mathfrak{su}(2)_A$. $N_+$ and $B_+$ are both singlets under 
$\mathfrak{su}(2)_S$ and $\mathfrak{su}(2)_A$. $Y^m$ and $Z^m$ have zero
$J$-charge, while $N_+$ and $B_+$ have $J$-charges $+1$ and $-1$,
respectively.

We list the charges of these  six operators with respect to $\left( S_0 , 
A_0 , J \right)$ in Table \ref{Table:Grade+1generators}.

\begin{small}
\begin{longtable}[c]{|c||r|r|r|r|r|r|}
\kill

\caption[Charges $\mathfrak{s}_3$, $\mathfrak{a}_3$ and $\mathfrak{J}$ of 
$\mathfrak{C}^+$ operators of $\mathfrak{so}^*(8)$ with respect to $S_0$, 
$A_0$ and $J$, respectively.]
{Charges $\mathfrak{s}_3$, $\mathfrak{a}_3$ and $\mathfrak{J}$ of 
$\mathfrak{C}^+$ operators of $\mathfrak{so}^*(8)$ with respect to $S_0$, 
$A_0$ and $J$, respectively.
\label{Table:Grade+1generators}} \\
\hline
& & & & & & \\
$\mathfrak{C}^+$ generator & $Y^1$ & $Y^2$ & $Z^1$ & $Z^2$ &
$N_+$ & $B_+$ \\
& & & & & & \\
\hline
& & & & & & \\
\endfirsthead
\caption[]{(continued)} \\
\hline
& & & & & & \\
$\mathfrak{C}^+$ generator & $Y^1$ & $Y^2$ & $Z^1$ & $Z^1$ &
$N_+$ & $B_+$ \\
& & & & & & \\
\hline
& & & & & & \\
\endhead
& & & & & & \\
\hline
\endfoot
& & & & & & \\
\hline
\endlastfoot

$\mathfrak{s}_3$ &
$+\frac{1}{2}$ & $+\frac{1}{2}$ & $-\frac{1}{2}$ & $-\frac{1}{2}$ & 0 & 0
\\[8pt]

\hline
 & & & & & & \\

$\mathfrak{a}_3$ &
$+\frac{1}{2}$ & $-\frac{1}{2}$ & $+\frac{1}{2}$ & $-\frac{1}{2}$ & 0 & 0
\\[8pt]

\hline
 & & & & & & \\

$\mathfrak{J}$ & 0 & 0 & 0 & 0 & $+1$ & $-1$
\\[8pt]

\end{longtable}
\end{small}

The $\mathfrak{C}^+$ operators commute with each other and satisfy  the 
following important relation:
\begin{equation}
Y^1 Z^2 - Y^2 Z^1 = N_+ B_+
\end{equation}
All the states that belong to a given $AdS$ energy level form an irrep of 
$SU(4)$. We give the $SU(2)_S \times SU(2)_A \times U(1)_J$ decomposition 
of these $SU(4)$ irreps  in Table \ref{Table:ScalarDoubleton} for the first 
three energy levels together with their respective $SU(4)$ Dynkin 
labels.\footnote{Our convention of Dynkin labels is such that, the 
fundamental representation $\mathbf{4}$ of $SU(4)$ corresponds to $(1,0,0)$.}

\begin{small}
\begin{longtable}[c]{|l|c|c|}
\kill

\caption[The $SU(2)_S \times SU(2)_A \times U(1)_J$ content of the first three energy levels of minimal
unitary representation of $SO^*(8)$]
{The $SU(2)_S \times SU(2)_A \times U(1)_J$ content of the first three energy levels of minimal unitary
representation of $SO^*(8)$ and their $SU(4)$ Dynkin labels. The irreps are labeled by their spins
$\left( \mathfrak{s} , \mathfrak{a} \right)$ and the $J$-charge as
$\left( \mathfrak{s} , \mathfrak{a} \right)^{\mathfrak{J}}$. $\mathfrak{s}$, $\mathfrak{a}$, $\mathfrak{J}$ and $E$ labels of the lowest energy 
state $ \ket{\psi_0^{(3/2)} \left( x \right) \,;\, 0 , 0 , 0 , 0}$ uniquely identify the $SO^*(8)$ irrep.
\label{Table:ScalarDoubleton}} \\
\hline
& & \\
Irrep $\left( \mathfrak{s} , \mathfrak{a} \right)^{\mathfrak{J}}$ &
$E$ & $SU(4)_{Dynkin}$ \\
 &
 &  \\
& & \\
\hline
& & \\
\endfirsthead
\caption[]{(continued)} \\
\hline
& & \\
Irrep $\left( \mathfrak{s} , \mathfrak{a} \right)^{\mathfrak{J}}$ &
$E$ & $SU(4)$ \\
 &
 & Dynkin \\
& & \\
\hline
& & \\
\endhead
& & \\
\hline
\endfoot
& & \\
\hline
\endlastfoot

$\left( 0 , 0 \right)^0$ &
2 & $(0,0,0)$ \\[8pt]

\hline
 & & \\

$\left( 0 , 0 \right)^{-1}
\oplus \left( 0 , 0 \right)^{+1}$ &
3 & $(0,1,0)$
\\[8pt]

$\oplus \left( \frac{1}{2} , \frac{1}{2} \right)^0$ &
 &
\\[8pt]

\hline
 & & \\

$\left( 0 , 0 \right)^{-2}
\oplus \left( 0 , 0 \right)^0
\oplus \left( 0 , 0 \right)^{+2}$ &
4 & $(0,2,0)$
\\[8pt]

$\oplus \left( \frac{1}{2} , \frac{1}{2} \right)^{-1}
\oplus \left( \frac{1}{2} , \frac{1}{2} \right)^{+1}$ &
 &
\\[8pt]

$\oplus \left( 1 , 1 \right)^0$ &
 &
\\[8pt]

\hline
 & & \\

\qquad\vdots & \vdots & \quad\vdots \\ [8pt]

\end{longtable}
\end{small}

By acting on the lowest weight state $\ket{\Omega}$ with $\mathfrak{C}^+$
generators $n$ times, one obtains a set of states that are eigenstates of $H$
with energy eigenvalues $n + 2$. They form an $SU(4)$ irrep with Dynkin
labels $(0,n,0)$, which decomposes into the following irreps of $SU(2)_S
\times SU(2)_A \times U(1)_J$ subgroup labelled by $\left( \mathfrak{s} , \mathfrak{a}
\right)^{\mathfrak{J}}$:
\begin{equation}
\begin{split}
\left( 0 , n , 0 \right)_{SU(4) \mathrm{~Dynkin}}
&= \left( 0 , 0 \right)^{-n} \oplus
   \left( 0 , 0 \right)^{-n+2} \oplus
   \dots \oplus \left( 0 , 0 \right)^{n}
\\
& \quad \oplus
   \left( \frac{1}{2} , \frac{1}{2} \right)^{-n+1} \oplus
   \left( \frac{1}{2} , \frac{1}{2} \right)^{-n+3} \oplus
   \dots \oplus \left( \frac{1}{2} , \frac{1}{2} \right)^{n-1}
\\
& \quad \oplus
   \left( 1 , 1 \right)^{-n+2} \oplus
   \left( 1 , 1 \right)^{-n+4} \oplus
   \dots \oplus \left( 1 , 1 \right)^{n-2}
\\
& \qquad \vdots
\\
& \quad \oplus
   \left( \frac{n}{2} , \frac{n}{2} \right)^0
\end{split}
\end{equation}

Comparing the $SU(4)$ content of the minrep of $SO^*(8)$ with that of the 
scalar doubleton representation of the seven dimensional $AdS$ group $SO(6,2) 
\simeq SO^*(8)$ obtained by the oscillator method 
\cite{Gunaydin:1984wc,Gunaydin:1999ci,Fernando:2001ak}, we
see that they coincide exactly. Thus the minrep of $SO^*(8)$ is simply the
scalar doubleton representation of $SO^*(8)$ whose Poincar\'{e} limit in
$AdS_7$ is  singular, just like Dirac's singletons of $SO(3,2)$ in $AdS_4$.
The doubleton representations correspond to massless representations of
$SO^*(8)$ considered as six dimensional conformal group
\cite{Gunaydin:1984wc,Fernando:2001ak}.

At this point we should stress one  important point. In the oscillator
construction of the scalar doubleton given in \cite{Gunaydin:1984wc}, one is
working in the Fock space of two sets of oscillators transforming in the
fundamental representation of the maximal compact subgroup  $U(4)$. The Fock
space of all eight oscillators decomposes into an infinite family of
doubleton representations that correspond to massless conformal fields of
ever increasing spin. In contrast, the minimal unitary representation we
constructed above is over the tensor product of Fock space of four
oscillators and the state space of a singular oscillator. 

In the subsequent sections we shall extend our construction of the minrep of
$SO^*(8)$ to the construction of  minimal unitary representation of the
supergroup $OSp(8^*|2N)$ which correspond to supermultiplets of massless
conformal fields in six dimensions.


\section{Construction of  Finite-Dimensional Representations of $USp(2N)$ in
terms of Fermionic Oscillators}
\label{USp(2N)}

We define two sets of $N$ fermionic oscillators $\alpha_r$, $\beta_r$ and
their hermitian conjugates $\alpha^r = \left( \alpha_r \right)^\dag$,
$\beta^r = \left( \beta_r \right)^\dag$ ($r = 1,2,\dots,N$), such that they
satisfy the usual anti-commutation relations:
\begin{equation}
\anticommute{\alpha_r}{\alpha^s}
= \anticommute{\beta_r}{\beta^s}
= \delta^s_r
\qquad \qquad
\anticommute{\alpha_r}{\alpha_s}
= \anticommute{\alpha_r}{\beta_s}
= \anticommute{\beta_r}{\beta_s}
= 0
\end{equation}

The Lie algebra $\mathfrak{usp}(2N)$ has a 3-graded decomposition with
respect to its subalgebra $\mathfrak{u}(N)$ as follows:
\begin{equation}
\begin{split}
\mathfrak{usp}(2N)
&= \mathfrak{g}^{(-1)} \, \oplus \, \mathfrak{g}^{(0)} \, \oplus \,
\mathfrak{g}^{(+1)}
\\
&= \, S_{rs} \, \oplus \, M^r_{~s} \, \oplus \, S^{rs}
\end{split}
\end{equation}
where
\begin{equation}
\begin{split}
S_{rs}
&= \alpha_r \beta_s + \alpha_s \beta_r
\\
M^r_{~s}
&= \alpha^r \alpha_s - \beta_s \beta^r
\\
S^{rs}
&= \beta^r \alpha^s + \beta^s \alpha^r
 = \left( S_{rs} \right)^\dag \,.
\end{split}
\label{USp(2N)generators}
\end{equation}
They satisfy the commutation relations:
\begin{equation}
\begin{split}
\commute{S_{rs}}{S^{tu}}
&= - \delta^t_s \, M^u_{~r} - \delta^t_r \, M^u_{~s}
   - \delta^u_s \, M^t_{~r} - \delta^u_r \, M^t_{~s}
\\
\commute{M^r_{~s}}{S_{tu}}
&= - \delta^r_u \, S_{st} - \delta^r_t \, S_{su}
\\
\commute{M^r_{~s}}{S^{tu}}
&= \delta^u_s \, S^{rt} + \delta^t_s \, S^{ru}
\\
\commute{M^r_{~s}}{M^t_{~u}}
&= \delta^t_s \, M^r_{~u} - \delta^r_u \, M^t_{~s}
\end{split}
\end{equation}
The quadratic Casimir of $\mathfrak{usp}(2N)$ is given by
\begin{equation}
\begin{split}
\mathcal{C}_2 \left[ \mathfrak{usp}(2N) \right]
&= M^r_{~s} M^s_{~r}
   + \frac{1}{2} \left( S_{rs} S^{rs} + S^{rs} S_{rs} \right)
\\
&= N \left( N + 2 \right)
   - \left( N_\alpha + N_\beta \right)
     \left[ \left( N_\alpha + N_\beta \right) + 2 \right]
   - 8 \, \alpha^{(r} \beta^{s)} \, \alpha_{(r} \beta_{s)}
\end{split}
\end{equation}
where ``$(rs)$'' represents symmetrization of weight one, $\alpha_{(r}
\beta_{s)} = \frac{1}{2} \left( \alpha_r \beta_s + \alpha_s \beta_r \right)$.

We choose the fermionic Fock vacuum such that
\begin{equation}
\alpha_r \ket{0}
= \beta_r \ket{0}
= 0 \,.
\end{equation}

To generate an irrep of $USp(2N)$ in the Fock space in a $U(N)$ basis, one
chooses a set of states $\ket{\Omega}$, transforming irreducibly under $U(N)$
and are annihilated by all grade $-1$ operators $S_{rs}$ (i.e. $S_{rs}
\ket{\Omega} = 0$), and act on them with grade $+1$ operators $S^{rs}$
\cite{Gunaydin:1990ag}:
\begin{equation}
\left\{
\ket{\Omega} \,,\,
S^{rs} \ket{\Omega} \,,\,
S^{rs} S^{tu} \ket{\Omega} \,,\,
\dots\dots
\right\}
\end{equation}
The possible sets of states $\ket{\Omega}$, that transform irreducibly under
$U(N)$ and are annihilated by $S_{rs}$, are of the form
\begin{equation}
\alpha^{r_1} \alpha^{r_2} \dots \alpha^{r_m} \ket{0}
\end{equation}
or of the equivalent form
\begin{equation}
\beta^{r_1} \beta^{r_2} \dots \beta^{r_m} \ket{0}
\end{equation}
where $m \le N$. They lead to irreps of $USp(2N)$ with Dynkin labels
\cite{Gunaydin:1990ag}
\begin{equation}
( \, \underbrace{0 , \dots , 0}_{(N-m-1)} , 1 ,
\underbrace{0 , \dots , 0}_{(m)} \, ) \,.
\end{equation}
In addition, we have the following states
\begin{equation}
\alpha^{[r} \beta^{s]} \ket{0}
= \frac{1}{2} \left( \alpha^r \beta^s - \alpha^s \beta^r \right) \ket{0}
\end{equation}
that are annihilated by all grade $-1$ operators $S_{tu}$. They lead to the
irrep of $USp(2N)$ with Dynkin labels
\begin{equation}
( \, \underbrace{0 , \dots , 0}_{(N-3)} , 1 , 0 , 0 \, ) \,.
\end{equation}
Note that in the special case of $\mathfrak{usp}(4)$, the states $\alpha^r
\alpha^s \ket{0}$, $\beta^r \beta^s \ket{0}$ and $\alpha^{[r} \beta^{s]}
\ket{0}$ all lead to the trivial representation.

It is important to also note that the following bilinears of fermionic
oscillators
\begin{equation}
F_+ = \alpha^r \beta_r
\qquad \qquad
F_- = \beta^r \alpha_r
\qquad \qquad
F_0 = \frac{1}{2} \left( N_\alpha - N_\beta \right)
\label{SU(2)F}
\end{equation}
where $N_\alpha = \alpha^r \alpha_r$ and $N_\beta = \beta^r \beta_r$ are the
respective number operators, generate a $\mathfrak{usp}(2)_F \simeq
\mathfrak{su}(2)_F$ algebra
\begin{equation}
\commute{F_+}{F_-} = 2 \, F_0
\qquad \qquad \qquad
\commute{F_0}{F_\pm} = \pm F_\pm
\end{equation}
that commutes with the $\mathfrak{usp}(2N)$ algebra defined above.
Nonetheless, the equivalent irreps of $USp(2N)$ constructed from the states
$\ket{\Omega}$ involving only $\alpha$-type excitations or $\beta$-type
excitations can form non-trivial representations of this $USp(2)_F$.

For example, the two irreps with Dynkin labels (1,0) of $USp(4)$ constructed 
from $\alpha^r \ket{0}$ and $\beta^r \ket{0}$ form spin $\frac{1}{2}$
representation (doublet) of $USp(2)_F$. The three singlet irreps of
$USp(4)$ corrsponding to $\alpha^r \alpha^s \ket{0}$, $\beta^r \beta^s
\ket{0}$ and $\alpha^{[r} \beta^{s]} \ket{0}$ form the spin 1 representation
(triplet) of $USp(2)_F$. The irrep of $USp(4)$ with Dynkin labels
(0,1) defined by the vacuum state $\ket{\Omega} = \ket{0}$ is a singlet of 
$USp(2)_F$.

We should note that the representations of $USp(2N)$ obtained above by
using two sets of fermionic oscillators transforming in the fundamental
representation of the subgroup $U(N)$ are the compact analogs of the
doubleton representations of $SO^*(2M)$ constructed using two sets of bosonic
oscillators transforming in the fundamental representation of $U(M)$
\cite{Gunaydin:1990ag}. By realizing the generators of $USp(2N)$ in terms of
an arbitrary (even) number of sets of oscillators, one can construct all the
finite dimensional representations of $USp(2N)$
\cite{Gunaydin:1990ag,Gunaydin:1984wc}.


\section{Minimal Unitary Representations of Supergroups $OSp(8^*|2N)$ with
Even Subgroups $SO^*(8) \times USp(2N)$  }
\label{minrepOSp(8*|2N)-5Gr}

The noncompact groups $SO^*(2M)$ with maximal compact subgroups $U(M)$ have
extensions to supergroups $OSp(2M^*|2N)$ with the even subgroups $SO^*(2M)
\times USp(2N)$ that admit positive energy unitary representations. In this
section we shall study the minimal unitary representations of  $OSp(8^*|2N)$,
leaving to a separate study the minimal representations of  $OSp(2M^*|2N)$
for arbitrary $M$ \cite{FGP2010}. We define minimal unitary representations
(supermultiplets) of $OSp(2M^*|2N)$ as those irreducible unitary representations ( supermultiplets) that
contain the minimal unitary representation of $SO^*(2M)$.

The superalgebra
$\mathfrak{osp}(8^*|2N)$  has a 5-grading
\begin{equation}
\mathfrak{osp}(8^*|2N)
= \mathfrak{g}^{(-2)} \oplus
  \mathfrak{g}^{(-1)} \oplus
  \mathfrak{g}^{(0)} \oplus
  \mathfrak{g}^{(+1)} \oplus
  \mathfrak{g}^{(+2)}
\end{equation}
with respect to the subsuperalgebra
\begin{equation}
\mathfrak{g}^{(0)}
= \mathfrak{osp}(4^*|2N) \oplus
  \mathfrak{su}(2) \oplus
  \mathfrak{so}(1,1)_\Delta \,.
\end{equation}

In the extension of the minimal unitary realization of $SO^*(8)$ to that of
$OSp(8^*|2N)$, the grade $-2$ generator $K_-$ remains unchanged. However, now
grade $-1$ subspace contains both even (bosonic) and odd (fermionic)
generators. More precisely, the grade $-1$ subspace $\mathfrak{g}^{(-1)}$ of
$\mathfrak{osp}(8^*|2N)$ contains 8 bosonic generators:
\begin{equation}
\begin{aligned}
U_m &= x \, a_m
\\
V_m &= x \, b_m
\end{aligned}
\qquad \qquad \qquad
\begin{aligned}
U^m &= x \, a^m
\\
V^m &= x \, b^m
\end{aligned}
\end{equation}
and $4 N$ supersymmetry generators:
\begin{equation}
\begin{aligned}
Q_r &= x \, \alpha_r
\\
S_r &= x \, \beta_r
\end{aligned}
\qquad \qquad \qquad
\begin{aligned}
Q^r &= x \, \alpha^r
\\
S^r &= x \, \beta^r
\end{aligned}
\label{5GsusyGr-1}
\end{equation}
They (anti-)commute into the single bosonic generator $K_-$ in grade $-2$
subspace $\mathfrak{g}^{(-2)}$ as follows:
\begin{equation}
\begin{split}
\commute{U_m}{U^n}
&= \commute{V_m}{V^n}
 = 2 \, \delta^n_m \, K_-
\\
\anticommute{Q_r}{Q^s}
&= \anticommute{S_r}{S^s}
 = 2 \, \delta^s_r \, K_-
\end{split}
\end{equation}
Even and odd generators in $\mathfrak{g}^{(-1)}$ commute with each other and together form a super Heisenberg algebra with
\begin{equation*}
K_- = \frac{1}{2} x^2
\end{equation*}
as its ``central charge.''

Now the $\mathfrak{su}(2)_S$ subalgebra in grade zero subspace
$\mathfrak{g}^{(0)}$ of $\mathfrak{so}^*(8)$ receives contributions from
fermionic oscillators in the supersymmetric extension of $\mathfrak{so}^*(8)$
to $\mathfrak{osp}(8^*|2N)$. The resultant $\mathfrak{su}(2)$ that commutes
with $\mathfrak{osp}(4^*|2N)$ of grade zero subalgebra is simply the diagonal
subalgebra of $\mathfrak{su}(2)_S $ and $\mathfrak{su}(2)_F$ defined earlier
in equation (\ref{SU(2)F}). We shall label it as  $\mathfrak{su}(2)_T$ in the
supersymmetric extension. Its generators are:
\begin{equation}
\begin{split}
T_+
&= S_+ + F_+
 = a^m b_m + \alpha^r \beta_r
\\
T_-
&= S_- + F_-
 = b^m a_m + \beta^r \alpha_r
\\
T_0
&= S_0 + F_0
 = \frac{1}{2}
   \left( N_a - N_b + N_\alpha - N_\beta \right)
\end{split}
\end{equation}
so that
\begin{equation}
\commute{T_+}{T_-} = 2 \, T_0
\qquad \qquad \qquad
\commute{T_0}{T_\pm} = \pm T_\pm \,.
\end{equation}

The subsuperalgebra $\mathfrak{osp}(4^*|2N)$ belonging to grade zero subspace
$\mathfrak{g}^{(0)}$ has an even subalgebra $\mathfrak{so}^*(4) \oplus
\mathfrak{usp}(2N)$. The generators of $\mathfrak{so}^*(4) =
\mathfrak{su}(2)_A \oplus \mathfrak{su}(1,1)_N$ were denoted as
$A_{\pm,0}$ and $N_{\pm,0}$ (see equation(\ref{SU(2)AN_generators})),
and the generators of $\mathfrak{usp}(2N)$ were denoted as $S_{rs}$,
$M^r_{~s}$, $S^{rs}$ ($r,s,\dots = 1,\dots,N$) (see equation
(\ref{USp(2N)generators})).

The $8 N$ supersymmetry generators of $\mathfrak{osp}(4^*|2N)$ are realized
by the following bilinears:
\begin{equation}
\begin{aligned}
\Pi_{mr}
&= a_m \beta_r - b_m \alpha_r
\\
\Sigma_m^{~r}
&= a_m \alpha^r + b_m \beta^r
\end{aligned}
\qquad \qquad
\begin{aligned}
\overline{\Pi}^{mr}
&= \left( \Pi_{mr} \right)^\dag
 = a^m \beta^r - b^m \alpha^r
\\
\overline{\Sigma}^m_{~r}
&= \left( \Sigma_m^{~r} \right)^\dag
 = a^m \alpha_r + b^m \beta_r
\end{aligned}
\end{equation}

Recalling that
\begin{equation}
\mathcal{C}_2 \left[ \mathfrak{su}(2)_A \right]
= \mathcal{C}_2 \left[ \mathfrak{su}(1,1)_N \right]
= \mathcal{C}_2 \left[ \mathfrak{su}(2)_S \right]
\end{equation}
we find that the quadratic Casimir of $\mathfrak{osp}(4^*|2N)$ can be
written as
\begin{equation}
\mathcal{C}_2 \left[ \mathfrak{osp}(4^*|2N) \right]
= \mathcal{C}_2 \left[ \mathfrak{su}(2)_S \right]
  - \frac{1}{8} \mathcal{C}_2 \left[ \mathfrak{usp}(2N) \right]
  + \frac{1}{4} \mathcal{F} \left( \Pi , \Sigma \right)
\end{equation}
where
\begin{equation}
\mathcal{F} \left( \Pi , \Sigma \right)
= \Pi_{mr} \, \overline{\Pi}^{mr}
  - \overline{\Pi}^{mr} \, \Pi_{mr}
  + \Sigma_m^{~r} \, \overline{\Sigma}^m_{~r}
  - \overline{\Sigma}^m_{~r} \, \Sigma_m^{~r} \,.
\end{equation}
Remarkably, it reduces  to the quadratic Casimir of $SU(2)_T$ modulo an additive constant for the minimal unitary realization
\begin{equation}
\mathcal{C}_2 \left[ \mathfrak{osp}(4^*|2N) \right]
= \mathcal{C}_2 \left[ \mathfrak{su}(2)_{T} \right]
  - \frac{N \left( N - 4 \right)}{16}
\end{equation}
 where
\begin{equation}
\mathcal{C}_2 \left[ \mathfrak{su}(2)_{T} \right]
= {T_0}^2
  + \frac{1}{2} \left( T_+ \, T_- + T_- \, T_+ \right)
\equiv \mathcal{T}^2 \,.
\end{equation}

Using this result, one can write the grade $+2$ generator in full generality
as
\begin{equation}
K_+
= \frac{1}{2} p^2
  + \frac{1}{4 \, x^2} \left( 8 \, \mathcal{T}^2 + \frac{3}{2} \right)
\end{equation}
that is valid for all $N$.
The generators in grade $+1$ subspace $\mathfrak{g}^{(+1)}$ are obtained from
the commutators of $\mathfrak{g}^{(-1)}$ generators with $K_+$:
\begin{equation}
\begin{aligned}
\widetilde{U}_m &= i \commute{U_m}{K_+}
\\
\widetilde{V}_m &= i \commute{V_m}{K_+}
\\
\widetilde{Q}_r &= i \commute{Q_r}{K_+}
\\
\widetilde{S}_r &= i \commute{S_r}{K_+}
\end{aligned}
\qquad \qquad \qquad
\begin{aligned}
\widetilde{U}^m &= i \commute{U^m}{K_+}
\\
\widetilde{V}^m &= i \commute{V^m}{K_+}
\\
\widetilde{Q}^r &= i \commute{Q^r}{K_+}
\\
\widetilde{S}^r &= i \commute{S^r}{K_+}
\end{aligned}
\end{equation}
The explicit form of the even and odd generators of $\mathfrak{g}^{(+1)}$ are
as follows:
\begin{equation}
\begin{split}
\widetilde{U}_m
&= - p \, a_m
   + \frac{2i}{x}
     \left[
      \left( T_0 + \frac{3}{4} \right) a_m + T_- b_m \right]
\\
\widetilde{U}^m
&= - p \, a^m
   - \frac{2i}{x}
     \left[
      \left( T_0 - \frac{3}{4} \right) a^m + T_+ b^m \right]
\\
\widetilde{V}_m
&= - p \, b_m
   - \frac{2i}{x}
     \left[
      \left( T_0 - \frac{3}{4} \right) b_m - T_+ a_m \right]
\\
\widetilde{V}^m
&= - p \, b^m
   + \frac{2i}{x}
     \left[
      \left( T_0 + \frac{3}{4} \right) b^m - T_- a^m \right]
\end{split}
\end{equation}
\begin{equation}
\begin{split}
\widetilde{Q}_r
&= - p \, \alpha_r
   + \frac{2i}{x}
     \left[
      \left( T_0 + \frac{3}{4} \right) \alpha_r + T_- \beta_r
     \right]
\\
\widetilde{Q}^r
&= - p \, \alpha^r
   - \frac{2i}{x}
     \left[
      \left( T_0 - \frac{3}{4} \right) \alpha^r + T_+ \beta^r
     \right]
\\
\widetilde{S}_r
&= - p \, \beta_r
   - \frac{2i}{x}
     \left[
      \left( T_0 - \frac{3}{4} \right) \beta_r - T_+ \alpha_r
     \right]
\\
\widetilde{S}^r
&= - p \, \beta^r
   + \frac{2i}{x}
     \left[
      \left( T_0 + \frac{3}{4} \right) \beta^r - T_- \alpha^r
     \right]
\end{split}
\label{5GsusyGr+1}
\end{equation}
They form a (super-)Heisenberg algebra together with $K_+$:
\begin{equation}
\begin{split}
\commute{\widetilde{U}_m}{\widetilde{U}^n}
&= \commute{\widetilde{V}_m}{\widetilde{V}^n}
 = 2 \, \delta^n_m \, K_+
\\
\anticommute{\widetilde{Q}_r}{\widetilde{Q}^s}
&= \anticommute{\widetilde{S}_r}{\widetilde{S}^s}
 = 2 \, \delta^s_r \, K_+
\end{split}
\end{equation}
The commutation relations of  grade $+1$ generators with the grade $-2$
generator $K_-$ are:
\begin{equation}
\begin{aligned}
\commute{\widetilde{U}_m}{K_-} &= i \, U_m
\\
\commute{\widetilde{V}_m}{K_-} &= i \, V_m
\\
\commute{\widetilde{Q}_r}{K_-} &= i \, Q_r
\\
\commute{\widetilde{S}_r}{K_-} &= i \, S_r
\end{aligned}
\qquad \qquad
\begin{aligned}
\commute{\widetilde{U}^m}{K_-} &= i \, U^m
\\
\commute{\widetilde{V}^m}{K_-} &= i \, V^m
\\
\commute{\widetilde{Q}^r}{K_-} &= i \, Q^r
\\
\commute{\widetilde{S}^r}{K_-} &= i \, S^r
\end{aligned}
\end{equation}

In terms of the generators defined above  the 5-graded decomposition  of the Lie superalgebra $\mathfrak{osp}(8^*|2N)$,
defined by the  generator $\Delta$, takes the form:
\begin{equation}
\begin{split}
\mathfrak{osp}(8^*|4)
&= \mathfrak{g}^{(-2)} \oplus
   \mathfrak{g}^{(-1)} \oplus
   \left[ \mathfrak{osp}(4^*|2N) \oplus
          \mathfrak{su}(2) \oplus
          \mathfrak{so}(1,1)_\Delta
   \right] \oplus
   \mathfrak{g}^{(+1)} \oplus
   \mathfrak{g}^{(+2)}
\\
&= K_-
   \oplus
   \left[ U_m \,,\, U^m \,,\, V_m \,,\, V^m \,,\,
          Q_r \,,\, Q^r \,,\, S_r \,,\, S^r \right]
\\
& \qquad
   \oplus
   \left[ A_{\pm,0} \,,\, N_{\pm,0} \,,\,
          S_{rs} \,,\, M^r_{~s} \,,\, S^{rs} \,,\,
          \Pi_{mr} \,,\, \overline{\Pi}^{mr} \,,\,
          \Sigma_m^{~r} \,,\, \overline{\Sigma}^m_{~r} \,,\,
          T_{\pm,0} \,,\, \Delta
   \right]
\\
& \qquad \qquad
   \oplus
   \left[ \widetilde{U}_m \,,\, \widetilde{U}^m \,,\,
          \widetilde{V}_m \,,\, \widetilde{V}^m \,,\,
          \widetilde{Q}_r \,,\, \widetilde{Q}^r \,,\,
          \widetilde{S}_r \,,\, \widetilde{S}^r
   \right]
   \oplus
   K_+
\end{split}
\end{equation}

We give the additional  (super-)commutation relations of this superalgebra
in the 5-graded basis in appendix \ref{OSp(8*|2N)-5Gr}.


\section{Compact 3-Grading of $OSp(8^*|2N)$ and its Minimal Unitary
Representation}
\label{minrepOSp(8*|2N)-3Gr}

The Lie superalgebra $\mathfrak{osp}(8^*|2N)$ can be given a 3-graded
decomposition with respect to its compact subsuperalgebra
$\mathfrak{u}(4|N) = \mathfrak{su}(4|N) \oplus \mathfrak{u}(1)_{\mathcal{H}}$
\begin{equation}
\mathfrak{osp}(8^*|2N)
= \mathfrak{C}^- \oplus \mathfrak{C}^0 \oplus \mathfrak{C}^+
\end{equation}
where
\begin{equation}
\begin{split}
\mathfrak{C}^-
&= \frac{1}{2} \left( U_m - i \, \widetilde{U}_m \right) \oplus
   \frac{1}{2} \left( V_m - i \, \widetilde{V}_m \right) \oplus
   N_- \oplus
   \frac{i}{2} \left[ \Delta + i \left( K_+ - K_- \right) \right] \oplus
   S_{rs}
\\
& \qquad \oplus
   \frac{1}{2} \left( Q_r - i \, \widetilde{Q}_r \right) \oplus
   \frac{1}{2} \left( S_r - i \, \widetilde{S}_r \right) \oplus
   \Pi_{mr}
\\
\mathfrak{C}^0
&= \left[
    T_{\pm,0} \oplus
    A_{\pm,0} \oplus
    \left[ N_0 - \frac{1}{2} \left( K_+ + K_- \right) \right] \oplus
    \frac{1}{2} \left( U_m + i \, \widetilde{U}_m \right) \oplus
    \frac{1}{2} \left( U^m - i \, \widetilde{U}^m \right)
   \right.
\\
& \qquad
   \left. \oplus
    \frac{1}{2} \left( V_m + i \, \widetilde{V}_m \right) \oplus
    \frac{1}{2} \left( V^m - i \, \widetilde{V}^m \right) \oplus
    M^r_{~s} \oplus
    \left[ \frac{1}{2} \left( K_+ + K_- \right) + \frac{2}{N} M_0 \right]
   \right] \oplus
   \mathcal{H}
\\
& \quad \oplus
   \frac{1}{2} \left( Q_r + i \, \widetilde{Q}_r \right) \oplus
   \frac{1}{2} \left( Q^r - i \, \widetilde{Q}^r \right) \oplus
   \frac{1}{2} \left( S_r + i \, \widetilde{S}_r \right) \oplus
   \frac{1}{2} \left( S^r - i \, \widetilde{S}^r \right) \oplus
   \Sigma_m^{~r} \oplus
   \overline{\Sigma}^m_{~r}
\\
\mathfrak{C}^+
&= \frac{1}{2} \left( U^m + i \, \widetilde{U}^m \right) \oplus
   \frac{1}{2} \left( V^m + i \, \widetilde{V}^m \right) \oplus
   N_+ \oplus
   - \frac{i}{2} \left[ \Delta - i \left( K_+ - K_- \right) \right] \oplus
   S^{rs}
\\
& \qquad \oplus
   \frac{1}{2} \left( Q^r + i \, \widetilde{Q}^r \right) \oplus
   \frac{1}{2} \left( S^r + i \, \widetilde{S}^r \right) \oplus
   \overline{\Pi}^{mr}
\end{split}
\end{equation}

The $U(1)$ generator $\mathcal{H}$ that defines the compact 3-grading of
$\mathfrak{osp}(8^*|2N)$ is given by
\begin{equation}
\mathcal{H}
= \frac{1}{2} \left( K_+ + K_- \right)
   + N_0 + M_0
\end{equation}
where $M_0$ is the  generator of the $U(1)$ factor of  $U(N)$ subgroup of $USp(2N)$ (equation (\ref{USp(2N)generators})):
\begin{equation}
M_0 = \frac{1}{2} \left( N_\alpha + N_\beta  - N \right)= \frac{1}{2} \left( N_\alpha - N_\beta  \right)
\end{equation}

Therefore
\begin{equation}
\mathcal{H}
= \frac{1}{4} \left( x^2 + p^2 \right)
   + \frac{1}{8 \, x^2}
     \left( 8 \, \mathcal{T}^2 + \frac{3}{2} \right)
   + \frac{1}{2} \left( N_a + N_b + N_\alpha + N_\beta \right)
   + \frac{2 - N}{2} \,.
\label{3-GrGenerator}
\end{equation}
 plays the role of the ``total energy'' operator.

Note that, in the supersymmetric extension, the $\mathfrak{u}(1)$ generator
$H$ in $\mathfrak{so}^*(8)$ (equation (\ref{BosonicHamiltonian2})), which is
the $AdS_7$ energy, that determines its compact 3-grading becomes
\begin{equation}
\begin{split}
H_B
&= \frac{1}{2} \left( K_+ + K_- \right) + N_0
\\
&= \frac{1}{4} \left( x^2 + p^2 \right)
   + \frac{1}{8 \, x^2} \left( 8 \, \mathcal{T}^2 + \frac{3}{2} \right)
   + \frac{1}{2} \left( N_a + N_b \right) + 1
\\
&= H_\odot + H_a + H_b \,.
\end{split}
\label{NonSusyH}
\end{equation}
The Hamiltonian of the singular oscillator now has contributions from the
fermionic oscillators:
\begin{equation}
H_\odot
= \frac{1}{4} \left( x^2 + p^2 \right)
  + \frac{1}{8 \, x^2} \left( 8 \, \mathcal{T}^2 + \frac{3}{2} \right)
\end{equation}
where $\mathcal{T}^2$ is the quadratic Casimir of $SU(2)_T$, the diagonal
subgroup of $SU(2)_S$ and $SU(2)_F$ which are realized in terms of purely
bosonic and purely fermionic oscillators, respectively. $H_a$ and $H_b$
remain unchanged in the supersymmetric extension:
\begin{equation*}
H_a = \frac{1}{2} \left( N_a + 1 \right)
\qquad \qquad \qquad
H_b = \frac{1}{2} \left( N_b + 1 \right)
\end{equation*}

The explicit expressions for the bosonic operators that belong to the
subspace $\mathfrak{C}^-$ of $\mathfrak{osp}(8^*|2N)$ in the compact
3-grading are as follows\footnote{ Note that we are using the same symbols for the generators of $SO^*(8)$ considered as a subgroup of $OSp(8^*|2N)$ that now includes contributions from the fermions. }:
\begin{equation}
\begin{split}
Y_m &= \frac{1}{2} \left( U_m - i \, \widetilde{U}_m \right)
     = \frac{1}{2} \left( x + i \, p \right) a_m
       + \frac{1}{x}
         \left[ \left( T_0 + \frac{3}{4} \right) a_m + T_- b_m \right]
\\
Z_m &= \frac{1}{2} \left( V_m - i \, \widetilde{V}_m \right)
     = \frac{1}{2} \left( x + i \, p \right) b_m
       - \frac{1}{x}
         \left[ \left( T_0 - \frac{3}{4} \right) b_m - T_+ a_m \right]
\\
N_- &= a_1 b_2 - a_2 b_1
\\
B_- &= \frac{i}{2} \left[ \Delta + i \left( K_+ - K_- \right) \right]
     = \frac{1}{4} \left( x + i \, p \right)^2
       - \frac{1}{8 \, x^2} \left( 8 \, \mathcal{T}^2 + \frac{3}{2} \right)
\\
S_{rs} &= \alpha_r \beta_s + \alpha_s \beta_r
\end{split}
\label{OSp(8*|N)Gr-1B}
\end{equation}
The $4 N$ supersymmetry generators in $\mathfrak{C}^-$ subspace are given by:
\begin{equation}
\begin{split}
\mathfrak{Q}_r
&= \frac{1}{2} \left( Q_r - i \, \widetilde{Q}_r \right)
 = \frac{1}{2} \left( x + i \, p \right) \alpha_r
   + \frac{1}{x}
     \left[
      \left( T_0 + \frac{3}{4} \right) \alpha_r + T_- \beta_r
     \right]
\\
\mathfrak{S}_r
&= \frac{1}{2} \left( S_r - i \, \widetilde{S}_r \right)
 = \frac{1}{2} \left( x + i \, p \right) \beta_r
   - \frac{1}{x}
     \left[
      \left( T_0 - \frac{3}{4} \right) \beta_r - T_+ \alpha_r
     \right]
\\
\Pi_{mr}
&= a_m \beta_r - b_m \alpha_r
\end{split}
\label{OSp(8*|N)Gr-1F}
\end{equation}

The operators that belong to $\mathfrak{C}^+$ subspace are the Hermitian
conjugates of those in $\mathfrak{C}^-$. The bosonic operators in
$\mathfrak{C}^+$ are:
\begin{equation}
\begin{split}
Y^m &= \frac{1}{2} \left( U^m + i \, \widetilde{U}^m \right)
     = \frac{1}{2} \left( x - i \, p \right) a^m
       + \frac{1}{x}
         \left[ \left( T_0 - \frac{3}{4} \right) a^m + T_+ b^m \right]
\\
Z^m &= \frac{1}{2} \left( V^m + i \, \widetilde{V}^m \right)
     = \frac{1}{2} \left( x - i \, p \right) b^m
       - \frac{1}{x}
         \left[ \left( T_0 + \frac{3}{4} \right) b^m - T_- a^m \right]
\\
N_+ &= a^1 b^2 - a^2 b^1
\\
B_+ &= - \frac{i}{2} \left[ \Delta - i \left( K_+ - K_- \right) \right]
     = \frac{1}{4} \left( x - i \, p \right)^2
       - \frac{1}{8 \, x^2} \left( 8 \, \mathcal{T}^2 + \frac{3}{2} \right)
\\
S^{rs} &= \alpha^r \beta^s + \alpha^s \beta^r
\end{split}
\label{OSp(8*|N)Gr+1B}
\end{equation}
The $4 N$ supersymmetry generators in $\mathfrak{C}^+$ subspace are:
\begin{equation}
\begin{split}
\mathfrak{Q}^r
&= \frac{1}{2} \left( Q^r + i \, \widetilde{Q}^r \right)
 = \frac{1}{2} \left( x - i \, p \right) \alpha^r
   + \frac{1}{x}
     \left[
      \left( T_0 - \frac{3}{4} \right) \alpha^r + T_+ \beta^r
     \right]
\\
\mathfrak{S}^r
&= \frac{1}{2} \left( S^r + i \, \widetilde{S}^r \right)
 = \frac{1}{2} \left( x - i \, p \right) \beta^r
   - \frac{1}{x}
     \left[
      \left( T_0 + \frac{3}{4} \right) \beta^r - T_- \alpha^r
     \right]
\\
\overline{\Pi}^{mr}
&= a^m \beta^r - b^m \alpha^r
\end{split}
\label{OSp(8*|N)Gr+1F}
\end{equation}
We find again the following relation
\begin{equation}
Y^1 \, Z^2 - Y^2 \, Z^1 = N_+ \, B_+
\end{equation}
among the generators in $\mathfrak{C}^+$ within the supersymmetric extension of the minrep.
We give the (super-)commutation relations between these $\mathfrak{C}^-$ and
$\mathfrak{C}^+$ operators and the explicit form of the generators of grade
zero subspace $\mathfrak{C}^0$ and their (super-)commutation relations in
appendix \ref{OSp(8*|2N)-3Gr}.

In the supersymmetric extension, the quadratic Casimirs of two $SU(2)$
subgroups are no longer identical. The generators of $SU(2)_A$ remain
unchanged, but $SU(2)_S$ generators get contributions from fermions and go
over to $SU(2)_T$. (See appendix \ref{OSp(8*|2N)-3Gr} for their explicit
forms.)

As we showed in section \ref{SU2SU2U1}, the minimal unitary representation of
$SO^*(8) \simeq SO(6,2)$ is a lowest weight representation with a unique
lowest weight vector $\ket{\psi^{(3/2)}_0 (x) \,;\, 0 , 0 , 0 , 0}$ that is 
annihilated by all the operators in $\mathfrak{C}^-$ subspace and corresponds 
to a conformal scalar in six dimensions. The lowest weight vector 
$\ket{\psi^{(3/2)}_0 (x) \,;\, 0 , 0 , 0 , 0}$ is a singlet of the 
semi-simple part of the little group, namely $SO(4) = SU(2)_S \times 
SU(2)_A$, of massless states in six dimensions. Now the minimal unitary representation of $OSp(8^*|2N)$ constructed above restricts to a
finite number of inequivalent unitary irreducible representations of
$SO^*(8)$, whose realization  involves 
fermionic as well as bosonic  oscillators. We shall refer to the resulting 
representations of $SO^*(8)$  as ``deformations'' of the minimal
unitary  representation.
These deformations of the minimal unitary representation of $SO^*(8)$ also
satisfy the Poincar\'{e} massless condition
\begin{equation}
\mathcal{M}^2 = \eta_{\mu\nu} P^\mu P^\nu = 0
\end{equation}
and hence correspond to massless conformal fields in six dimensions. Note
that $SO(4)$ is the six dimensional analog of the little group $SO(2)$ of
massless states in four dimensions. The minimal unitary representation of $4D$ conformal group 
$SO(4,2) = SU(2,2) / \mathbf{Z}_2$ corresponds to a massless conformal field
in four dimensions \cite{Fernando:2009fq}, and its deformations, which are
labeled by a real parameter $\zeta$, also describe massless conformal
fields. For physical fields, this parameter is simply twice the helicity of a
massless unitary representation of the Poincar\'e subgroup of $SO(4,2)$. They
are the doubleton representations of $SU(2,2)$, whose supersymmetric
extensions were studied in
\cite{Gunaydin:1984fk,Gunaydin:1998jc,Gunaydin:1998sw}. It was shown  long
time ago that the corresponding representations of the conformal group
$SU(2,2)$ remain irreducible under the restriction to the four dimensional
Poincar\'e subgroup \cite{Mack:1969dg}.\footnote{These representations are
sometimes referred to as the ladder representations in the literature.} We
expect the  massless doubleton representations of $SO^*(8)$ to  remain
irreducible  under restriction to  $6D$ Poincar\'e subgroup as well.


\section{Minimal Unitary Supermultiplet of $\mathfrak{osp}(8^*|2N)$}
\label{minrepsupermultiplet}

Since the subgroup $SU(2)_S$ of $SO^*(8)$ is replaced by $SU(2)_T$ when 
$SO^*(8)$ is extended to the supergroup $OSp(8^*|2N)$, the parameter $\alpha$ 
in the wave functions $\psi_n^{(\alpha)} \left( x \right)$ (as defined in 
equation (\ref{singularwavefunctions})) now depends on $\mathfrak{t}$, 
instead of $\mathfrak{s}$, where $\mathfrak{t}$ is the $SU(2)_T$ spin.

In this section, for the sake of simplicity, we shall denote the tensor 
product of the lowest energy state of the ``singular'' part $H_\odot$ of the 
bosonic Hamiltonian $H$ with the coordinate wave function
\begin{equation}
\psi_0^{(\alpha_\mathfrak{t} = 3/2)} \left( x \right)
= C_0 \, x^{\frac{3}{2}} \, e^{- x^2 / 2}
\end{equation}
and the vacuum state of all the bosonic and fermionic oscillators $a^m$,
$b^m$, $\alpha^r$ and $\beta^r$ simply as $\ket{\psi_0^{(3/2)}}$:
\begin{equation}
\begin{split}
a_m \, \ket{\psi_0^{(3/2)}}
 = b_m \, \ket{\psi_0^{(3/2)}}
&= 0
\\
\alpha_\mu \, \ket{\psi_0^{(3/2)}}
 = \beta_\mu \, \ket{\psi_0^{(3/2)}}
&= 0
\end{split}
\end{equation}
Note that for a general state involving bosonic and fermionic excitations,
\begin{equation}
\alpha_\mathfrak{t} = 2 \mathfrak{t} +3/2
\end{equation}
if $\mathfrak{t} \left( \mathfrak{t} + 1 \right)$ is the eigenvalue of the 
quadratic Casimir $\mathcal{T}^2$ of $SU(2)_T$ on that state.


\subsection{Minimal unitary supermultiplet of $\mathfrak{osp}(8^*|4)$}
\label{minsupermultipletN=2}

First we shall present the results for the case $N = 2$ (i.e. for $USp(4)$),
which is relevant to the symmetry supergroup of the $S^4$ compactification of 
the eleven dimensional supergravity. The oscillator construction of the 
unitary supermultiplets of $OSp(8^*|4)$ has been studied in
\cite{Gunaydin:1984wc,Gunaydin:1999ci,Fernando:2001ak}. It has 32 supersymmetry generators, 
16 of which belong to grade zero subspace $\mathfrak{C}^0$ and 8 each belong 
to grade $\pm 1$ subspaces $\mathfrak{C}^{\pm}$.

The state $\ket{\psi_0^{(3/2)}}$ is the unique normalizable lowest energy
state annihilated by all 9 bosonic operators as well as all 8 supersymmetry
generators in $\mathfrak{C}^-$ subspace. It is a singlet of $SU(4\,|\,2)$
subalgebra. By acting on it with grade $+1$ operators in the subspace
$\mathfrak{C}^+$, one obtains an infinite set of states which forms a basis
for the minimal unitary irreducible representation of
$\mathfrak{osp}(8^*|4)$. This infinite set of states can be decomposed into a
finite number of irreducible representations of the even subgroup $SO^*(8)
\times USp(4)$, with each irrep of $SO^*(8) \simeq SO(6,2)$ corresponding to
a massless conformal field in six dimensions.

In Table \ref{Table:minrepsupermultipletN=2}, we present the supermultiplet
that is obtained by starting from this unique lowest weight vector
\begin{equation}
 \ket{\psi_0^{(3/2)}}
\end{equation}
and acting on it with the generators of grade $+1$ subspace $\mathfrak{C}^+$.

The resulting minimal unitary supermultiplet is the ultra-short doubleton supermultiplet of $AdS_7$ supergroup $OSp(8^*|4)$ which does not have a Poincar\'e limit in seven dimensions and whose field theory lives on the boundary of $AdS_7$ on which $SO^*(8)$ acts as the conformal group \cite{Gunaydin:1984wc}.  It describes a
massless  $(2,0)$ conformal supermultiplet whose interacting field theory is believed to be dual to M-theory on $AdS_7\times S^4$ \cite{Maldacena:1997re}. The corresponding minimal supermultiplet of $4D$ superconformal algebra $SU(2,2|4)$ is 
the $\mathcal{N} = 4$ Yang-Mills supermultiplet in four dimensions
\cite{Fernando:2009fq}. In earlier literature, it was called
the CPT self-conjugate doubleton supermultiplet. In the twistorial oscillator
approach, the lowest weight vector $\ket{\Omega}$ for this supermultiplet is
the vacuum vector $\ket{0}$ of all the oscillators in the $SU(4\,|\,2) \times
U(1)$ basis \cite{Gunaydin:1984wc,Gunaydin:1999ci}.

Recalling that the positive energy unitary irreducible representations of
$SO^*(8)$ are uniquely labeled  by their lowest energy $SU(4)$ irreps, we 
note that each such $SU(4)$ irrep can in turn be uniquely labeled by an irrep 
of its subgroup $SU(2)_T \times SU(2)_A \times U(1)_J$ with respect to which 
it admits a compact three grading. Denoting the $SU(2)_T \times SU(2)_A$ 
spins as $\mathfrak{t}, \mathfrak{a}$ and the eigenvalue of $J$ as 
$\mathfrak{J}$, the Table \ref{Table:minrepsupermultipletN=2} also gives the 
decompositions of the lowest energy $SU(4)$ irreps of $SO^*(8)$ in the 
minimal supermultiplet.
The $USp(4)$ transformation properties of these $SO^*(8)$ irreps follow from 
the results of section \ref{USp(2N)}.

In  Table \ref{Table:minrepsupermultipletN=2}, $\left( \mathcal{Q}\right)^n \ket{\Omega}$ denotes symbolically a lowest energy
irrep of $SO^*(8)$ obtained by acting on the lowest weight state 
$\ket{\Omega}$ with $ n$ copies of 
supersymmetry generators  $\mathcal{Q} = \left\{ \mathfrak{Q}^r , 
\mathfrak{S}^r , \Pi^{mr} \right\}$.

\begin{small}
\begin{longtable}[c]{|r||c|c||l||c|c|}
\kill

\caption[The minimal unitary supermultiplet of $\mathfrak{osp}(8^*|4)$]
{The minimal unitary supermultiplet of $\mathfrak{osp}(8^*|4)$ defined by the lowest weight vector $\ket{\psi_0^{(3/2)}}$ .
The decomposition of $SU(4)$ irreps with respect to $SU(2)_T \times SU(2)_A 
\times U(1)_J$ is denoted by 
$\left( \mathfrak{t} , \mathfrak{a} \right)^{\mathfrak{J}}$. $H$ is the $AdS$
energy (negative conformal dimension), and $\mathcal{H}$ is the total energy.
The Dynkin labels of the lowest energy $SU(4)$ representations of $SO^*(8)$ coincide with the Dynkin labels of the corresponding massless $6D$ conformal fields under the Lorentz group $SU^*(4)$. $USp(4)$ Dynkin labels of these fields are also  given.
\label{Table:minrepsupermultipletN=2}} \\
\hline
& & & & & \\
States~~~~~ &
$H$ & $\mathcal{H}$ &
$( \mathfrak{t} , \mathfrak{a} )^{\mathfrak{J}}$ & $SU(4)=SU^*(4)$ & $USp(4)$ \\
 &
 &  &
 & Dynkin & Dynkin \\
& & & & & \\
\hline
& & & & & \\
\endfirsthead
\caption[]{(continued)} \\
\hline
& & & & & \\
States~~~~~ &
$H$ & $\mathcal{H}$ &
$( \mathfrak{t} , \mathfrak{a} )^{\mathfrak{J}}$ & $SU(4)=SU^*(4)$ & $USp(4)$ \\
 &
 &  &
 & Dynkin & Dynkin \\
& & & & & \\
\hline
& & & & & \\
\endhead
& & & & & \\
\hline
\endfoot
& & & & & \\
\hline
\endlastfoot

$\ket{\psi_0^{(3/2)}}$ &
2 & 1 &
$(0,0)^0$
 & (0,0,0) & (0,1) \\[8pt]

\hline
& & & & & \\

$\mathcal{Q} \ket{\psi_0^{(3/2)}}$ &
$\frac{5}{2}$ & 2 &
$\left( \frac{1}{2} , 0 \right)^{-\frac{1}{2}}
\oplus \left( 0 , \frac{1}{2} \right)^{+\frac{1}{2}}$
 & (1,0,0) & (1,0) \\[8pt]

\hline
& & & & & \\

$\left( \mathcal{Q} \right)^2 \ket{\psi_0^{(3/2)}}$ &
3 & 3 &
$\left( 1 , 0 \right)^{-1}
\oplus \left( \frac{1}{2} , \frac{1}{2} \right)^{0}
\oplus \left( 0 , 1 \right)^{+1}$
 & (2,0,0) & (0,0) \\[8pt]

\end{longtable}
\end{small}


\subsection{Minimal unitary supermultiplet of $OSp(8^*|2N)$}
\label{minsupermultipletN}

The supergroup  $OSp(8^*|2N)$
has  $16N$ supersymmetry generators, $8N$ of which belong
to grade zero subspace $\mathfrak{C}^0$ and $4N$ each belong to grade $\pm
1$ subspaces $\mathfrak{C}^{\pm}$ in the compact three grading.
Once again, the state $\ket{\psi_0^{(3/2)}}$ is the unique normalizable
lowest energy state annihilated by all $6 + N(N+1)/2$ bosonic operators as
well as all $4N$ supersymmetry generators in $\mathfrak{C}^-$ subspace. It is
a singlet of the subsuperalgebra $\mathfrak{su}(4\,|\,N)$. By acting on it
with grade $+1$ operators in the subspace $\mathfrak{C}^+$, one obtains an
infinite set of states which forms a basis for the minimal unitary
irreducible representation of $\mathfrak{osp}(8^*|2N)$. This infinite set of
states can be decomposed into a finite number of irreducible representations
of the even subgroup $SO^*(8) \times USp(2N)$, with each irrep of $SO^*(8)
\simeq SO(6,2)$ corresponding to a massless conformal field in six
dimensions.

In Table \ref{Table:minrepsupermultipletN}, we present the minimal unitary
supermultiplet of $\mathfrak{osp}(8^*|2N)$ obtained by starting from the
lowest weight state
\[
 \ket{\psi_0^{(3/2)}} \,.
\]
$\left( \mathcal{Q}\right)^n \ket{\psi_0^{(3/2)}}$ denotes, symbolically, the set of states  obtained by acting
on the lowest weight state $\ket{\psi_0^{(3/2)}}$ $n$ times with supersymmetry generators
where $\mathcal{Q} = \left\{ \mathfrak{Q}^r , \mathfrak{S}^r , \Pi^{mr}
\right\}$ that determine the $SO^*(8)$ irreps and their $USp(2N)$ transformation properties uniquely.

\begin{small}
\begin{longtable}[c]{|r||c|c||l||c|c|}
\kill

\caption[The minimal unitary supermultiplet of $\mathfrak{osp}(8^*|2N)$]
{Below we give the minimal unitary supermultiplet of $\mathfrak{osp}(8^*|2N)$ defined by the lowest weight vector $\ket{\psi_0^{(3/2)}}$ .
 The decomposition of lowest energy  $SU(4)$ irreps of $SO^*(8)$ with respect to 
 $SU(2)_T \times SU(2)_A 
\times U(1)_J$  subgroup
 is given in column 4. $H$ is the $AdS$
energy (negative conformal dimension), and $\mathcal{H}$ is the total energy.
The Dynkin labels of the lowest energy $SU(4)$ representations of $SO^*(8)$ coincide with the Dynkin labels of the corresponding massless $6D$ conformal fields with respect to  the Lorentz group $SU^*(4)$.  $USp(2N)$ Dynkin labels of these fields are also  given.
\label{Table:minrepsupermultipletN}} \\
\hline
& & & & & \\
State~~~~~ &
$H$ & $\mathcal{H}$ &
$( \mathfrak{t} , \mathfrak{a} )^{\mathfrak{J}}$ & $SU(4)$ & $USp(2N)$ \\
 &
 &  &
 & Dynkin & Dynkin \\
& & & & & \\
\hline
& & & & & \\
\endfirsthead
\caption[]{(continued)} \\
\hline
& & & & & \\
State~~~~~ &
$H$ & $\mathcal{H}$ &
$( \mathfrak{t} , \mathfrak{a} )^{\mathfrak{J}}$ & $SU(4)$ & $USp(2N)$ \\
 &
 &  &
 & Dynkin & Dynkin \\
& & & & & \\
\hline
& & & & & \\
\endhead
& & & & & \\
\hline
\endfoot
& & & & & \\
\hline
\endlastfoot

$\ket{\Omega}$ &
2 & $2 - \frac{N}{2}$ &
$(0,0)^0$
 & (0,0,0) & $(\underbrace{0,\dots,0}_{(N-1)},1)$ \\[8pt]

\hline
& & & & & \\

$\mathcal{Q} \ket{\Omega}$ &
$\frac{5}{2}$ & $3 - \frac{N}{2}$ &
$\left( \frac{1}{2} , 0 \right)^{-\frac{1}{2}}
\oplus \left( 0 , \frac{1}{2} \right)^{+\frac{1}{2}}$
 & (1,0,0) & $(\underbrace{0,\dots,0}_{(N-2)},1,0)$ \\[8pt]

\hline
& & & & & \\

$\left( \mathcal{Q} \right)^2 \ket{\Omega}$ &
3 & $4 - \frac{N}{2}$ &
$\left( 1 , 0 \right)^{-1}
\oplus \left( \frac{1}{2} , \frac{1}{2} \right)^{0}
\oplus \left( 0 , 1 \right)^{+1}$
 & (2,0,0) & $(\underbrace{0,\dots,0}_{(N-3)},1,0,0)$ \\[8pt]

\hline
& & & & & \\

\vdots~~~~ & \vdots & \vdots & \qquad \qquad \qquad \qquad \vdots & \vdots & \vdots \\[8pt]

\hline
& & & & & \\

$\left( \mathcal{Q} \right)^n \ket{\Omega}$ &
$2 + \frac{n}{2}$ & $2 + n - \frac{N}{2}$ &
$\left( \frac{n}{2} , 0 \right)^{-\frac{n}{2}}
\oplus \left( \frac{n-1}{2} , \frac{1}{2} \right)^{-\frac{n}{2}+1}$
 & $(n,0,0)$ & $(\underbrace{0,\dots,0}_{(N-n-1)},1,\underbrace{0,\dots,0}_{(n)})$ \\[8pt]

 &
 &
 &
$\oplus \dots \dots$
 & & \\[8pt]

 &
 &
 &
$\oplus \left( \frac{1}{2} , \frac{n-1}{2} \right)^{\frac{n}{2}-1} \oplus 
\left( 0 , \frac{n}{2} \right)^{\frac{n}{2}}$
 & & \\[8pt]

\hline
& & & & & \\

\vdots~~~~ & \vdots & \vdots & \qquad \qquad \qquad \qquad \vdots & \vdots & \vdots \\[8pt]

\hline
& & & & & \\

$\left( \mathcal{Q} \right)^N \ket{\Omega}$ &
$2 + \frac{N}{2}$ & $2 + \frac{N}{2}$ &
$\left( \frac{N}{2} , 0 \right)^{-\frac{N}{2}}
\oplus \left( \frac{N-1}{2} , \frac{1}{2} \right)^{-\frac{N}{2}+1}$
 & $(N,0,0)$ & $(0,\dots,0)$ \\[8pt]

 &
 &
 &
$\oplus \dots \dots$
 & & \\[8pt]

 &
 &
 &
$\oplus \left( \frac{1}{2} , \frac{N-1}{2} \right)^{\frac{N}{2}-1}
\oplus \left( 0 , \frac{N}{2} \right)^{\frac{N}{2}}$
 & & \\[8pt]

\end{longtable}
\end{small}


\section{Deformations of the Minimal Unitary Representation of $SO^*(8)$}
\label{SO*(8)deformations}

Above we showed that the minrep of $SO^*(8)$ is simply the scalar doubleton 
representation that describes a conformal scalar field in six dimensions. The 
group $SO^*(8)$ admits infinitely many doubleton representations 
corresponding to $6D$ massless conformal fields of arbitrary spin 
\cite{Gunaydin:1984wc,Gunaydin:1999ci,Fernando:2001ak}. They all can be 
constructed by the oscillator method over the Fock space of two pairs of 
twistorial oscillators transforming in the spinor representation of 
$SO^*(8)$. One would like to know  whether the doubleton representations 
corresponding to massless conformal fields of higher spin can all be obtained 
from the minimal unitary representation by a ``deformation'' in a manner 
similar to what happens in the case of $4D$ conformal group  $SU(2,2)$
\cite{Fernando:2009fq}. Remarkably, once again, we find that there exists 
infinitely many deformations of the minrep labeled by the spin ($t$) of an 
$SU(2)$  subgroup. By allowing this spin $t$ to take on all possible values, 
we obtain all the doubleton irreps as deformations of the minrep of 
$SO^*(8)$.  

To realize the deformations of the minimal representation of $SO^*(8)$, we 
first introduce an arbitrary number $P$ pairs of fermionic oscillators 
$\xi_x$ and $\chi_x$ and their hermitian conjugates $\xi^x = \left( \xi_x 
\right)^\dag$ and $\chi^x = \left( \chi_x \right)^\dag$ ($x = 1,2,\dots,P$) 
that satisfy the usual anti-commutation relations
\begin{equation}
\anticommute{\xi_x}{\xi^y}
= \anticommute{\chi_x}{\chi^y}
= \delta^x_y
\qquad \qquad
\anticommute{\xi_x}{\xi_y}
= \anticommute{\xi_x}{\chi_y}
= \anticommute{\chi_x}{\chi_y}
= 0 \,.
\end{equation}
Note that the following bilinears of these fermionic oscillators
\begin{equation}
G_+ = \xi^x \chi_x
\qquad \qquad
G_- = \chi^x \xi_x
\qquad \qquad
G_0 = \frac{1}{2} \left( N_\xi - N_\chi \right)
\end{equation}
where $N_\xi = \xi^x \xi_x$ and $N_\chi = \chi^x \chi_x$ are the respective
number operators, generate an $\mathfrak{su}(2)_G$ algebra:
\begin{equation}
\commute{G_+}{G_-} = 2 \, G_0
\qquad \qquad \qquad
\commute{G_0}{G_\pm} = \pm G_0
\end{equation}

We choose the Fock vacuum of these fermionic oscillators such that
\begin{equation}
\xi_x \ket{0} = \chi_x \ket{0} = 0
\end{equation}
for all $x = 1,2,\dots,P$. Clearly a state of the form
\begin{equation*}
\chi^{[1} \chi^2 \chi^3 \dots \chi^{P]} \ket{0}
\end{equation*}
has a definite eigenvalue of $G_0$ and is annihilated by the operator $G_-$.
Note that square bracketing of fermionic indices imply complete 
anti-symmetrization of weight one. By acting on this state with the operator 
$G_+$, one can obtain $P$ other states, namely:
\begin{equation*}
\xi^{[1} \chi^2 \chi^3 \dots \chi^{P]} \ket{0}
\, \oplus \,
\xi^{[1} \xi^2 \chi^3 \dots \chi^{P]} \ket{0}
\, \oplus \,
\dots \dots
\, \oplus \,
\xi^{[1} \xi^2 \xi^3 \dots \xi^{P]} \ket{0}
\end{equation*}

This set of $P+1$ states
transforms irreducibly under $\mathfrak{su}(2)_G$ in the spin $t = 
\frac{P}{2}$ representation.


Recall that the ``undeformed'' minimal unitary realization  of 
$\mathfrak{so}^*(8)$ has a 5-graded decomposition with respect to the  
subalgebra $\mathfrak{g}^{(0)} = \mathfrak{su}(2)_A \oplus 
\mathfrak{su}(1,1)_N \oplus \mathfrak{su}(2)_S \oplus \mathfrak{so}(1,1)$, as 
given in equation (\ref{so*(8)5-grading}). Now to deform the minimal unitary 
realization of $\mathfrak{so}^*(8)$, we extend the subalgebra 
$\mathfrak{su}(2)_S$ to the diagonal subalgebra 
$\mathfrak{su}(2)_{\buildrel _\circ \over {T}}$ of $\mathfrak{su}(2)_S$ and
$\mathfrak{su}(2)_G$. In other words, the generators of $\mathfrak{su}(2)_S$ 
receive contributions from the $\xi$- and $\chi$-type fermionic oscillators 
as follows:
\begin{equation}
\begin{split}
\buildrel _\circ \over {T}_+
&= S_+ + G_+
 = a^m b_m + \xi^x \chi_x
\\
\buildrel _\circ \over {T}_-
&= S_- + G_-
 = b^m a_m + \chi^x \xi_x
\\
\buildrel _\circ \over {T}_0
&= S_0 + G_0
 = \frac{1}{2} \left( N_a - N_b + N_\xi - N_\chi \right)
\end{split}
\end{equation}
The quadratic Casimir of this subalgebra
$\mathfrak{su}(2)_{\buildrel _\circ \over {T}}$ is given by
\begin{equation}
\mathcal{C}_2 \left[ \mathfrak{su}(2)_{\buildrel _\circ \over {T}} \right]
= \,\, \buildrel _\circ \over {\mathcal{T}}^2
= \,\, \buildrel _\circ \over {T}_0 \buildrel _\circ \over {T}_0
  + \frac{1}{2}
    \left( \, \buildrel _\circ \over {T}_+ \buildrel _\circ \over {T}_-
           + \buildrel _\circ \over {T}_- \buildrel _\circ \over {T}_+
    \right) \,.
\end{equation}
The  5-graded decomposition of the deformed minimal unitary realization, 
which we denote as $\mathfrak{so}^*(8)_D$, is now with respect to the 
subalgebra $\mathfrak{su}(2)_A \oplus \mathfrak{su}(1,1)_N \oplus
\mathfrak{su}(2)_{\buildrel _\circ \over {T}} \oplus \mathfrak{so}(1,1)$,
where, once again, the $\mathfrak{so}(1,1)$ generator $\Delta$ defines the
5-grading:
\begin{equation}
\mathfrak{so}^*(8)_D
 = \mathfrak{g}^{(-2)}_D \oplus
   \mathfrak{g}^{(-1)}_D \oplus
   \left[
    \mathfrak{su}(2)_A \oplus \mathfrak{su}(1,1)_N \oplus
    \mathfrak{su}(2)_{\buildrel _\circ \over {T}} \oplus \Delta
   \right] \oplus
   \mathfrak{g}^{(+1)}_D \oplus
   \mathfrak{g}^{(+2)}_D
\end{equation}

The rest of the grade zero subspace, $\mathfrak{su}(2)_A \oplus
\mathfrak{su}(1,1)_N \oplus \Delta$, remains unchanged under this deformation
(see equation (\ref{SU(2)AN_generators})). However, it should be noted that
the quadratic Casimir of $\mathfrak{su}(2)_{\buildrel _\circ \over {T}}$ is
no longer equal to those of $\mathfrak{su}(2)_A$ and $\mathfrak{su}(1,1)_N$.

Grade $-2$ and $-1$ generators of $\mathfrak{so}^*(8)_D$ are also the same as
those of the undeformed $\mathfrak{so}^*(8)$:
\begin{equation}
\buildrel _\circ \over {K}_- = K_- = \frac{1}{2} x^2
\end{equation}
\begin{equation}
\begin{aligned}
\buildrel _\circ \over {U}_m &= U_m = x \, a_m
\\
\buildrel _\circ \over {V}_m &= V_m = x \, b_m
\end{aligned}
\qquad \qquad \qquad \qquad
\begin{aligned}
\buildrel _\circ \over {U}^m &= U^m = x \, a^m
\\
\buildrel _\circ \over {V}^m &= V^m = x \, b^m
\end{aligned}
\end{equation}

However, since $\mathfrak{su}(2)_S$ has now been extended to 
$\mathfrak{su}(2)_{\buildrel _\circ \over {T}}$, the grade $+2$ generator,
which previously contained the quadratic Casimir $\mathcal{S}^2$ of
$\mathfrak{su}(2)_S$, now depends on $\buildrel _\circ \over 
{\mathcal{T}}^2$:
\begin{equation}
\buildrel _\circ \over {K}_+
= \frac{1}{2} p^2
  + \frac{1}{4 \, x^2} \left( 8 \, \buildrel _\circ \over {\mathcal{T}}^2
                              + \frac{3}{2} \right) \,.
\end{equation}
The  generators in grade $+1$ subspace are also modified since they are  
obtained from the commutators of the form 
$\commute{\mathfrak{g}_D^{(-1)}}{\mathfrak{g}_D^{(+2)}}$:
\begin{equation}
\begin{split}
\buildrel _\circ \over {\widetilde{U}}_m
= i \commute{\,\buildrel _\circ \over {U}_m}{\buildrel _\circ \over {K}_+}
& \qquad \qquad \qquad \qquad
\buildrel _\circ \over {\widetilde{U}}^m
= \left( \buildrel _\circ \over {\widetilde{U}}_m \right)^\dag
= i \commute{\,\buildrel _\circ \over {U}^m}{\buildrel _\circ \over {K}_+}
\\
\buildrel _\circ \over {\widetilde{V}}_m
= i \commute{\,\buildrel _\circ \over {V}_m}{\buildrel _\circ \over {K}_+}
& \qquad \qquad \qquad \qquad
\buildrel _\circ \over {\widetilde{V}}^m
= \left( \buildrel _\circ \over {\widetilde{V}}_m \right)^\dag
= i \commute{\,\buildrel _\circ \over {V}^m}{\buildrel _\circ \over {K}_+}
\end{split}
\end{equation}
The explicit form of these grade $+1$ generators are as follows:
\begin{equation}
\begin{split}
\buildrel _\circ \over {\widetilde{U}}_m
&= - p \, a_m
   + \frac{2i}{x}
     \left[
      \left( \buildrel _\circ \over {T}_0 + \frac{3}{4} \right) a_m
      + \buildrel _\circ \over {T}_- b_m
     \right]
\\
\buildrel _\circ \over {\widetilde{U}}^m
&= - p \, a^m
   - \frac{2i}{x}
     \left[
      \left( \buildrel _\circ \over {T}_0 - \frac{3}{4} \right) a^m
      + \buildrel _\circ \over {T}_+ b^m
     \right]
\\
\buildrel _\circ \over {\widetilde{V}}_m
&= - p \, b_m
   - \frac{2i}{x}
     \left[
      \left( \buildrel _\circ \over {T}_0 - \frac{3}{4} \right) b_m
      - \buildrel _\circ \over {T}_+ a_m
     \right]
\\
\buildrel _\circ \over {\widetilde{V}}^m
&= - p \, b^m
   + \frac{2i}{x}
     \left[
      \left( \buildrel _\circ \over {T}_0 + \frac{3}{4} \right) b^m
      - \buildrel _\circ \over {T}_- a^m
     \right]
\end{split}
\end{equation}
The deformed generators of $\mathfrak{so}^*(8)_D$ with ``$\circ$'' over them
satisfy the same commutation relations as the corresponding  ``undeformed''
generators of $\mathfrak{so}^*(8)$. Therefore, the 5-grading of
$\mathfrak{so}^*(8)_D$, defined by $\Delta$, takes the form:
\begin{equation}
\begin{split}
\mathfrak{so}^*(8)_D
&= ~ \mathbf{1} ~~ \oplus
   ~~ \left( \mathbf{4} , \mathbf{2} \right) ~ \oplus
   \left[ \mathfrak{su}(2)_A \oplus
          \mathfrak{su}(1,1)_N \oplus
          \mathfrak{su}(2)_{\buildrel _\circ \over {T}} \oplus
          \mathfrak{so}(1,1)_{\Delta}
   \right] \oplus
   ~ \left( \mathbf{4} , \mathbf{2} \right) ~ \oplus
   ~ \mathbf{1}
\\
&= \buildrel _\circ \over {K}_-
   \oplus
   \left[ \,\buildrel _\circ \over {U}_m \,,\,
          \buildrel _\circ \over {U}^m \,,\,
          \buildrel _\circ \over {V}_m \,,\,
          \buildrel _\circ \over {V}^m \,
   \right]
   \oplus
   \left[ ~ A_{\pm,0} ~ \oplus ~ N_{\pm,0} ~ \oplus
          ~ \buildrel _\circ \over {T}_{\pm,0} ~ \oplus
          ~ \Delta ~ \right]
\\
& \qquad \qquad \qquad \qquad \qquad \qquad \qquad \qquad \qquad \qquad \quad
   \oplus
   \left[ \,\buildrel _\circ \over {\widetilde{U}}_m \,,\,
          \buildrel _\circ \over {\widetilde{U}}^m \,,\,
          \buildrel _\circ \over {\widetilde{V}}_m \,,\,
          \buildrel _\circ \over {\widetilde{V}}^m \,
   \right]
   \oplus
   \buildrel _\circ \over {K}_+ \,
\end{split}
\end{equation}
The quadratic Casimir of $\mathfrak{so}^*(8)_D$ is given by
\begin{equation}
\begin{split}
\mathcal{C}_2 \left[ \mathfrak{so}^*(8)_D \right]
&= \mathcal{C}_2 \left[ \mathfrak{su}(2)_{\buildrel _\circ \over {T}} \right]
   + \mathcal{C}_2 \left[ \mathfrak{su}(2)_A \right]
   + \mathcal{C}_2 \left[ \mathfrak{su}(1,1)_N \right]
   + \mathcal{C}_2 \left[ \mathfrak{su}(1,1)_{\buildrel _\circ \over {K}} \right]
\\
& \quad
  - \frac{i}{4} \, \mathcal{F} \left( \buildrel _\circ \over {U} \,,\,
                                      \buildrel _\circ \over {V} \right)
\end{split}
\end{equation}
where
\begin{equation}
\begin{split}
\mathcal{F} \left( \buildrel _\circ \over {U} \,,\,
                   \buildrel _\circ \over {V} \right)
&= \left( \buildrel _\circ \over {U}_m
          \buildrel _\circ \over {\widetilde{U}}^m
          + \buildrel _\circ \over {V}_m
            \buildrel _\circ \over {\widetilde{V}}^m
          + \buildrel _\circ \over {\widetilde{U}}^m
            \buildrel _\circ \over {U}_m
          + \buildrel _\circ \over {\widetilde{V}}^m
            \buildrel _\circ \over {V}_m \right)
\\
& \qquad
   - \left( \buildrel _\circ \over {U}^m
            \buildrel _\circ \over {\widetilde{U}}_m
            + \buildrel _\circ \over {V}^m
              \buildrel _\circ \over {\widetilde{V}}_m
            + \buildrel _\circ \over {\widetilde{U}}_m
              \buildrel _\circ \over {U}^m
            + \buildrel _\circ \over {\widetilde{V}}_m
              \buildrel _\circ \over {V}^m \right)
\end{split}
\end{equation}
and reduces to
\begin{equation}
\mathcal{C}_2 \left[ \mathfrak{so}^*(8)_D \right]
= 2 \, \mathcal{G}^2 - 4
\end{equation}
where $\mathcal{G}^2$ is the quadratic Casimir of $\mathfrak{su}(2)_G$.


\subsection{The 3-grading of $SO^*(8)_D$ with respect to the subgroup $SU(4)
\times U(1)$}
\label{3GrSO*(8)deformed}

The Lie algebra of $\mathfrak{so}^*(8)_D$ can be given a compact 3-grading
\begin{equation}
\mathfrak{so}^*(8)_D
= \mathfrak{C}^-_D \oplus \mathfrak{C}^0_D \oplus \mathfrak{C}^+_D
\end{equation}
with respect to its maximal compact subalgebra $\mathfrak{su}(4) \oplus
\mathfrak{u}(1)$, determined by the $\mathfrak{u}(1)$ generator
\begin{equation}
\buildrel _\circ \over {H}
= N_0
  + \frac{1}{2} \left( \buildrel _\circ \over {K}_+
                       + \buildrel _\circ \over {K}_- \right) \,.
\label{deformedH}
\end{equation}
The generators that belong to the grade 0,$\pm1$ subspaces
are as follows:
\begin{equation}
\begin{split}
\mathfrak{C}^-_D
&= \left( \buildrel _\circ \over {U}_m
          - \, i \buildrel _\circ \over {\widetilde{U}}_m \right) \oplus
   \left( \buildrel _\circ \over {V}_m
          - \, i \buildrel _\circ \over {\widetilde{V}}_m \right) \oplus
   N_- \oplus
   \left[ \Delta
          + i \left( \buildrel _\circ \over {K}_+
                     - \buildrel _\circ \over {K}_- \right) \right]
\\
\mathfrak{C}^0_D
&= \left[
    \buildrel _\circ \over {T}_{\pm,0} \oplus
    A_{\pm,0} \oplus
    \left( N_0
           - \frac{1}{2} \left( \buildrel _\circ \over {K}_+
                                + \buildrel _\circ \over {K}_- \right)
                         \right)
   \right.
\\
& \qquad
   \left.
    \oplus
    \left( \buildrel _\circ \over {U}_m
           + \, i \buildrel _\circ \over {\widetilde{U}}_m \right) \oplus
    \left( \buildrel _\circ \over {V}_m
           + \, i \buildrel _\circ \over {\widetilde{V}}_m \right) \oplus
    \left( \buildrel _\circ \over {U}^m
           - \, i \buildrel _\circ \over {\widetilde{U}}^m \right) \oplus
    \left( \buildrel _\circ \over {V}^m
           - \, i \buildrel _\circ \over {\widetilde{V}}^m \right)
   \right]
   \oplus \buildrel _\circ \over {H}
\\
\mathfrak{C}^+_D
&= \left( \buildrel _\circ \over {U}^m
          + \, i \buildrel _\circ \over {\widetilde{U}}^m \right) \oplus
   \left( \buildrel _\circ \over {V}^m
          + \, i \buildrel _\circ \over {\widetilde{V}}^m \right) \oplus
   N_+ \oplus
   \left[ \Delta
          - i \left( \buildrel _\circ \over {K}_+
                     - \buildrel _\circ \over {K}_- \right) \right]
\end{split}
\label{3Gr-SO*8deformed}
\end{equation}


\subsection{Deformed minreps of  $SO^*(8)$ as massless $6D$ conformal fields}
\label{LWVSO*(8)deformed}

Consider the vacuum state $\ket{0}$ that is annihilated by the bosonic
oscillators $a_m$, $b_m$ ($m = 1,2$) and the fermionic oscillators $\xi_x$,
$\chi_x$ ($x = 1,2,\dots,P$):
\begin{equation}
a_m \ket{0} = b_m \ket{0} = \xi_x \ket{0} = \chi_x \ket{0} = 0
\end{equation}
The tensor products of the states of the form $\left( a^m \right)^{n_{a,m}}
\ket{0}$, $\left( b^m  \right)^{n_{b,m}} \ket{0}$, $\xi^x \ket{0}$ and
$\chi^x \ket{0}$, where $n_{a,m}$ and $n_{b,m}$ are non-negative integers,
form a ``particle basis'' of states in this Fock space.

As the ``particle basis'' of the Hilbert space of the deformed minimal
unitary representation of $SO^*(8)$, we take the tensor product of the above
states with the state space of the singular (isotonic) oscillator:
\begin{equation*}
\left( a^1 \right)^{n_{a,1}} \ket{0} \otimes
\left( a^2 \right)^{n_{a,2}} \ket{0} \otimes
\left( b^1 \right)^{n_{b,1}} \ket{0} \otimes
\left( b^2 \right)^{n_{b,2}} \ket{0} \otimes
\xi^{[x_1} \dots \xi^{x_k} \chi^{x_{k+1}} \dots \chi^{x_P]} \ket{0} \otimes
\ket{\psi_n^{(\alpha_t)}}
\end{equation*}
where square brackets imply full antisymmetrization and denote them as
\begin{equation*}
\left( a^1 \right)^{n_{a,1}} \left( a^2 \right)^{n_{a,2}} 
\left( b^1 \right)^{n_{b,1}} \left( b^2 \right)^{n_{b,2}}
\xi^{[x_1} \dots \xi^{x_k} \chi^{x_{k+1}} \dots \chi^{x_P]}
\ket{\psi_n^{(\alpha_t)}}
\end{equation*}
or simply as
\begin{equation*}
\ket{\psi_n^{(\alpha_t)} \,;\, n_{a,1} , n_{a,2} , n_{b,1} , n_{b,2} \,;\, 
     \frac{P}{2} , k - \frac{P}{2}}
\end{equation*}
where $k = 0,1,2,\dots,P$. Note that  $\alpha_t$ now depends on the eigenvalue $t$ of the quadratic Casimir of $SU(2)_{\buildrel _\circ \over {T}}$.

Note that the $(P+1)$ states 
\begin{equation}
\ket{\psi_n^{(\alpha_t)} \,;\, 0 , 0 , 0 , 0 \,;\, 
     \frac{P}{2} , k - \frac{P}{2}} \qquad k=0,1,\cdots ,P 
\end{equation}
are  annihilated by all grade $-1$ operators in $\mathfrak{C}^-_D$ and
transforms in the spin $t = \frac{P}{2}$ representation of
$\mathfrak{su}(2)_{\buildrel _\circ \over {T}}$, if $\alpha_t$ satisfies
\begin{equation}
\alpha_t = 2 \, t + \frac{3}{2} \,.
\end{equation}
These states have a definite eigenvalue of $t + 2$ with respect to
$\buildrel _\circ \over {H}$ (given in equation (\ref{deformedH})):
\begin{equation}
\buildrel _\circ \over {H} \ket{\psi_n^{(2t+3/2)} \,;\, 0 , 0 , 0 , 0 \,;\, 
     \frac{P}{2} , k - \frac{P}{2}}
= \left( t + 2 \right) \ket{\psi_n^{(2t+3/2)} \,;\, 0 , 0 , 0 , 0 \,;\, 
     \frac{P}{2} , k - \frac{P}{2}}
\end{equation}
By acting on these $(P+1)$ states with the coset generators $(C^{1m},C^{2m})$
of
\begin{equation*}
SU(4) \,/\, \left[ SU(2)_{\buildrel _\circ \over {T}} \times SU(2)_A \times U(1) \right]
\end{equation*}
one obtains a set of states, which we denote collectively as $\ket{\Omega^{(2t+3/2)}}$, transforming in an irrep of $SU(4)$ with Dynkin labels  $(2t,0,0)$ that are eigenstates of $\buildrel _\circ \over {H}$  with eigenvalue $( t+2)$. The states $\ket{\Omega^{(2t+3/2)}}$ are annihilated by all the  operators in $\mathfrak{C}^-_D$. 
Therefore, the deformed minimal unitary representation of $SO^*(8)$ is a
unitary lowest weight representation. All the other states of the ``particle
basis'' of the deformed minrep can be obtained from the set of states 
$\ket{\Omega^{(2t+3/2)}}$ by
repeatedly acting on it with grade $+1$ operators in the $\mathfrak{C}^+_D$
subspace of $SO^*(8)_D$.

In Table \ref{Table:deformedSO*(8)minrep}, we present the deformed minrep of
$SO^*(8)$. The notation $\left( \mathfrak{C}^+_D \right)^n 
\ket{\Omega^{(\alpha_t)}}$ represents all the states obtained by acting on the 
lowest weight state $\ket{\Omega^{(\alpha_t)}}$ with $n$ grade $+1$ generators
$( \buildrel _\circ \over {Y}^m \,,\, \buildrel _\circ \over {Z}^m \,,\,
\buildrel _\circ \over {N}_+ \,,\, \buildrel _\circ \over {B}_+ )$.
The deformed minrep with parameter $t$ corresponds to a  massless conformal field in six dimensions whose transformation under the $6D$ Lorentz group $SU^*(4)$ coincides with the transformation of the states $\ket{\Omega^{(\alpha_t)}}$ under the $SU(4)$ 
subgroup of $SO^*(8)_D$ and its conformal dimension is equal to $-(t+2)$ .

\begin{small}
\begin{longtable}[c]{|l||c|c|}
\kill

\caption[The deformed minimal unitary representation of $SO^*(8)$]
{$SU(4) \times U(1)$ decomposition of the  deformed minimal unitary representation of $SO^*(8)$. For the deformed minrep with deformation parameter $t$, $\alpha_t = 2 t + 3/2$. It corresponds to a massless $6D$ conformal field transforming in the $(2t,0,0)_\mathrm{Dynkin}$ representation of the Lorentz group $SU^*(4)$ with conformal dimension $-(t+2)$.  
\label{Table:deformedSO*(8)minrep}} \\
\hline
& & \\
State & $E$ & $SU(4)$ \\
 &  & Dynkin \\
& & \\
\hline
& & \\
\endfirsthead
\caption[]{(continued)} \\
\hline
& & \\
State & $E$ & $SU(4)$ \\
 &  & Dynkin \\
& & \\
\hline
& & \\
\endhead
& & \\
\hline
\endfoot
& & \\
\hline
\endlastfoot

$\ket{\Omega^{(\alpha_t)}}$
 & $t+2$ & $(2t,0,0)$ \\[8pt]

\hline
& & \\

$\mathfrak{C}^+_D \ket{\Omega^{(\alpha_t)}}$
 & $t+3$ & $(2t,1,0)$
\\[8pt]

\hline
& & \\

$\left( \mathfrak{C}^+_D \right)^2 \ket{\Omega^{(\alpha_t)}}$
 & $t+4$ & $(2t,2,0)$
\\[8pt]

\hline
& & \\

\qquad \vdots & \vdots & \vdots
\\[8pt]

\hline
& & \\

$\left( \mathfrak{C}^+_D \right)^n \ket{\Omega^{(\alpha_t)}}$
 & $t+n+2$ & $(2t,n,0)$
\\[8pt]

\hline
& & \\

\qquad \vdots & \vdots & \vdots
\\[8pt]

\end{longtable}
\end{small}


\section{Conclusions}
\label{conclusions}

In this paper we first studied the minimal unitary representation of $SO^*(8)
\simeq SO(6,2)$, which is the seven dimensional $AdS$ group or the six
dimensional conformal group, obtained by quantizing its quasiconformal
realization.  The resulting minrep coincides with  the scalar
doubleton representation of $SO^*(8)$, whose Poincar\'{e} limit in $AdS_7$ is
singular.

We then introduced supersymmetry and extended these results to construct the
minimal unitary supermultiplet of $OSp(8^*|2N)$, and, in particular, the
minimal unitary supermultiplet of $OSp(8^*|4)$. The minimal unitary
supermultiplet of $OSp(8^*|4)$ is simply the massless supermultiplet of (2,0)
conformal field theory in six dimensions  that is believed to be dual to M-theory on $AdS_7
\times S^4$.

Finally, we presented a method to introduce a  family of
deformations of the minrep of $SO^*(8)$, with respect to one of its $SU(2)$
subgroups. For each non-negative integer or half-integer value of the
deformation parameter corresponding to the spin $t$ of $SU(2)$  one obtains a unique positive energy unitary irreducible 
representation of $SO^*(8)$ which describes a massless conformal field of
higher spin in six dimensions and coincide with the infinite family of
doubletons studied in \cite{Gunaydin:1984wc,Gunaydin:1999ci,Fernando:2001ak}.

One can also obtain the ``deformed'' minimal unitary supermultiplets of $OSp(8^*|2N)$ by 
first deforming the minrep of $SO^*(8)$ and then extending it to the 
superalgebras $OSp(8^*|2N)$. These deformed minimal unitary supermultiplets 
correspond to six dimensional massless superconformal multiplets  involving 
higher spin fields than the undeformed minimal unitary supermultiplet and 
will be given in a separate study \cite{workinprogress_sfmg}.


{\bf Acknowledgements:}  We would like to thank Oleksandr Pavlyk for many stimulating discussions and his generous help with Mathematica. S.F. would like to thank the Center for Fundamental Theory of  the
Institute for Gravitation and the Cosmos at Pennsylvania State University, where part of this work was done, for their warm hospitality. \\
  This work was supported in part by the National
Science Foundation under grants numbered PHY-0555605 and PHY-0855356. Any opinions,
findings and conclusions or recommendations expressed in this
material are those of the authors and do not necessarily reflect the
views of the National Science Foundation.


\appendix

\numberwithin{equation}{section}

\section*{Appendix}


\section{The decomposition of $SO(6,2)$ with respect to the subgroup $SO(4) 
\times SO(2,2)$}
\label{SO(4)xSO(2,2)}

The generators  $\widetilde{M}_{ij} = i \, M_{ij}$ ($i,j,\dots = 1,2,3,4$)
form the $\mathfrak{so}(4)$ subalgebra of $\mathfrak{so}(6,2) \supset
\mathfrak{so}(4) \oplus \mathfrak{so}(2,2)$. This $\mathfrak{so}(4)$ can be
decomposed as a direct sum
\begin{equation}
\mathfrak{so}(4) = \mathfrak{su}(2)_L \oplus \mathfrak{su}(2)_R
\end{equation}
where the generators of the two $\mathfrak{su}(2)$ subalgebras are given
by:
\begin{equation}
\begin{aligned}
L_1 &= \frac{1}{2} \left( \widetilde{M}_{23} - \widetilde{M}_{14} \right)
\\
R_1 &= \frac{1}{2} \left( \widetilde{M}_{23} + \widetilde{M}_{14} \right)
\end{aligned}
\qquad
\begin{aligned}
L_2 &= \frac{1}{2} \left( \widetilde{M}_{13} + \widetilde{M}_{24} \right)
\\
R_2 &= \frac{1}{2} \left( \widetilde{M}_{13} - \widetilde{M}_{24} \right)
\end{aligned}
\qquad
\begin{aligned}
L_3 &= \frac{1}{2} \left( \widetilde{M}_{12} - \widetilde{M}_{34} \right)
\\
R_3 &= \frac{1}{2} \left( \widetilde{M}_{12} + \widetilde{M}_{34} \right)
\end{aligned}
\end{equation}

They satisfy the commutation relations
\begin{equation}
\begin{split}
\commute{L_+}{L_-} = 2 \, L_3
& \qquad \qquad \qquad
\commute{L_3}{L_\pm} = \pm \, L_\pm
\\
\commute{R_+}{R_-} = 2 \, R_3
& \qquad \qquad \qquad
\commute{R_3}{R_\pm} = \pm \, R_\pm
\end{split}
\end{equation}
where
\begin{equation}
\begin{split}
L_{\pm} &= L_1 \pm i \, L_2
\\
R_{\pm} &= R_1 \pm i \, R_2 \,.
\end{split}
\end{equation}
The quadratic Casimir operators of the two $\mathfrak{su}(2)$'s are equal:
\begin{equation}
L^2
= L_1^2 + L_2^2 + L_3^2
= R^2
= R_1^2 + R_2^2 + R_3^2
=\frac{1}{8} \mathcal{I}_4 + 1
\end{equation}

The centralizer of $SU(2)_L \times SU(2)_R$ within $SO(6,2)$ is $SO(2,2)$,
which also decomposes as
\begin{equation}
SO(2,2) = SU(1,1)_{\mathfrak{L}} \times SU(1,1)_{\mathfrak{R}}
\end{equation}
where $SU(1,1)_{\mathfrak{L}}$ is generated by $K_+ , K_-$ and $\Delta$ and
$SU(1,1)_{\mathfrak{R}}$ is generated by $J_{\pm}$ and $J_0$.

In their compact bases the generators of $SU(1,1)_{\mathfrak{L}}$ and
$SU(1,1)_{\mathfrak{R}}$ take the form
\begin{equation}
\begin{aligned}
\mathfrak{L}_+
&= - \frac{1}{2}
   \left[ \Delta - i \left( K_+ - K_- \right) \right]
\\
\mathfrak{L}_-
&= - \frac{1}{2}
   \left[ \Delta + i \left( K_+ - K_- \right) \right]
\\
\mathfrak{L}_3
&= \frac{1}{2} \left( K_+ + K_- \right)
\end{aligned}
\qquad \qquad
\begin{aligned}
\mathfrak{R}_+
&= - \frac{1}{2}
   \left[ J_0 + \frac{i}{2} \left( J_+ - J_- \right) \right]
\\
\mathfrak{R}_-
&= - \frac{1}{2}
   \left[ J_0 - \frac{i}{2} \left( J_+ - J_- \right) \right]
\\
\mathfrak{R}_3
&= - \frac{1}{4} \left( J_+ + J_- \right)
\end{aligned}
\end{equation}
and  satisfy the commutation relations:
\begin{equation}
\begin{split}
\commute{\mathfrak{L}_+}{\mathfrak{L}_-} = - 2 \, \mathfrak{L}_3
& \qquad \qquad \qquad
\commute{\mathfrak{L}_3}{\mathfrak{L}_\pm} = \pm \, \mathfrak{L}_\pm
\\
\commute{\mathfrak{R}_+}{\mathfrak{R}_-} = - 2 \, \mathfrak{R}_3
& \qquad \qquad \qquad
\commute{\mathfrak{R}_3}{\mathfrak{R}_\pm} = \pm \, \mathfrak{R}_\pm
\end{split}
\end{equation}

Their  quadratic Casimir operators
\begin{equation}
\mathfrak{L}^2
= {\mathfrak{L}_3}^2
  - \frac{1}{2} \left( \mathfrak{L}_+ \mathfrak{L}_-
                       + \mathfrak{L}_- \mathfrak{L}_+
                \right)
\qquad  \qquad \qquad
\mathfrak{R}^2
= {\mathfrak{R}_3}^2
  - \frac{1}{2} \left( \mathfrak{R}_+ \mathfrak{R}_-
                       + \mathfrak{R}_- \mathfrak{R}_+
                \right)
\end{equation}
coincide:
\begin{equation}
\mathfrak{L}^2 = \mathfrak{R}^2 = \frac{1}{8} \mathcal{I}_4 + 1
\end{equation}
Thus the quadratic Casimirs of $SU(2)_L$, $SU(2)_R$, $SU(1,1)_{\mathfrak{L}}$
and $SU(1,1)_{\mathfrak{R}}$ are all equal within the minimal unitary
realization of $SO(6,2)$.


\section{The compact 3-grading of $SO^*(8)$ with respect to the subgroup
$SU(4) \times U(1)$}
\label{C3GrSO*(8)}

The Lie algebra $\mathfrak{so}^*(8)$ can be given a compact 3-grading
\begin{equation}
\mathfrak{so}^*(8)
= \mathfrak{C}^- \oplus \mathfrak{C}^0 \oplus \mathfrak{C}^+
\end{equation}
with respect to its maximal compact subalgebra $\mathfrak{su}(4) \oplus
\mathfrak{u}(1)$, determined by the $\mathfrak{u}(1)$ generator
\begin{equation}
H = N_0 + \frac{1}{2} \left( K_+ + K_- \right) \,.
\label{BosonicHamiltonian2}
\end{equation}
The operators that belong to  the grade 0,$\pm1$ subspaces
are as follows:
\begin{equation}
\begin{split}
\mathfrak{C}^-
&= \left( U_m - i \, \widetilde{U}_m \right) \oplus
   \left( V_m - i \, \widetilde{V}_m \right) \oplus
   N_- \oplus
   \left[ \Delta + i \left( K_+ - K_- \right) \right]
\\
\mathfrak{C}^0
&= \left[
    S_{\pm,0} \oplus
    A_{\pm,0} \oplus
    \left( N_0 - \frac{1}{2} \left( K_+ + K_- \right) \right)
   \right.
\\
& \qquad
   \left.
    \oplus
    \left( U_m + i \, \widetilde{U}_m \right) \oplus
    \left( V_m + i \, \widetilde{V}_m \right) \oplus
    \left( U^m - i \, \widetilde{U}^m \right) \oplus
    \left( V^m - i \, \widetilde{V}^m \right)
   \right]
   \oplus H
\\
\mathfrak{C}^+
&= \left( U^m + i \, \widetilde{U}^m \right) \oplus
   \left( V^m + i \, \widetilde{V}^m \right) \oplus
   N_+ \oplus
   \left[ \Delta - i \left( K_+ - K_- \right) \right]
\end{split}
\label{SO*8Gr3}
\end{equation}

It is convenient to express the generators of the $\mathfrak{su}(4)$
subalgebra in $\mathfrak{C}^0$ subspace in its $\mathfrak{su}(4) \supset
\mathfrak{su}(2)_S \oplus \mathfrak{su}(2)_A \oplus \mathfrak{u}(1)_J$
decomposition, where
\begin{equation}
J = N_0 - \frac{1}{2} \left( K_+ + K_- \right)
\end{equation}
determines a 3-grading of $\mathfrak{su}(4)$. We note that $S_{\pm,0}$ and
$A_{\pm,0}$ were given in equations (\ref{SU(2)S_generators}) and
(\ref{SU(2)AN_generators}).
The remaining
generators of $\mathfrak{su}(4)$ are given by
\begin{equation}
\begin{aligned}
C_{1m} &= \frac{1}{2} \left( U_m + i \, \widetilde{U}_m \right)
\\
C^{1m} &= \frac{1}{2} \left( U^m - i \, \widetilde{U}^m \right)
\end{aligned}
\qquad\qquad
\begin{aligned}
C_{2m} &= \frac{1}{2} \left( V_m + i \, \widetilde{V}_m \right)
\\
C^{2m} &= \frac{1}{2} \left( V^m - i \, \widetilde{V}^m \right) \,.
\end{aligned}
\end{equation}
Then the $\mathfrak{su}(4)$ algebra becomes
\begin{equation}
\begin{split}
&\commute{S^{m^\prime}_{~n^\prime}}{S^{k^\prime}_{~l^\prime}}
= \delta^{k^\prime}_{n^\prime} \, S^{m^\prime}_{~l^\prime}
  - \delta^{m^\prime}_{l^\prime} \, S^{k^\prime}_{~n^\prime}
\qquad \qquad \qquad
\commute{A^m_{~n}}{A^k_{~l}}
= \delta^k_n \, A^m_{~l} - \delta^m_l \, A^k_{~n}
\\
&\commute{C^{m^\prime m}}{C_{n^\prime n}}
= \delta^m_n \, S^{m^\prime}_{~n^\prime}
  + \delta^{m^\prime}_{n^\prime} \, A^m_{~n}
  + \delta^{m^\prime}_{n^\prime} \delta^m_n \, J
\\
&\commute{S^{m^\prime}_{~n^\prime}}{C^{k^\prime m}}
= \delta^{k^\prime}_{n^\prime} \, C^{m^\prime m}
  - \frac{1}{2} \delta^{m^\prime}_{n^\prime} \, C^{k^\prime m}
\qquad
\commute{A^m_{~n}}{C^{m^\prime k}}
= \delta^k_n \, C^{m^\prime m}
  - \frac{1}{2} \delta^m_n \, C^{m^\prime k} \,.
\end{split}
\end{equation}
where we have labeled the generators of $\mathfrak{su}(2)_S$ and
$\mathfrak{su}(2)_A$ as $S^m_{~n}$ and $A^m_{~n}$, respectively:
\begin{equation}
\begin{aligned}
S^1_{~1} &= - S^2_{~2} = S_0
\\
A^1_{~1} &= - A^2_{~2} = A_0
\end{aligned}
\qquad \qquad \qquad
\begin{aligned}
S^1_{~2} &= S_+
\\
A^1_{~2} &= A_+
\end{aligned}
\qquad \qquad \qquad
\begin{aligned}
S^2_{~1} = \left( {S^1_{~2}} \right)^\dag &= S_-
\\
A^2_{~1} = \left( {A^1_{~2}} \right)^\dag &= A_-
\end{aligned}
\end{equation}

We shall label $\mathfrak{C}^\pm$ operators as
\begin{equation}
\begin{aligned}
Y_m &= \frac{1}{2} \left( U_m - i \, \widetilde{U}_m \right)
\\
Z_m &= \frac{1}{2} \left( V_m - i \, \widetilde{V}_m \right)
\\
N_- &
\\
B_- &= \frac{i}{2} \left[ \Delta + i \left( K_+ - K_- \right) \right]
\end{aligned}
\qquad \qquad
\begin{aligned}
Y^m &= \frac{1}{2} \left( U^m + i \, \widetilde{U}^m \right)
\\
Z^m &= \frac{1}{2} \left( V^m + i \, \widetilde{V}^m \right)
\\
N_+ &
\\
B_+ &= - \frac{i}{2} \left[ \Delta - i \left( K_+ - K_- \right) \right] \,.
\end{aligned}
\label{SO*8Grpm1}
\end{equation}
The commutators $\commute{\mathfrak{C}^-}{\mathfrak{C}^+}$ close into
$\mathfrak{C}^0$:
\begin{equation}
\begin{aligned}
\commute{Y_m}{Y^n}
&= \delta^n_m \, H
   + \delta^n_m \, S_0
   + A^n_{~m}
\\
\commute{Y_m}{Z^n}
&= \delta^n_m \, S_-
\\
\commute{Y_m}{N_+}
&= + \epsilon_{mn} \, C^{2n}
\\
\commute{Y_m}{B_+}
&= C_{1m}
\\
\commute{N_-}{N_+}
&= H + J
\end{aligned}
\qquad \qquad
\begin{aligned}
\commute{Z_m}{Z^n}
&= \delta^n_m \, H
   - \delta^n_m \, S_0
   + A^n_{~m}
\\
\commute{N_-}{B_+}
&= 0
\\
\commute{Z_m}{N_+}
&= - \epsilon_{mn} C^{1n}
\\
\commute{Z_m}{B_+}
&= C_{2m}
\\
\commute{B_-}{B_+}
&= H - J
\end{aligned}
\end{equation}

The quadratic Casimir of this subalgebra $\mathfrak{su}(4)$ is given by
\begin{equation}
\mathcal{C}_2 \left[ \mathfrak{su}(4) \right]
= S^{m^\prime}_{~n^\prime} S^{n^\prime}_{~m^\prime}
  + A^m_{~n} A^n_{~m}
  + \left( C^{m^\prime m} C_{m^\prime m}
           + C_{m^\prime m} C^{m^\prime m} \right)
  + J^2 \,.
\end{equation}


\section{Transformations between $SO(6,2)$ oscillators $c_i$ and
$SO^*(8)$ oscillators $a_m$, $b_m$}
\label{app:bogoliubov}

The minrep of $SO(6,2)$ can be related to the minrep of $SO^*(8)$ very simply
by rewriting the oscillators $a_m$, $b_m$ and $a^m$, $b^m$ in terms of $c_i$
and $c_i^\dag$ as follows:
\begin{equation}
\begin{aligned}
a_1 &= - \frac{i}{\sqrt{2}} \left( c_1 + i \, c_2 \right)
\\
a_2 &= \frac{1}{\sqrt{2}} \left( c_3 + i \, c_4 \right)
\\
b_1 &= \frac{1}{\sqrt{2}} \left( c_3 - i \, c_4 \right)
\\
b_2 &= - \frac{i}{\sqrt{2}} \left( c_1 - i \, c_2 \right)
\end{aligned}
\qquad \qquad \qquad
\begin{aligned}
a^1 &= \frac{i}{\sqrt{2}} \left( c_1^\dag - i \, c_2^\dag \right)
\\
a^2 &= \frac{1}{\sqrt{2}} \left( c_3^\dag - i \, c_4^\dag \right)
\\
b^1 &= \frac{1}{\sqrt{2}} \left( c_3^\dag + i \, c_4^\dag \right)
\\
b^2 &= \frac{i}{\sqrt{2}} \left( c_1^\dag + i \, c_2^\dag \right)
\end{aligned}
\end{equation}

Then it is easy to see that we have the following mapping between the
subalgebra $\mathfrak{su}(2)_L \oplus \mathfrak{su}(2)_R \oplus
\mathfrak{su}(1,1)_{\mathfrak{R}}$ of $\mathfrak{so}(6,2)$ and the subalgebra
$\mathfrak{su}(2)_A \oplus \mathfrak{su}(2)_S \oplus \mathfrak{su}(1,1)_N$ of
$\mathfrak{so}^*(8)$:
\begin{equation}
\begin{aligned}
L_3 &\longrightarrow A_0
\\
L_+ &\longrightarrow i \, A_+
\\
L_- &\longrightarrow - i \, A_-
\end{aligned}
\qquad \qquad
\begin{aligned}
R_3 &\longrightarrow S_0
\\
R_+ &\longrightarrow i \, S_+
\\
R_- &\longrightarrow - i \, S_-
\end{aligned}
\qquad \qquad
\begin{aligned}
\mathfrak{R}_3 &\longrightarrow N_0
\\
\mathfrak{R}_+ &\longrightarrow - i \, N_+
\\
\mathfrak{R}_- &\longrightarrow i \, N_-
\end{aligned}
\end{equation}
The relation between $\mathfrak{su}(1,1)_{\mathfrak{L}}$ of
$\mathfrak{so}(6,2)$ and $\mathfrak{su}(1,1)_K$ of $\mathfrak{so}^*(8)$ is
quite straight forward.


\section{The superalgebra $\mathfrak{osp}(8^*|2N)$ in the 5-grading with
respect to the subsuperalgebra $\mathfrak{osp}(4^*|2N)$}
\label{OSp(8*|2N)-5Gr}

We gave the explicit realization of the superalgebra $\mathfrak{osp}(8^*|2N)$
in the 5-grading with respect to the subsuperalgebra $\mathfrak{osp}(4^*|2N)$
in section \ref{minrepOSp(8*|2N)-5Gr}. In this appendix, we provide the
commutation relations between the generators of $\mathfrak{osp}(8^*|2N)$ in
this basis.

The (super-)commutation relations between the generators of grade zero
subspace $\mathfrak{g}^{(0)} = \mathfrak{osp}(4^*|2N)$ are as follows:
\begin{equation}
\begin{aligned}
\anticommute{\Pi_{mr}}{\overline{\Pi}^{ns}}
&= \delta^s_r \, A^n_{~m}
   - \delta^n_m \, M^s_{~r}
   + \delta^s_r \, \delta^n_m \, N_0
\\
\anticommute{\Sigma_m^{~r}}{\overline{\Sigma}^n_{~s}}
&= \delta^r_s \, A^n_{~m}
   + \delta^n_m \, M^r_{~s}
   + \delta^r_s \, \delta^n_m \, N_0
\\
\commute{A^m_{~n}}{\Pi_{kr}}
&= - \delta^m_k \, \Pi_{nr}
   + \frac{1}{2} \delta^m_n \, \Pi_{kr}
\\
\commute{A^m_{~n}}{\Sigma_k^{~r}}
&= - \delta^m_k \, \Sigma_n^{~r}
   + \frac{1}{2} \delta^m_n \, \Sigma_k^{~r}
\\
\commute{S_{rs}}{\overline{\Pi}^{mt}}
&= \delta^t_s \, \overline{\Sigma}^m_{~r}
   + \delta^t_r \, \overline{\Sigma}^m_{~s}
\\
\commute{S_{rs}}{\Sigma_m^{~t}}
&= - \delta^t_r \, \Pi_{ms}
   - \delta^t_s \, \Pi_{mr}
\\
\commute{M^r_{~s}}{\Pi_{mt}}
&= - \delta^r_t \, \Pi_{ms}
\\
\commute{M^r_{~s}}{\Sigma_m^{~t}}
&= \delta^t_s \, \Sigma_m^{~r}
\\
\commute{S_{rs}}{\Pi_{mt}}
&= 0
\end{aligned}
\qquad \qquad
\begin{aligned}
\anticommute{\Pi_{mr}}{\Sigma_n^{~s}}
&= \epsilon_{mn} \, \delta^s_{~r} \, N_-
\\
\anticommute{\Pi_{mr}}{\overline{\Sigma}^n_{~s}}
&= - \delta^n_m \, S_{rs}
\\
\commute{N_+}{\Pi_{mr}}
&= - \epsilon_{mn} \, \overline{\Sigma}^n_{~r}
\\
\commute{N_+}{\Sigma_m^{~r}}
&= \epsilon_{mn} \, \overline{\Pi}^{nr}
\\
\commute{N_-}{\Pi_{mr}}
&= 0
\\
\commute{N_-}{\Sigma_m^{~r}}
&= 0
\\
\commute{N_0}{\Pi_{mr}}
&= - \frac{1}{2} \, \Pi_{mr}
\\
\commute{N_0}{\Sigma_m^{~r}}
&= - \frac{1}{2} \, \Sigma_m^{~r}
\\
\commute{S_{rs}}{\overline{\Sigma}^m_{~t}}
&= 0
\end{aligned}
\end{equation}

The anticommutators between the supersymmetry generators in
$\mathfrak{g}^{(-1)}$ and  $\mathfrak{g}^{(+1)}$ (given in equations
(\ref{5GsusyGr-1}) and (\ref{5GsusyGr+1})) close into the bosonic generators
in $\mathfrak{g}^{(0)}$:
\begin{equation}
\begin{aligned}
\anticommute{Q_r}{\widetilde{Q}_s}
&= 0
\\
\anticommute{Q_r}{\widetilde{Q}^s}
&= - \delta^s_r \, \Delta
   - 2 i \, \delta^s_r \, T_0
   + 2 i \, M^s_{~r}
\\
\anticommute{S_r}{\widetilde{S}_s}
&= 0
\\
\anticommute{S_r}{\widetilde{S}^s}
&= - \delta^s_r \, \Delta
   + 2 i \, \delta^s_r \, T_0
   + 2 i \, M^s_{~r}
\end{aligned}
\qquad \qquad
\begin{aligned}
\anticommute{Q_r}{\widetilde{S}_s}
&= - 2 i \, S_{rs}
\\
\anticommute{Q_r}{\widetilde{S}^s}
&= - 2 i \, \delta^s_r \, T_-
\\
\anticommute{S_r}{\widetilde{Q}_s}
&= + 2 i \, S_{rs}
\\
\anticommute{S_r}{\widetilde{Q}^s}
&= - 2 i \, \delta^s_r \, T_+
\end{aligned}
\end{equation}

Finally, the commutators between the bosonic (even) and fermionic (odd)
generators of $\mathfrak{g}^{(-1)}$ and $\mathfrak{g}^{(+1)}$ subspaces close
into the fermionic (odd) generators of $\mathfrak{g}^{(0)}$:
\begin{equation}
\begin{aligned}
\commute{U_m}{\widetilde{Q}_r}
&= 0
\\
\commute{U_m}{\widetilde{Q}^r}
&= - 2 i \, \Sigma_m^{~r}
\\
\commute{U_m}{\widetilde{S}_r}
&= - 2 i \, \Pi_{mr}
\\
\commute{U_m}{\widetilde{S}^r}
&= 0
\end{aligned}
\qquad \qquad
\begin{aligned}
\commute{V_m}{\widetilde{Q}_r}
&= + 2 i \, \Pi_{mr}
\\
\commute{V_m}{\widetilde{Q}^r}
&= 0
\\
\commute{V_m}{\widetilde{S}_r}
&= 0
\\
\commute{V_m}{\widetilde{S}^r}
&= - 2 i \, \Sigma_m^{~r}
\end{aligned}
\end{equation}
\begin{equation}
\begin{aligned}
\commute{Q_r}{\widetilde{U}_m}
&= 0
\\
\commute{Q^r}{\widetilde{U}_m}
&= - 2 i \, \Sigma_m^{~r}
\\
\commute{S_r}{\widetilde{U}_m}
&= - 2 i \, \Pi_{mr}
\\
\commute{S^r}{\widetilde{U}_m}
&= 0
\end{aligned}
\qquad \qquad
\begin{aligned}
\commute{Q_r}{\widetilde{V}_m}
&= + 2 i \, \Pi_{mr}
\\
\commute{Q^r}{\widetilde{V}_m}
&= 0
\\
\commute{S_r}{\widetilde{V}_m}
&= 0
\\
\commute{S^r}{\widetilde{V}_m}
&= - 2 i \, \Sigma_m^{~r}
\end{aligned}
\end{equation}


\section{The superalgebra $\mathfrak{osp}(8^*|2N)$ in the 3-grading with
respect to the subsuperalgebra $\mathfrak{u}(4|N)$}
\label{OSp(8*|2N)-3Gr}

As shown in section \ref{minrepOSp(8*|2N)-3Gr}, the superalgebra
$\mathfrak{osp}(8^*|2N)$ has a 3-graded decomposition with respect to the
subsuperalgebra $\mathfrak{u}(4|N)$. In this appendix we give the explicit
form of the remaining bosonic and supersymmetry generators and some useful
(super-)commutation relations among them.

The commutators $\commute{\mathfrak{C}^-}{\mathfrak{C}^+}$ close into
$\mathfrak{C}^0$:
\begin{equation}
\begin{aligned}
\commute{Y_m}{Y^n}
&= \delta^n_m \, H_B
   + \delta^n_m \, T_0
   + A^n_{~m}
\\
\commute{Y_m}{Z^n}
&= \delta^n_m \, T_-
\\
\commute{Y_m}{N_+}
&= + \epsilon_{mn} \, C^{2n}
\\
\commute{Y_m}{B_+}
&= C_{1m}
\\
\commute{N_-}{N_+}
&=  H_B + J
\end{aligned}
\qquad \qquad
\begin{aligned}
\commute{Z_m}{Z^n}
&= \delta^n_m \, H_B
   - \delta^n_m \, T_0
   + A^n_{~m}
\\
\commute{N_-}{B_+}
&= 0
\\
\commute{Z_m}{N_+}
&= - \epsilon_{mn} C^{1n}
\\
\commute{Z_m}{B_+}
&= C_{2m}
\\
\commute{B_-}{B_+}
&=  H_B - J
\end{aligned}
\end{equation}
where the generators $C^{mn}$ and $C_{mn}$ are coset generators $SU(4) \,/\,
\left[ SU(2)_T \times SU(2)_A \times U(1)_J \right]$ defined below.

The $\mathfrak{su}(4|N)$ part of $\mathfrak{C}^0$ has an even subalgebra
$\mathfrak{su}(4|N) \supset \mathfrak{su}(4) \oplus \mathfrak{su}(N) \oplus
\mathfrak{u}(1)_D$, where the $\mathfrak{u}(1)_D$ charge and
$\mathfrak{su}(N)$ generators are given by
\begin{equation}
\begin{split}
D
&= \frac{1}{2} \left( K_+ + K_- \right) + \frac{2}{N} \, M_0
\\
\widetilde{M}^r_{~s}
&= \alpha^r \alpha_s - \beta_s \beta^r - \frac{2}{N} \, \delta^r_s \, M_0
\end{split}
\end{equation}

The generators of $\mathfrak{su}(4)$, in its $\mathfrak{su}(4) \supset
\mathfrak{su}(2)_T \oplus \mathfrak{su}(2)_A \oplus \mathfrak{u}(1)_J$
decomposition, are realized as follows:
\begin{subequations}
\begin{equation}
\begin{aligned}
T_+ &= a^m b_m + \alpha^r \beta_r
\\
T_- &= b^m a_m + \beta^r \alpha_r
\\
T_0 &= \frac{1}{2} \left( N_a - N_b + N_\alpha - N_\beta \right)
\end{aligned}
\qquad \qquad
\begin{aligned}
A_+ &= a^1 a_2 + b^1 b_2
\\
A_- &= a_1 a^2 + b_1 b^2
\\
A_0 &= \frac{1}{2} \left( a^1 a_1 - a^2 a_2 + b^1 b_1 - b^2 b_2 \right)
\end{aligned}
\end{equation}
\begin{equation}
\begin{split}
J &= N_0 - \frac{1}{2} \left( K_+ + K_- \right)
\\
  &= - \frac{1}{4} \left( x^2 + p^2 \right)
     - \frac{1}{8 \, x^2} \left( 8 \, \mathcal{T}^2 + \frac{3}{2} \right)
     + \frac{1}{2} \left( N_a + N_b \right) + 1
\end{split}
\end{equation}
\begin{equation}
\begin{split}
C_{1m} &= \frac{1}{2} \left( U_m + i \, \widetilde{U}_m \right)
        = \frac{1}{2} \left( x - i \, p \right) a_m
          - \frac{1}{x}
            \left[ \left( T_0 + \frac{3}{4} \right) a_m + T_- b_m \right]
\\
C^{1m} &= \frac{1}{2} \left( U^m - i \, \widetilde{U}^m \right)
        = \frac{1}{2} \left( x + i \, p \right) a^m
          - \frac{1}{x}
            \left[ \left( T_0 - \frac{3}{4} \right) a^m + T_+ b^m \right]
\\
C_{2m} &= \frac{1}{2} \left( V_m + i \, \widetilde{V}_m \right)
        = \frac{1}{2} \left( x - i \, p \right) b_m
          + \frac{1}{x}
            \left[ \left( T_0 - \frac{3}{4} \right) b_m - T_+ a_m \right]
\\
C^{2m} &= \frac{1}{2} \left( V^m - i \, \widetilde{V}^m \right)
        = \frac{1}{2} \left( x + i \, p \right) b^m
          + \frac{1}{x}
            \left[ \left( T_0 + \frac{3}{4} \right) b^m - T_- a^m \right]
\end{split}
\end{equation}
\end{subequations}

One half of total supersymmetry generators of $\mathfrak{osp}(8^*|2N)$
belong to grade zero space, as part of the subsuperalgebra
$\mathfrak{su}(4|N)$. Below we list these $8 N$ supersymmetry generators:
\begin{equation}
\begin{split}
\widetilde{\mathfrak{Q}}_r
&= \frac{1}{2} \left( Q_r + i \, \widetilde{Q}_r \right)
 = \frac{1}{2} \left( x - i \, p \right) \alpha_r
   - \frac{1}{x}
     \left[
      \left( T_0 + \frac{3}{4} \right) \alpha_r + T_- \beta_r
     \right]
\\
\widetilde{\mathfrak{Q}}^r
&= \frac{1}{2} \left( Q^r - i \, \widetilde{Q}^r \right)
 = \frac{1}{2} \left( x + i \, p \right) \alpha^r
   - \frac{1}{x}
     \left[
      \left( T_0 - \frac{3}{4} \right) \alpha^r + T_+ \beta^r
     \right]
\\
\widetilde{\mathfrak{S}}_r
&= \frac{1}{2} \left( S_r + i \, \widetilde{S}_r \right)
 = \frac{1}{2} \left( x - i \, p \right) \beta_r
   + \frac{1}{x}
     \left[
      \left( T_0 - \frac{3}{4} \right) \beta_r - T_+ \alpha_r
     \right]
\\
\widetilde{\mathfrak{S}}^r
&= \frac{1}{2} \left( S^r - i \, \widetilde{S}^r \right)
 = \frac{1}{2} \left( x + i \, p \right) \beta^r
   + \frac{1}{x}
     \left[
      \left( T_0 + \frac{3}{4} \right) \beta^r - T_- \alpha^r
     \right]
\\
\widetilde{\Sigma}_m^{~r}
&= \Sigma_m^{~r}
 = a_m \alpha^r + b_m \beta^r
\\
\widetilde{\Sigma}^m_{~r}
&= \overline{\Sigma}^m_{~r}
 = a^m \alpha_r + b^m \beta_r
\end{split}
\end{equation}

Under supercommutation, they close in to the bosonic generators of
$\mathfrak{su}(4|N)$:
\begin{equation}
\begin{aligned}
\anticommute{\widetilde{\mathfrak{Q}}_r}{\widetilde{\mathfrak{Q}}^s}
&= - \delta^s_r \, T_0 + \widetilde{M}^s_{~r} + \delta^s_r \, D
\\
\anticommute{\widetilde{\mathfrak{S}}_r}{\widetilde{\mathfrak{S}}^s}
&= + \delta^s_r \, T_0 + \widetilde{M}^s_{~r} + \delta^s_r \, D
\\
\anticommute{\widetilde{\Sigma}_m^{~r}}{\widetilde{\Sigma}^n_{~s}}
&= \delta^r_s \, A^n_{~m}
   + \delta^n_m \, \widetilde{M}^r_{~s}
   + \delta^r_s \, \delta^n_m \, D
   + \delta^r_s \, \delta^n_m \, J
\end{aligned}
\qquad \qquad
\begin{aligned}
\anticommute{\widetilde{\mathfrak{Q}}_r}{\widetilde{\mathfrak{S}}^s}
&= - \delta^s_r \, T_-
\\
\anticommute{\widetilde{\Sigma}_m^{~r}}{\widetilde{\mathfrak{Q}}_s}
&= \delta^r_s \, C_{1m}
\\
\anticommute{\widetilde{\Sigma}_m^{~r}}{\widetilde{\mathfrak{S}}_s}
&= \delta^r_s \, C_{2m}
\end{aligned}
\end{equation}

These supersymmetry generators in $\mathfrak{C}^{\pm}$ satisfy the following
(anti-)commutation relations:
\begin{equation}
\begin{split}
\anticommute{\mathfrak{Q}_r}{\mathfrak{Q}^s}
&= + \delta^s_r \, T_0
   - \widetilde{M}^s_{~r}
   + \delta^s_r \, H_\odot
   - \frac{2}{N} \, \delta^s_r \, M_0
\\
\anticommute{\mathfrak{S}_r}{\mathfrak{S}^s}
&= - \delta^s_r \, T_0
   - \widetilde{M}^s_{~r}
   + \delta^s_r \, H_\odot
   - \frac{2}{N} \, \delta^s_r \, M_0
\\
\anticommute{\Pi_{mr}}{\overline{\Pi}^{ns}}
&= \delta^s_r \, A^n_m
   - \delta^n_m \, \widetilde{M}^s_{~r}
   + \delta^n_m \, \delta^s_r \, N_0
   - \frac{2}{N} \, \delta^n_m \, \delta^s_r \, M_0
\\
\anticommute{\mathfrak{Q}_r}{\mathfrak{S}^s}
&= \delta^s_r \, T_-
\\
\anticommute{\Pi_{mr}}{\mathfrak{Q}^s}
&= - \delta^s_r \, C_{2m}
\\
\anticommute{\Pi_{mr}}{\mathfrak{S}^s}
&= + \delta^s_r \, C_{1m}
\end{split}
\end{equation}

The commutators between bosonic operators in $\mathfrak{C}^-$ and
supersymmetry generators in $\mathfrak{C}^+$ are as follows:
\begin{equation}
\begin{aligned}
\commute{Y_n}{\mathfrak{Q}^r}
 &= \widetilde{\Sigma}_n^{~r}
\\
\commute{Z_n}{\mathfrak{Q}^r}
 &= 0
\\
\commute{N_-}{\mathfrak{Q}^r}
 &= 0
\\
\commute{B_-}{\mathfrak{Q}^r}
 &= \widetilde{\mathfrak{Q}}^r
 \\
\commute{S_{st}}{\mathfrak{Q}^r}
 &= - 2 \, \delta^r_{(s} \, \widetilde{\mathfrak{S}}_{t)}
\end{aligned}
\quad
\begin{aligned}
\commute{Y_n}{\mathfrak{S}^r}
 &= 0
\\
\commute{Z_n}{\mathfrak{S}^r}
 &= \widetilde{\Sigma}_n^{~r}
\\
\commute{N_-}{\mathfrak{S}^r}
 &= 0
\\
\commute{B_-}{\mathfrak{S}^r}
 &= \widetilde{\mathfrak{S}}^r
 \\
\commute{S_{st}}{\mathfrak{S}^r}
 &= + 2 \, \delta^r_{(s} \, \widetilde{\mathfrak{Q}}_{t)}
\end{aligned}
\quad
\begin{aligned}
\commute{Y_n}{\overline{\Pi}^{mr}}
 &= + \delta^m_n \, \widetilde{\mathfrak{S}}^r
\\
\commute{Z_n}{\overline{\Pi}^{mr}}
 &= - \delta^m_n \, \widetilde{\mathfrak{Q}}^r
\\
\commute{N_-}{\overline{\Pi}^{mr}}
 &= \epsilon^{mn} \, \widetilde{\Sigma}_n^{~r}
\\
\commute{B_-}{\overline{\Pi}^{mr}}
 &= 0
 \\
\commute{S_{st}}{\overline{\Pi}^{mr}}
 &= 2 \, \delta^r_{(s} \, \widetilde{\Sigma}^m_{~t)}
\end{aligned}
\end{equation}
Note that we have used the notation ``$(st)$'' to indicate symmetrization of
indices $s$ and $t$ with weight $1$.

The anticommutators of supersymmetry generators in $\mathfrak{C}^0$ and
those in $\mathfrak{C}^+$ can be written as
\begin{equation}
\begin{aligned}
\anticommute{\widetilde{\mathfrak{Q}}_r}{\mathfrak{Q}^s}
&= \delta^s_r \, B_+
\\
\anticommute{\widetilde{\mathfrak{Q}}^r}{\mathfrak{Q}^s}
&= 0
\\
\anticommute{\widetilde{\mathfrak{S}}_r}{\mathfrak{Q}^s}
&= 0
\\
\anticommute{\widetilde{\mathfrak{S}}^r}{\mathfrak{Q}^s}
&= + S^{rs}
\\
\anticommute{\widetilde{\Sigma}^m_{~r}}{\mathfrak{Q}^s}
&= + \delta^s_r \, Y^m
\\
\anticommute{\widetilde{\Sigma}_m^{~r}}{\mathfrak{Q}^s}
&= 0
\end{aligned}
\qquad \quad
\begin{aligned}
\anticommute{\widetilde{\mathfrak{Q}}_r}{\mathfrak{S}^s}
&= 0
\\
\anticommute{\widetilde{\mathfrak{Q}}^r}{\mathfrak{S}^s}
&= - S^{rs}
\\
\anticommute{\widetilde{\mathfrak{S}}_r}{\mathfrak{S}^s}
&= \delta^s_r \, B_+
\\
\anticommute{\widetilde{\mathfrak{S}}^r}{\mathfrak{S}^s}
&= 0
\\
\anticommute{\widetilde{\Sigma}^m_{~r}}{\mathfrak{S}^s}
&= + \delta^s_r \, Z^m
\\
\anticommute{\widetilde{\Sigma}_m^{~r}}{\mathfrak{S}^s}
&= 0
\end{aligned}
\qquad \quad
\begin{aligned}
\anticommute{\widetilde{\mathfrak{Q}}_r}{\overline{\Pi}^{ns}}
&= - \delta^s_r \, Z^n
\\
\anticommute{\widetilde{\mathfrak{Q}}^r}{\overline{\Pi}^{ns}}
&= 0
\\
\anticommute{\widetilde{\mathfrak{S}}_r}{\overline{\Pi}^{ns}}
&= + \delta^s_r \, Y^n
\\
\anticommute{\widetilde{\mathfrak{S}}^r}{\overline{\Pi}^{ns}}
&= 0
\\
\anticommute{\widetilde{\Sigma}^m_{~r}}{\overline{\Pi}^{ns}}
&= - \epsilon^{mn} \, \delta^s_r \, N_+
\\
\anticommute{\widetilde{\Sigma}_m^{~r}}{\overline{\Pi}^{ns}}
&= - \delta^n_m \, S^{rs}
\end{aligned}
\end{equation}


\providecommand{\href}[2]{#2}\begingroup\raggedright\endgroup


\end{document}